\documentclass[useAMS,usenatbib]{mn2e}
\usepackage{times}
\usepackage{graphicx}
\usepackage{amssymb}
\usepackage{natbib}
\usepackage{color,ulem}
\bibpunct{(}{)}{;}{a}{}{,}

\title[Galaxy populations in Antlia - III]{Galaxy populations in the Antlia cluster - 
III. Properties of faint early-type galaxies
\thanks{Based on observations carried out at the Cerro Tololo Inter--American 
  Observatory (Chile), at Las Campanas Observatory (Chile), and 
  at the European Southern Observatory, Paranal (Chile). 
  Also based on observations obtained at the Gemini Observatory, which is 
  operated by the Association of Universities for Research in Astronomy, Inc., 
  under a cooperative agreement with the NSF on behalf of the Gemini 
  partnership: the National Science Foundation (United States), the Science 
  and Technology Facilities Council (United Kingdom), the National Research 
  Council (Canada), CONICYT (Chile), the Australian Research Council 
  (Australia), Ministerio da Ciencia e Tecnologia (Brazil) and Ministerio de 
  Ciencia, Tecnolog\'ia e Innovaci\'on Productiva (Argentina).
  }
}

\author[Smith Castelli et al.]{Anal\'ia V. Smith Castelli$^{1,2,3}$
\thanks{E-mail:\,asmith@fcaglp.unlp.edu.ar}
, Sergio A. Cellone$^{1,2,3}$, Favio R. Faifer$^{1,2,3}$, Lilia P. Bassino$^{1,2,3}$,
\newauthor Tom Richtler$^{4}$, Gisela A. Romero$^{3,5}$, Juan Pablo Calder\'on$^{1,2,3}$
and Juan Pablo Caso$^{1,2,3}$\\
$^{1}$Consejo Nacional de Investigaciones Cient\'ificas y T\'ecnicas, Rivadavia 1917, Buenos Aires, Argentina\\
$^{2}$Instituto de Astrof\'isica de La Plata (CCT La Plata - CONICET - UNLP), La Plata, Argentina \\
$^{3}$Facultad de Ciencias Astron\'omicas y Geof\'{\i}sicas,
      Universidad Nacional de La Plata,
      Paseo del Bosque, B1900FWA La Plata,
      Argentina\\ 
$^{4}$Departamento de Astronom\'ia, Universidad de Concepci\'on, Casilla 160-C, 
Concepci\'on, Chile\\
$^{5}$Departamento de F\'isica y Astronom\'ia, Universidad de Valpara\'iso, Valpara\'iso, Chile}

\begin{document}

\date{Accepted . Received ; in original form }


\maketitle

\label{firstpage}

\begin{abstract}

We present a new analysis of the early-type galaxy population in the central
region of the Antlia cluster, focusing on the faint systems like dwarf
ellipticals (dE) and dwarf spheroidals (dSph).  The colour--magnitude
relation (CMR) and the relation between luminosity and mean effective
surface brightness for galaxies in the central region of Antlia have been
previously studied in Paper\,I of the present series. Now we confirm 22
early-type galaxies as Antlia members, using GEMINI--GMOS and MAGELLAN--MIKE
spectra. Among them, 15 are dEs from the FS90 Antlia Group catalogue, 2
belong to the rare type of compact ellipticals (cE), and 5 are new faint
dwarfs that had never been catalogued before. In addition, we present 16
newly identified low surface brightness galaxy candidates, almost half of
them displaying morphologies consistent with being Antlia's counterparts of
Local Group dSphs, that extend the faint luminosity limit of our study down
to $M_B=-10.1~(B_{\rm T} = 22.6)$ mag.  With these new data, we built 
an improved CMR in the Washington photometric system, i.e. integrated $T_1$ 
magnitudes versus $(C-T_1)$ colours, which extends $\sim 4$\,mag faintwards 
the limit of spectroscopically confirmed Antlia members. When only confirmed 
early-type members are considered, this relation extends over 10 mag in 
luminosity with no apparent change in slope or increase in colour dispersion 
towards its faint end. The intrinsic colour scatter of the relation is 
compared with those reported for other clusters of galaxies; we argue that 
it is likely that the large scatter of the CMR, usually reported at faint 
magnitudes, is mostly due to photometric errors along with an improper 
membership/morphological classification.
The distinct behaviour of the luminosity versus mean effective surface
brightness relation at the bright and faint ends is analyzed, while it is
confirmed that dE galaxies on the same relation present a very similar
effective radius, regardless of their colour.  The projected spatial
distribution of the member sample confirms the existence of two groups in
Antlia, each one dominated by a giant elliptical galaxy and with one cE
located close to each giant. Size and position, with respect to massive 
galaxies, of the dSph candidates are estimated and compared to Local Group 
couterparts.

\end{abstract}



\section{Introduction}

\label{intro}

The fact that early-type galaxies in clusters and groups define sequences in
the colour-magnitude and surface brightness-luminosity diagrams has been
known for a long time (\citealp[e.g.][]{B59,S72,VS77}; \citealp{SV78a};
\citealp[][b]{K77a}; \citealp{CB87,FB94}). These relations are expected to
provide clues about the evolutionary status of the galaxies that follow them
and they seem to be universal. Such universality led several authors to
suggest their use as reliable distance indicators
\citep[e.g.][]{S72,BJ98,C99} and to set membership criteria
\citep[e.g.][]{Chi06}.

The faint end of these photometric relations is populated by very faint  
dwarf elliptical (dE) and dwarf spheroidal (dSph) galaxies.
Despite morphological criteria have been proven
useful for identifying dwarf galaxies in nearby clusters and groups  
\citep*[e.g.][hereafter FS90]{B85,F89,FS90}, background  
galaxies may mimic dwarf morphologies if they are not observed with  
sufficient spatial resolution (see, for example, object 27 in  
\citealp*{Con03} and \citealp{PC08}). This fact may introduce a significant  
contamination into the observed relations and may cause disagreements in  
their interpretation. Examples of discrepancies that deserve some analysis  
are those related with the linearity and the scatter of the  
colour--magnitude relation (CMR).  

Most CMRs show no perceptible change of slope from luminous galaxies to the  
dwarf regime (\citealp*{S97,TK99,LC04,An06}; \citealp{SC08a}, hereafter  
Paper\,I; \citealp*{P09}) except, perhaps, at the very bright end (see  
fig.\,8b in \citealp*{M08} and fig.\,3 in \citealp*{M09}). However, non-linear  
trends for the CMR of the Virgo cluster have been found \citep{F06,JL09}. In  
addition, there is a considerable increase in the scatter of all CMRs towards  
their faint end in comparison to that displayed by the brightest galaxies. It  
is still under discussion whether such increase has a physical origin  
pointing to differences in ages, metal abundances or formation scenarios 
among dwarf galaxies \citep*[e.g.][]{C02,Con03,PC08,J11}, or whether it can 
be  accounted for by photometric errors \citep[e.g.][]{S97,M08,M09,JL09} 
and/or background galaxies contamination \citep[e.g.][]{LC04}. 

Now, we have spectroscopically confirmed new dwarf Antlia members and 
we have  identified previously non catalogued dwarf candidates.  
We  revisit the relations analyzed in Paper\,I, i.e. the colour--magnitude  
and the luminosity--mean effective surface brightness relations,  
regarding particularly the behaviour of their faint ends. In order to  
analyze the structure of the Antlia cluster, we study the projected  
spatial distribution of the whole galaxy sample.    

In addition, we have confirmed the existence of two compact elliptical (cE)  
galaxies in Antlia (\citealp{SC08b}, hereafter Paper\,II; \citealp{SC09}). 
These are rare low-mass systems; about only a dozen have been identified 
within a distance of $\sim 100$ Mpc (\citealp{Chilli10} and
references therein; \citealp{Huxor11}) and a
similar amount at larger distances \citep{Chilli09}. We include these  
two peculiar  confirmed members in our analysis thus contributing to the 
discussion of the 
alleged dichotomy between bright and faint elliptical galaxies 
\citep[e.g.][]{Graham03,Kormendy09,Graham10}, as cEs are thought to be
either the extension to low luminosities of the family of giant ellipticals 
(\citealp{Kormendy09} and references therein) or not (\citealp{Graham02}). 

Finally, physical characteristics of the dSph candidates are compared  
to galaxies of the same type in other groups and clusters. The dSphs are  
particularly interesting as they are supposed to be the faintest systems  
containing dark matter and can be used to constrain cold dark matter (CDM)  
models of galaxy formation (e.g. \citealp*{Penarrubia08}, 
\citealp{Penarrubia09}). Due to their 
low surface brightness, they have mostly been studied in detail within the 
Local Group, like the ones surrounding the Milky Way or Andromeda galaxies  
\citep[e.g.][]{Belokurov08,Kalirai10}. 

This paper is based on photometric data obtained from CTIO-MOSAIC  
and VLT-FORS1 images of the central region of Antlia, and radial velocities   
measured from GEMINI-GMOS and MAGELLAN-MIKE spectra. Antlia  
is the third nearest well populated galaxy cluster after Virgo and  
Fornax. Before our studies on Antlia (\citealp*{dir03,Bass08}; Paper\,I;  
Paper\,II), the photographic work of FS90 was the last major effort  
devoted to study the faint galaxy content in this cluster. They identified, 
by visual inspection on photographic plates, 375 galaxies in their Antlia 
Group Catalogue down to $M_B=-14.7~(B_T=18)$ mag. Among them, 252 are 
classified as dwarf galaxies or probable dwarf objects, 71 of which are found 
in our frames. Only 15 of these 71 galaxies had measured radial velocities at 
that moment. Our spectra have allowed us to add 23 new radial velocities to 
this sample (16 early-type and 3 late-type members plus 4 background galaxies), and to detect 5 previously unclassified new Antlia members. In addition, 
we have visually identified 16 unclassified galaxies displaying dE and dSph  
morphologies, extending the luminosity range of probable Antlia members  
down to $M_B=-10.1~(B_{\rm T}=22.6)$ mag.  

Throughout this paper we will adopt $(m-M)=32.73$ as the Antlia distance  
modulus \citep{dir03}. It corresponds to a distance of  
35.2 Mpc at which 1 arcsec subtends 170 pc. The paper is organized  
as follows. Section\,\ref{data} describes our observational data and  
Section\,\ref{sample} presents our galaxy sample. In Section\,\ref{results}  
we revisit the photometric relations of early-type Antlia galaxies  
including the new confirmed members and new dwarf galaxy candidates,  
and analyze the projected spatial distribution of the whole sample.  
Section\,\ref{discussion} presents a discussion of  
the results and Section\,\ref{conclusions} our conclusions.

\section{Observational Data}
\label{data}

\begin{figure}
\includegraphics[scale=0.43]{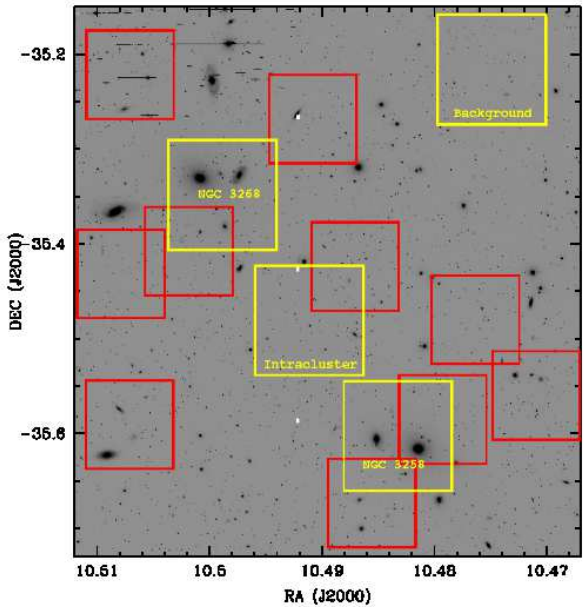}\\
\caption{$R$ image of the MOSAIC field corresponding to the centre of 
the Antlia cluster. Yellow squares show the positions of the four 
FORS1--VLT frames used to check part of our galaxy identifications. 
Red squares indicate the location of the ten GEMINI--GMOS fields 
used for spectroscopy (see text). North is up and east to the left.}
\label{mosaic_ch4}
\end{figure}

We have performed a visual search for new dwarf galaxy candidates on a 
Kron-Cousins $R$ image obtained with the MOSAIC camera (8 CCDs mosaic imager) 
mounted at the prime focus of the 4-m Blanco telescope at the Cerro 
Tololo Inter--American Observatory (CTIO), during 2002 April 4--5. This image, 
as well as a similar one taken in the $C$ band of the Washington photometric 
system, covers the central region of the Antlia cluster (see  
Fig.\,\ref{mosaic_ch4}). This observational material has already been used in  
Paper\,I and in Paper\,II. 

One pixel of the MOSAIC camera subtends 0.27 arcsec on the sky. This results 
in a field of view of $36 \times 36$ arcmin$^2$ (i.e. about $370 \times 370$ 
kpc$^2$ at the Antlia distance). The final $R$/$C$ images were obtained 
from the combination of three/seven exposures of 600 sec. The seeing FWHM on  
these final images is 1 arcsec for the $R$-image and 1.1 arcsec for the 
$C$-image. $R$ magnitudes were transformed into Washington $T_1$ magnitudes 
through the calibration given by \citet{dir03}.

Part of our galaxy identification was checked with FORS1--VLT images   
in the $V$ and $I$ bands. These images correspond to two fields centred  
on each of the Antlia dominant galaxies, NGC\,3258 and NGC\,3268,   
an {\it intracluster} field located between both giant galaxies, and a  
{\it background} field further away from them (see Fig.\,\ref{mosaic_ch4}).  
We refer to \citet{Bass08} for more details on the observations and reduction.  
In addition, we examined ten GEMINI--GMOS images obtained in the r\_G0326  
filter for designing spectroscopic masks (programs ID: GS-2008A-Q-56, PI: T.  
Richtler; ID: GS-2009A-Q-25, PI: L. Bassino; ID: GS-2010A-Q-21, PI: L.  
Bassino). Each image covers a field of view of $5.5 \times 5.5$ arcmin$^2$ and  
their scale is 0.146 arcsec pixel$^{-1}$. The total exposure time of
each image is 500 sec and the seeing ranges from 0.6 to 0.7 arcsec. 

We have also obtained GEMINI--GMOS multi-object spectra for the ten Antlia  
fields mentioned above (see Fig.~\ref{mosaic_ch4}), from which we measured  
radial velocities for several galaxies. The B600\_G5303 grating blazed at  
5000 \AA\,was used  with three different central wavelengths (5000, 5050  
and 5100 \AA) in order to fill in the CCD gaps. A slit width of 1 arcsec  
was selected; seeing was 0.5 - 0.6 arcsec in average. This configuration  
gives a wavelength coverage of 3500 - 7200 \AA\,depending on the positions  
of the slits, and a resolution (fwhm) of $\sim 4.6$\,\AA.  The total exposure  
times ranged between 2.5 and 3.7 hours. Data reduction was performed in a  
standard manner using the GEMINI.GMOS package within
IRAF\footnote{IRAF is distributed by the National Optical Astronomy  
Observatories, which are operated by the Association of Universities for  
Research in Astronomy, Inc., under cooperative agreement with the National  
Science Foundation.}.

In addition, MAGELLAN-MIKE echelle spectra of the two compact elliptical  
galaxy candidates presented in Paper\,II (namely, FS90\,110 and FS90\,192) 
were  obtained at the CLAY telescope of Las Campanas Observatory in 2009 March 
27 and 28 (program LCO-CNTAC09A\_042). Slits of $1 \times 5$ arcsec and binning 
$2 \times 3$ were used. The spectral coverage at the red side of the echelle 
spectrograph was 4900-10000 \AA, with a resolution (fwhm) of $\sim 0.35$\,\AA. 
For FS90\,110, two spectra of 900 sec and one of 2400 sec were obtained, and 
only one of 2400 sec for FS90\,192. The reduction was performed using a 
combination of the IRAF \textsf{imred.ccdred.echelle} package, and the 
\textsf{mtools} package written by Jack Baldwin for dealing with the tilted 
slits in MIKE spectra (available at the LCO website).

Using the GEMINI-GMOS spectra, radial velocities were measured 
with the IRAF task \textsf{fxcor} by means of cross-correlation against simple 
stellar population (SSP) templates suitably selected from the MILES database 
\citep{MILES}. In Table\,\ref{velocidades} we list the error weighted 
average values, and the corresponding errors, obtained for each galaxy with 
the four templates used ([Z/H]= -0.71, 0.0; Ages 7.4 and 10.0 Gyrs). In the 
case of galaxies showing bright emission lines, these were used to check and 
refine the previously measured radial velocities. For the cE galaxy FS90\,192 
we obtained the radial velocity value by fitting gaussian profiles with the 
task \textsf{splot} within IRAF, to H$_\alpha$ and the Ca\,II triplet lines.

\section{Characteristics of the Galaxy Sample} 
\label{sample} 

\subsection{Confirmed Antlia members from the FS90 catalogue} 
\label{FS90_members} 
 
We have obtained new radial velocities for 23 FS90 galaxies. Among them,  
20 have become new confirmed members of the Antlia cluster, and 3 
are background galaxies. We recall that we consider as Antlia members all 
galaxies with radial velocities in the range 1200 -- 4200 km s$^{-1}$ (see  
Section\,2 of Paper\,I for a justification). In Appendix~\ref{FS90_vr} we 
present their coordinates, FS90 morphological classification and membership 
status as well as the measured radial velocities (Table\,\ref{velocidades}). 
Logarithmically scaled $R$ images of these galaxies are shown in 
Fig.\,\ref{fig1_ch4}.
     
We recall that FS90 assigned to each galaxy listed in their Antlia catalogue 
a membership status: 1 for definite members, 2 for likely members and 3 for 
possible members. This assignment was based on morphological criteria as only 
6 per cent of the listed galaxies had radial velocity determinations at the 
moment of the catalogue publication. 

In general, we confirm the morphological/membership classification assigned
by FS90. The exceptions are two cE galaxies (FS90\,110 and FS90\,192)
that are new members of Antlia and were originally considered just as 
``possible" members, two galaxies originally classified as ``possible" and 
``likely" dE members, which turned out to be an Antlia member (FS90\,103) and
a background systems (FS90~205), and a confirmed member but with a different 
(irregular) morphology (FS90~221). As expected, in some objects we are able to 
see features like nuclei, spiral arms or knots of star formation, thanks to the 
spatial resolution of our CCD images. 

All status 1 galaxies of the spectroscopic sample became confirmed members,
while only two of the 20 spectroscopically confirmed Antlia members were
classified as status 3 galaxies by FS90 (FS90\,110 and FS90\,192, the new cE
Antlia galaxies). Adding up the 23 FS90 galaxies with new radial velocities 
presented in this paper with the 37 already presented in Paper I, we have the 
following statistics: cluster membership is confirmed for 100\% (31/31) of 
FS90 status 1 galaxies, 81\% (13/16) of status 2 ones, and 54\% (7/13) 
of status 3 objects. This agrees with previous works in the NGC\,5044 Group 
\citep{CB05, Men08}, showing that morphologically assigned memberships are 
still reliable (mainly for diffuse dwarfs) out to distances of $\sim 35$\,Mpc.

With the addition of the new data, we will revisit the photometric relations 
of early-type galaxies in the central region of Antlia considering 
spectroscopically confirmed FS90 members and FS90 galaxies with membership 
status 1 displaying early-type morphologies. This criterion will allow us to 
extend the CMR relation down to $M_{T_1} \approx -13$\,mag, minimizing 
contamination by background galaxies lacking radial velocities in this 
subsample. 

We have measured with the task ELLIPSE within IRAF some galaxies 
presented with SExtractor \citep{ba96} photometry in Paper\,I (see 
Table\,\ref{tabla_ELL}) in order to compare the measurements obtained with 
these two softwares (see Appendix\,\ref{ELL_SEX}). As bright galaxies were 
all measured with ELLIPSE due to saturation in their central regions 
(see Paper\,I for details), we will consider these new photometry in our 
diagrams and in the different fits performed to the photometric relations.  

\subsection{New dE members} 
\label{nuevas_confirmadas} 

We have obtained radial velocities for five new  Antlia members that were 
not in the FS90 catalogue. Their radial velocities and photometric parameters 
are given in Table\,\ref{tabla_ndgm}. In Fig.\,\ref{fig2_ch4} we show the  
logarithmically  scaled $R$ images of these galaxies. 

All of them show what seems to be a nucleus. This is a selection effect, as 
a bright central nucleus turns out to  be necessary for obtaining reliable 
radial velocities from absorption-line spectra of faint diffuse galaxies. Our 
spectroscopic data include other low-surface brightness objects without nuclei 
for which it was not possible to obtain a redshift. Their photometric data
were obtained through SExtractor or ELLIPSE, depending if the object was
detected by the former software or not.

The new members are not catalogued in NED. Therefore, we will refer  
to them with the acronym ANTL followed by the J2000  
coordinates, that is {\it ANTL Jhhmmss$-$ddmmss}, according to the IAU  
recommendations for designating new sources (see  
{\it http://vizier.u-strasbg.fr/Dic/how.html}).  

\subsection{New dE and dSph candidates without redshifts} 
\label{nuevas_candidatas}    

We have performed a visual search on the $R$ MOSAIC image for new dwarf
galaxy candidates based on morphological criteria. We looked for extended
objects displaying low-surface brightnesses, that do not show substructure 
or evidence of star formation in the $C$ MOSAIC frame. For those objects 
falling into the deeper FORS and GMOS frames, we have also checked their 
morphologies by examining their appearance in these deeper images.

As these galaxies are not included in NED, we will designate them in the
same way as new dE Antlia members.  We present their images in
Fig.\,\ref{fig_new_dwarfs}. We obtained integrated magnitudes and 
colours for all but one of the 16 new dwarf systems (see 
Table\,\ref{candidates}).  These measurements were
performed in the way described in section\,2 of Paper\,I through 
SExtractor and ELLIPSE, or with PHOT within IRAF. The seven candidates 
that have no structural parameters determinations are either too faint and 
diffuse, or are located near bright galaxies or stars which prevented
 a more exhaustive photometric analysis.   

In particular, we are interested in identifying counterparts of Local Group
dSph galaxies as they will represent the very faint end of the dwarf galaxy
population in Antlia, certainly missed in the FS90 Antlia catalogue.
Dwarf spheroidals are low luminosity ($M_V \gtrsim -13$) galaxies with
smooth appearance, that present no evidence of star formation 
\citep{Kalirai10}, have low optical surface brightness ($\mu_V>$ 22 mag
arcsec$^{-1}$) and no nucleus \citep{Gallagher94}.  We have identified 16
galaxies with morphologies consistent with faint early-type galaxies
belonging to the cluster. Among them, 7 match the dSphs description; they
are the ones designated with order numbers 4, 8, 12, 13, 14, 15 and 16 in
Table\,\ref{candidates}.

\section{Results} 
\label{results} 

\subsection{Colour-Magnitude Relation} 
\label{CMR} 

\begin{table*}
\caption{Results of least-square biweight fits $T_{1_0}=a+b\cdot(C-T_1)_0$ 
performed to the absorption and extinction corrected CMR of early-type 
definite members of Antlia (i.e. FS90 early-type status 1 objects and 
early-type galaxies with radial velocities). The first column indicates the 
different samples and the second column gives the number of data points. 
Slopes and zero-points given in the third and four columns correspond to 
reddening and extinction corrected mean CMRs. In the 
fifth column we show the observed scatter around the mean relation for each 
sample. In the sixth column we indicate the corresponding mean colour error
$\langle\epsilon\rangle$. Following \citet*[][see their eq.\,1]{T01}, we 
calculate the intrinsic scatter, given in the last column, as 
$\sigma_{intr}=\sqrt{\sigma_{obs}^2-\langle\epsilon\rangle^2}$, where 
$\langle\epsilon\rangle$ is the mean colour error of the sample. The adopted 
limiting magnitude to separate bright and dwarf galaxies ($T_1=14$ mag) 
corresponds to $M_V\sim -18$ mag \citep{G05}.} 
\begin{tabular}{lcccccc}
\\ 
\hline 
\\ 
\multicolumn{1}{c}{Sample} & Data & \multicolumn{1}{c}{$a$}  & $b$ & $\sigma_{\small(C-T_1)_0}$ & $\langle\epsilon_{(C-T_1)_0}\rangle$ & $\sigma_{(C-T_1)_0}$\\ 
\multicolumn{1}{c}{} &  & \multicolumn{1}{c}{}  &  & \scriptsize {\it observed} & & \scriptsize{\it intrinsic} \\ 
\hline 
\\ 
All definite+confirmed members           & 58 & $43.2\pm2.7$     & $-16.1\pm1.5$ & 0.10 & 0.03 & 0.09\\
Dwarf definite+confirmed members         & 43 & $44.7\pm5.7$     & $-17.0\pm3.5$ & 0.11 & 0.04 & 0.10\\
E definite+confirmed members             & 42 & $41.6\pm3.3$     & $-15.1\pm1.9$ & 0.10 & 0.04 & 0.09\\
All confirmed members          		 & 42 & $43.0\pm3.4$     & $-16.0\pm1.9$ & 0.09 & 0.03 & 0.08\\
Bright confirmed members                 & 15 & $~~44.5\pm10.4$  & $-16.7\pm4.8$ & 0.06 & 0.02 & 0.06\\
Dwarf confirmed members                  & 27 & $43.6\pm8.0$     & $-16.4\pm4.3$ & 0.10 & 0.03 & 0.09\\
Dwarf confirmed members no cEs           & 25 & $42.2\pm5.9$     & $-15.8\pm3.3$ & 0.08 & 0.03 & 0.07\\
E confirmed members                      & 26 & $41.2\pm4.2$     & $-14.8\pm2.3$ & 0.09 & 0.03 & 0.08\\
E confirmed members no cEs               & 24 & $39.7\pm2.9$     & $-14.2\pm1.6$ & 0.07 & 0.03 & 0.06\\
S0 confirmed members                     & 16 & $40.4\pm8.5$     & $-14.7\pm4.1$ & 0.08 & 0.02 & 0.07\\
\\
\hline 
\end{tabular} 
\label{ajustes} 
\end{table*}

\begin{figure}
\begin{center}
\includegraphics[scale=0.435]{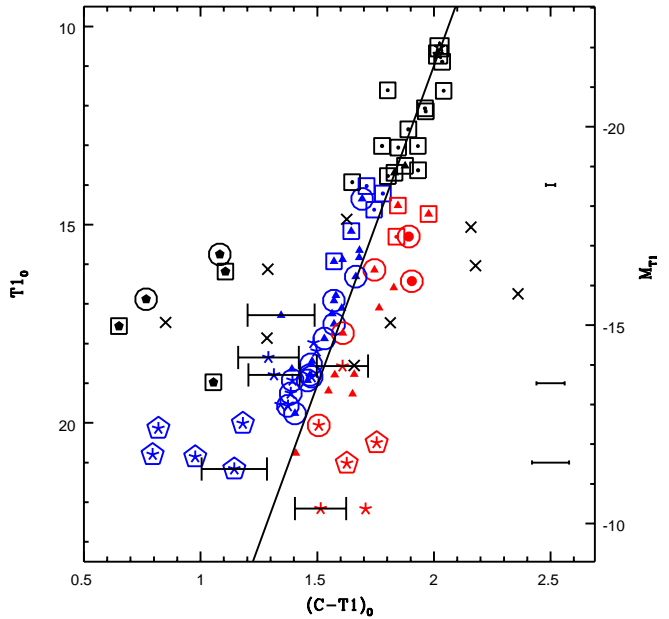}
\caption{Reddening and extinction corrected 
colour--magnitude diagram for FS90 Antlia definite and confirmed members
plus new dwarf galaxy candidates and members. {\it Big stars:} Antlia's giant 
ellipticals; {\it filled circles:} cEs; {\it filled triangles:} ellipticals;
{\it dots:} S0s; {\it filled pentagons:} dIs and BCDs; {\it asterisks:} new 
dwarf candidates, with those displaying dSph morphologies within open 
pentagons; {\it open squares:} Antlia confirmed members from Paper\,I; 
{\it open circles:} confirmed members from GEMINI and MAGELLAN data. The 
solid line shows the fit to all early-type galaxies in the sample. We have 
also plotted the location of FS90 confirmed background galaxies ({\it crosses}) 
in order to see the scatter introduced by these 
objects to the relation. The isolated error bars show typical colour errors 
corresponding to the range 10 $< T_1 <$ 18 (small, 0.02 mag), 18 $< T_1 <$ 20 
(medium, 0.06 mag) and 20 $< T_1 <$ 22 (large, 0.08 mag). For clarity, we only 
show individual errorbars for those objects with colour uncertainties larger 
than 0.1 mag. Red and blue symbols depict dwarf galaxies showing redder and 
bluer colours than the mean CMR, respectively.}
\label{CMR_esferoides}
\end{center}
\end{figure}

In Fig.\,\ref{CMR_esferoides} we show the CMR defined by FS90 early-type
galaxies that are spectroscopically confirmed Antlia members or were
considered as definite members (status 1) in the FS90 Antlia catalogue. We
have added the new dwarf galaxies presented in the previous section.
Several biweight fits to the observed relation have been performed, some of
them considering only spectroscopically confirmed members (see
Table\,\ref{ajustes}).  Following \citet{G05}, we will consider as dwarf
galaxies those with $M_V\sim -18$ mag, which corresponds to $T_1>14$ mag at
the Antlia distance \citep*[see table 3a from][]{F95}. In order to test the
results found by \citet{Barazza09} in Abell 901/902 regarding the existence
of a colour--density relation in the projected radial distribution, we
divided our dwarf galaxy sample into systems displaying redder and bluer
colours than the mean CMR.
 
From Fig.\,\ref{CMR_esferoides} and Table\,\ref{ajustes}, we can see that
spectroscopically confirmed members define a tight relation down to
$T_{1_0}=20$ mag with no change of slope or increase in the scatter.  The
new dwarf galaxy candidates seem to extend the relation down to $T_{1_0}>22$
mag, although introducing a considerable dispersion in the CMR at its very
faint end. In particular, dSph candidates tend to increase the scatter
towards bluer colours than the mean relation. In comparison with the results
presented in Paper\,I, the CMR gets slightly steeper with the addition of
the new confirmed members. When FS90 non confirmed members are included in
the fit, both the slope of the CMR and its scatter increase.
 
The selection of the spectroscopic targets was performed independently of
their location in the photometric relations analyzed in this paper. The
GEMINI-GMOS fields were chosen
 to include as many FS90 galaxy candidates as possible. Therefore, it is
 remarkable that all early-type confirmed members suite so well in the
 CMR. In addition, we can see that spectroscopically confirmed background
 galaxies included in the FS90 Antlia catalogue, as well as dI and BCD
 confirmed members, would introduce a substantial dispersion to the relation
 if they were wrongly considered as early-type cluster members.
 
The new confirmed cE galaxies share the same CMR as ``normal'' early-type dwarf 
galaxies. However, as can be seen from the data included in  
Table~\ref{ajustes}, they increase the dispersion of the relation defined by 
dwarf confirmed members towards the red side of the colour-magnitude diagram.  
When these peculiar galaxies are excluded from the fit, faint early-type  
members define a CMR with a similar scatter to that traced by bright
ones. The location of both cEs in the CMR diagram is consistent with a
luminosity fading ($\approx 2.8$\,mag for FS90\,110, and $\approx 3.7$\,mag
for FS90\,192) at constant colour.

\subsection{Surface brightness-Luminosity Relation}
\label{mueff}

\begin{figure}
\begin{center}
\includegraphics[scale=0.45]{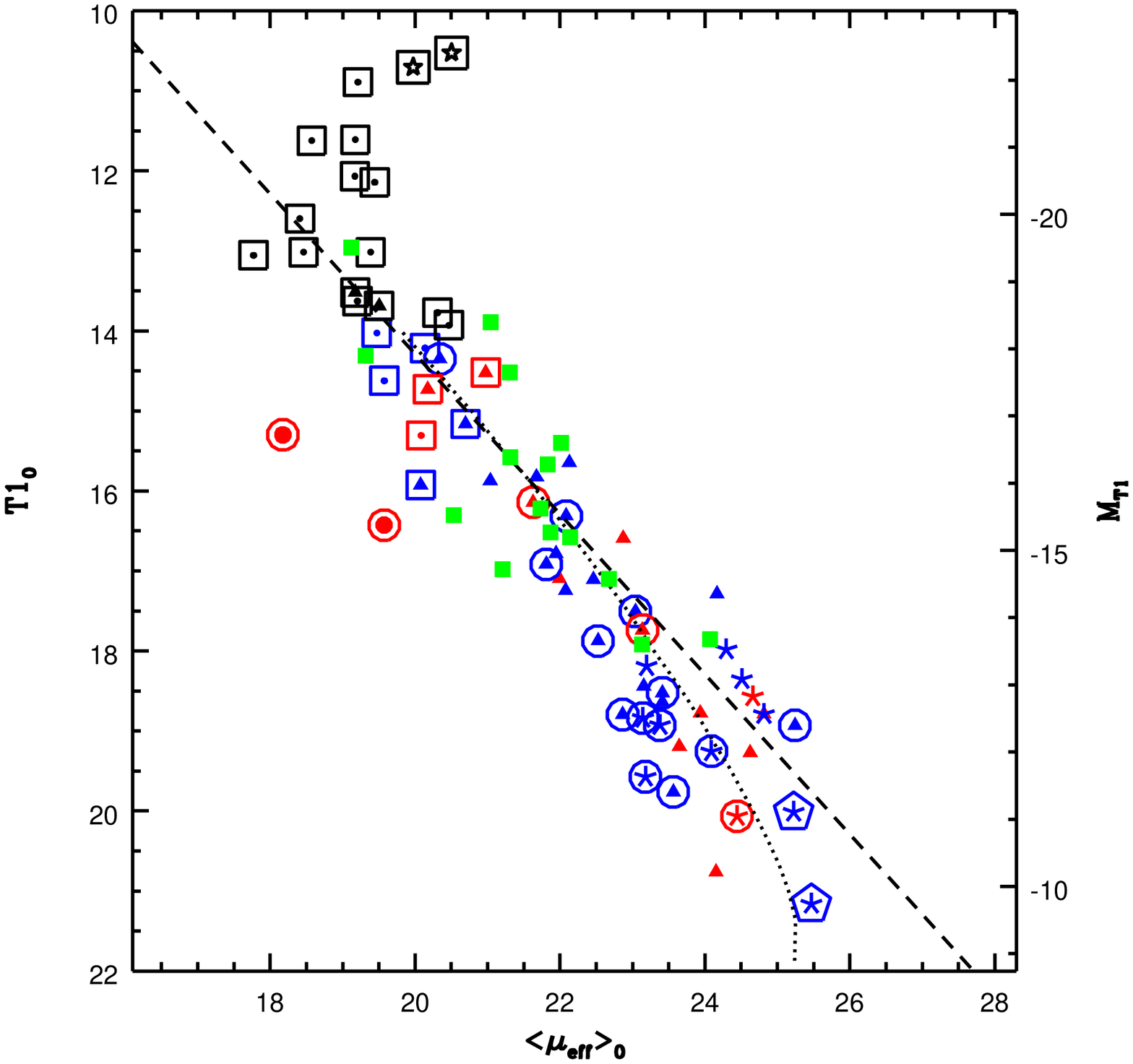}
\caption{Extinction corrected $T_1$ magnitude versus mean 
effective surface brightness diagram for 
Antlia definite and confirmed members plus new dwarf galaxy candidates
and members. The dashed line shows the locus of constant effective radius 
($\sim$ 1 kpc) followed by FS90 galaxies fainter than $T_1=13$\,mag (Paper\,I). 
New dwarf galaxy candidates tend to depart towards smaller $r_{\rm eff}$. 
The dotted line shows the theoretical effect of isophotal truncation (at 
26.0 mag\,arcsec$^{-2}$) on the galaxy surface brightness distributions at 
the faint end of the relation. The symbol code for Antlia galaxies is the same 
as in Fig.\,\ref{CMR_esferoides}. Red and blue symbols depict dwarf galaxies
showing redder and bluer colours than the mean CMR, respectively. Green 
squares represent the early-type population of poor groups recently studied
by \citet{An11}.}
\label{mueff_esferoides}
\end{center}
\end{figure}

In Fig.\,\ref{mueff_esferoides} we present the revised  
luminosity--$\langle\mu_{\rm eff}\rangle$ relation of FS90 Antlia definite 
and confirmed members, and new dwarf candidates and members. We show, as a  
dashed line, the locus of constant effective radius ($r_{\rm eff}\sim 1$\,kpc)  
found for galaxies fainter than $T_{1_0}=13$ mag in Paper\,I. Recall that  
lines parallel to this locus towards fainter luminosities correspond to smaller 
effective radii (see eq.\,1 in Paper\,I).  
 
For our new sample of dwarf early-type members and candidates, we obtain  
a mean effective radius of $\langle r_{\rm eff} \rangle = 0.81$ 
(rms 0.31) kpc, considering 57 galaxies with $T_{1_0} > 13$ mag and 
excluding the two cEs which represent extreme cases. However, we see that both 
FS90 and new confirmed members fainter than $T_{1_0} \sim 18$ mag tend to 
depart from the linear relation towards smaller effective radii. When only 
dwarf galaxies in the range $13 < T_{1_0} < 18$ are considered, the mean 
effective radius increases to $\langle r_{\rm eff} \rangle = 0.93$ (rms 
0.28) kpc.     
 
One possible explanation for the faint break is that it arises due to the  
isophotal limit of our photometry, which causes different fractions of galaxy  
luminosity to be lost outside the limiting isophote for galaxies with  
different profile shapes. In order to quantify this effect, we considered a  
set of S\'ersic models spanning an appropriate magnitude range, with shape  
parameters ($n$) following the luminosity -- shape relation given by  
\citet[][see their fig.\,10]{Graham03}. We fixed the effective radii of all  
the models at 1\,kpc (5.87 arcsec at the Antlia cluster distance), and we then  
computed for each model the fraction of light lost outside the limiting  
radius, and the effective radius ``measured'' from the truncated profile  
\citep*[see][for the relevant expressions]{TGC01}. We thus obtained for each  
model the values of isophotal magnitude and mean effective surface brightness  
that would be measured from a truncated profile. 
 
We show the results in Fig.\,\ref{mueff_esferoides}, where the dotted line
follows the isophotal magnitude versus mean effective surface brightness
relations for S\'ersic models truncated at isophotal radii corresponding to
26.0 mag arcsec$^{-2}$. While profile truncation always leads to a magnitude
dimming and a lower measured $r_\mathrm{eff}$, low $n$ (i.e., fainter)
galaxies are more strongly affected than galaxies with $n\ge 1$. This gives
a curved relation in the observed magnitude vs. mean effective surface
brightness plane, which should be a straight line for galaxies with the same
effective radius. Then, the downturn of the relation for our dwarfs at faint
galaxy magnitudes may be attributed to this effect, corresponding to our
limiting isophote of about 26.0 mag\,arcsec$^{-2}$.  The same departure
towards fainter magnitudes of the dwarfs at the faint end of this relation,
with respect to the general trend, is present in the compilation from
different environments by \citet{deRij09}.
 
Both cE galaxies clearly depart from the locus of constant effective radius  
but, at first glance, in a different manner. FS90\,110 seems to extend the  
break defined by the brightest galaxies towards fainter magnitudes, but  
FS90\,192 is located with the bulk of early-type dwarf galaxies defining 
what could be their low limiting $r_{\rm eff}$. If we calculate for their  
luminosities the corresponding $\langle\mu_{\rm eff}\rangle$ on the  
locus of $r_{\rm eff} = 1$\,kpc, we obtain that they differ from the measured 
values in $\Delta\langle\mu_{\rm eff}\rangle=2.9$ mag arcsec$^{-2}$ for  
FS90\,110, and $\Delta\langle\mu_{\rm eff}\rangle=2.7$ mag arcsec$^{-2}$ for  
FS90\,192. Therefore both cE galaxies depart from the trend defined by  
early-type dwarf galaxies in a similar manner, defining what could be a  
photometric criterion to identify candidate cE galaxies. 
Alternatively, the location of the cEs with respect to the surface
brightness-luminosity relation can be judged as consistent with a stripping
scenario.

The luminosity--effective radius diagram presented by \citet{Barazza09} in
Abell 901/902 (their fig. 2a), shows that their galaxies scatter about a
mean constant effective radius. In Figure\,\ref{mueff_esferoides}, the galaxies 
located below the dashed line have $r_{\rm eff} < 1$ kpc and those found above 
the line, $r_{\rm eff} > 1$. We  see that early-type dwarfs in the central 
region of Antlia, either redder or bluer than the colour of the mean CMR, show 
similar effective radii, in agreement with Barazza et al.'s findings.  
In particular, red dwarfs (excluding the two cEs, 16 galaxies) display a  
mean effective radius $\langle r_{\rm eff} \rangle = 0.74$ (rms 0.34) 
kpc, while blue dwarfs (33 systems), $\langle r_{\rm eff} \rangle = 0.79$ 
(rms 0.30) kpc. 
 
We have also plotted, with green squares, the data corresponding to the  
study of the early-type galaxy population in poor groups performed by  
\citet{An11}. The bright and faint early-type systems in these groups,  
follow the same luminosity--$\langle\mu_{\rm eff}\rangle$ relation  
as  Antlia galaxies with a similar scatter. 
 
\subsection{Projected Spatial Distribution}
\label{distribution}

\begin{figure}
\begin{center}
\includegraphics[scale=0.45]{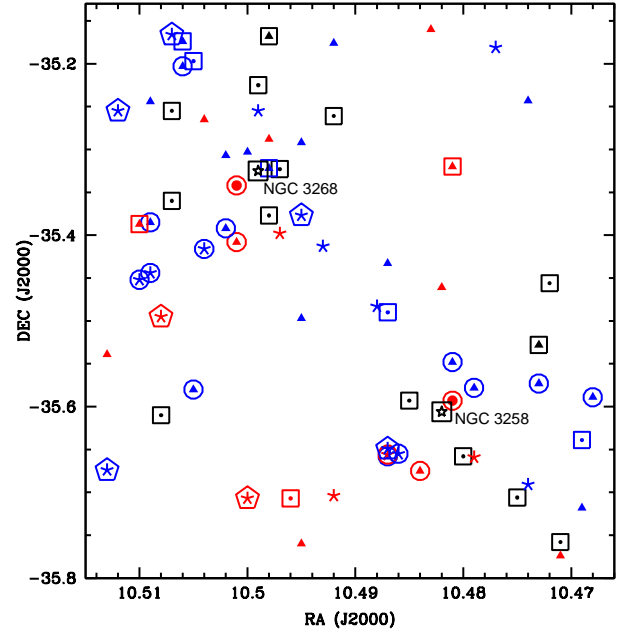}
\caption{Distribution of the galaxy sample in the MOSAIC field. The symbol 
code is the same as in Fig.\,\ref{CMR_esferoides}. Red and 
blue symbols depict dwarf galaxies displaying redder and bluer colours than
the mean CMR, respectively. North is up and east to the left.}
\label{posiciones}
\end{center}
\end{figure}

In Fig.\,\ref{posiciones} we show the location of our galaxy sample in the
MOSAIC field. Two concentrations of galaxies are present around NGC\,3258 and  
NGC\,3268, respectively, with a poorly populated zone between them. Both 
concentrations are similar in number of galaxies, contain several bright  
lenticulars, and in each of them there is a cE galaxy placed in the  
neighbourhood of the dominant system.  
 
\citet{Barazza09} have shown that redder dwarf early-type galaxies in
Abell 901/902 are located closer to the central cluster galaxies than bluer 
ones of similar luminosity. In our case the mean distances of the dwarfs  
to each one of the two dominant galaxies (the closer one) are: 8.0 
(rms 4.8) arcmin for the 20 dwarfs redder than the mean CMR at the same 
luminosity, and 7.7 (rms 4.3) arcmin for 37 dwarfs bluer than the mean 
CMR. If we consider the luminosity ranges $14 < T_{1_0} < 17$, $17 < T_{1_0} 
< 20$ and $T_{1_0} > 20$, we obtain, respectively, mean distances of 8.2 
(rms 6.1) arcmin (7 red dwarfs) and 8.4 (rms 4.8) arcmin (12 blue 
dwarfs), 8.3 (rms 5.6) arcmin (7 red dwarfs) and 7.6 (rms 4.5) 
arcmin (20 blue dwarfs), and 7.4 (rms 2.0) arcmin (6 red dwarfs) and 
5.9 (rms 1.1) arcmin (5 blue dwarfs). Therefore, we do not find a clear 
trend in the sense of that reported by Barazza et al. 
However, this differences are rather marginal, and should be tested with a 
larger sample including the outer regions of Antlia.

\section{Discussion} 
\label{discussion} 
 
\subsection{Colour-Magnitude Relation}  
\label{discusion_CMR}

\subsubsection{Interpretation} 
 
The CMR of early-type galaxies in groups and clusters of galaxies is 
a well known photometric relation that has been studied for long. In the 
colour-magnitude diagram, early-type galaxies define a sequence in the sense 
that bright galaxies are redder than fainter ones. The physical 
interpretation of this feature has been the subject of numerous observational 
and theoretical studies. While luminosity or brightness can be easily 
associated with the (stellar) mass of a galaxy, the well known 
{\it age-metallicity} degeneracy prevents a clear physical identification 
for integrated colours. 

From the study of photometric indices correlated with line strenghts for
31 elliptical galaxies belonging to pairs or small groups, 
\citet{Faber73} has found that, for the redder galaxies of the sample, the 
differences in integrated colours are correlated with variations in 
abundances. In a more recent work, \citet{TK99} have studied the 
$\rm C_{2}4668$, $\rm Fe4383$, $\rm H\gamma_{\rm A}$ and 
$\rm H\delta_{\rm A}$ spectral absorption line indices of 101 galaxies in 
the Coma cluster. They found that the CMR in Coma is mainly driven by a 
luminosity--metallicity effect. This is consistent with the results of 
\citet{Trager00} about Fornax and Virgo elliptical galaxies. They are 
basically old stellar systems and define a mass--metallicity trend, in 
contrast with field ellipticals, where age and $\alpha$-element fraction 
may be more strongly correlated with stellar mass than overall metallicity. 
This finding for field galaxies is confirmed by \citet{Howell05}; however, 
his sample covers a small luminosity range which strongly limits any analysis 
of correlations involving absolute magnitudes.

\citet{Clemens11} show the mid-IR CMR for early-type galaxies in the 
Virgo cluster with absolute K-band magnitudes between -26 and -19. By 
comparing their results with those of optical studies, they conclude that 
the CMR is driven by metallicity effects, as a mass--age relation would 
produce a slope in the opposite sense to that observed. Observations
of clusters at intermediate redshift showed that the slope of the CMR 
display small variations with redshift \citep[][and references therein]{J11}.

Hence, although other stellar population parameters like age and
$\alpha$-element fraction are shown to depend on the luminosity of
early-type galaxies, and this behaviour is shown to depend on the
environment, a clear mass-metallicity relation holds through different
environments, from rich clusters to the Local Group, spanning more than 5
decades in $\log(M_\star)$ \citep[e.g.][]{Mendel09}.

Within the discussion of the CMR in Antlia it may be thus interesting to
turn, as a comparison, to the NGC\,5044 group, a relaxed galaxy aggregate
intermediate between small groups and rich clusters \citep{Forbes07}, where
early-type galaxies are also found to follow a well-defined CMR spanning
$\sim 10$ mag in luminosity \citep{CB05}. A spectroscopic analysis relying 
on Lick indices \citep{Mendel09} shows that, among different stellar population
parameters, metallicity displays the strongest correlation with stellar
mass. There is, however, a clear decoupling between Fe and the
$\alpha$-elements (mostly traced by Mg and Ca) which extends down to the dE
mass range \citep[see also][]{Buzzoni11}. This behaviour
confirms and extends previous results which point at different enrichment
channels acting along the mass sequence of early-type galaxies in groups and
clusters, as a consequence of different star-formation efficiencies taking 
place at different masses. Age effects, although playing a minor role, are 
certainly present, as witnessed by a small number of ``blue-nucleus" dEs with 
Balmer emission lines in their spectra \citep[e.g.][]{CB01,Lisker06}.

At the lowest stellar-mass extreme, it is expected that dSph's, like their
Local Group counterparts, have experienced complex star-formation histories
\citep[e.g.][and references therein]{Tolstoy09}. A low metal content is,
however, their general characteristic, and should be the main driver of
their optical integrated colours.

Therefore, despite the identification of the physical mechanisms 
involved in defining the CMR is still an open issue, there is strong evidence 
that the CMR should be mainly interpreted as a mass--metallicity relation.

\subsubsection{Observed characteristics} 
 
It is widely accepted that the CMR is linear along its extension, and that 
it shows no perceptible change of slope from the bright galaxies to the dwarf  
regime (\citealp{S97,T01,LC04,An06}; Paper I; \citealp{M08,M09}). In addition 
there is strong evidence for the universality of this relation in clusters  
of galaxies.  
 
\citet*{B92} have found that early-type galaxies in Virgo and Coma present  
{\it identical} $(U-V)$ and $(V-K)$ colours, following the same linear CMR.  
\citet{C02} found, for the Perseus cluster, a fit to the bright end of the  
CMR in agreement to that derived by \citet{S97} for the Coma cluster. In  
Paper\,I we have obtained a linear fit to the CMR of the Antlia cluster with  
a slope in agreement with those found for dwarf galaxies in Virgo  
\citep*{LGB08}, Fornax \citep*{H03,M07}, Perseus \citep{C02} and Coma  
\citep{LC04}. \citet{deRij09} have shown that early-type galaxies belonging to  
different clusters and groups do follow the same colour-magnitude relation.  
Moreover, in Perseus \citep{P10} and Coma \citep{T01} it has been found that  
the CMR keeps constant as a function of radius within the cluster. In  
particular, the CMR of Perseus has slope and zero point consistent with  
those reported for Fornax \citep{M07}, Hydra\,I \citep{M08} and Centaurus  
\citep{M09}. Such linearity and universality led several authors to use the  
CMR as a reliable distance indicator \citep{S72} and to suggest that the  
origin of this relation in galaxy clusters is independent of the  
environment (Paper\,I; \citealp{M08,deRij09}).   
 
However, for the Virgo cluster, \citet{JL09} have obtained an S-shaped CMR  
that seems to be consistent with the non-linear relation found by \citet{F06}  
for the same cluster. In addition, it is still under discussion whether there  
is a genuine (i.e., with a physical origin) increase in the scatter of the CMR 
towards its faint end or, instead, the larger scatter can be accounted for by  
photometric or classification errors and/or background galaxies  
contamination. 
 
In the Perseus cluster, \citet{C02} have found a significant 
spread ($\sigma_{(B-R)}\sim$ 0.5 mag) around the mean CMR, in the colours  
of galaxies with $M_B > -16$ mag. They argued in favor of a physical origin of  
this scatter, pointing to differences in metal abundances and formation  
scenarios among blue and red low mass cluster galaxies. From a more recent  
spectroscopic study of the faint population of Perseus, \citet{PC08} have 
found that most of the red and blue galaxies responsible of the scatter are 
in the background. However, there is still an increase in the colour 
dispersion of confirmed members with $M_B > -15$ mag, in comparison with  
brighter ones. \citet{JL09} found an increase of the scatter about the CMR 
at intermediate brightness, while for the brighter and fainter galaxies  
the scatter is interpreted as due to photometric errors. The authors conclude 
that the increase at intermediate luminosities confirms the fact that, in  
Virgo, dwarf and bright early-type galaxies do not share one common linear CMR. 
   
On the contrary, from a thorough morphological identification of the galaxies  
included in the Coma CMR, \citet{T01} have found no detectable increase in the  
colour dispersion all along the early-type galaxies' relation. When spiral and 
irregular galaxies, as well as unclassified objects, are plotted in the same 
colour--magnitude diagram, they introduce a considerable scatter towards the 
blue side of the relation at intermediate and low luminosities. \citet{An06}  
have obtained a thin ($\sigma_{(B-R)}=$ 0.04 mag) CMR in Abell\,1185, with a  
scatter that does not increase with magnitude. \citet{LC04} have found, for  
57 X-ray detected Abell clusters in the redshift range $0.02\leq z \leq 0.18$,  
a universal CMR with an average colour dispersion of $\sigma_{(B-R)}=0.074$  
mag. They noticed that the cluster displaying the largest scatter in its CMR  
(0.5 mag at $R=18$ mag) presents background contamination from higher redshift  
clusters. These authors also pointed out that photometric errors could become  
an important source of dispersion in the relation \citep[see, also,][]{S97}.  
In the Hydra \citep{M08} and Centaurus \citep{M09} clusters a larger colour  
scatter has also been found for faint early-type galaxies, in  
comparison to luminous ones. However, such increase is considered as due to  
photometric errors and it is shown that dubious members of the cluster 
and background galaxies can introduce significant dispersion in the relation. 
Recently, \citet{J11} have found that 172 early-type confirmed members of 13  
clusters and groups with redshifts $0.4 \lesssim z \lesssim 0.8$, define tight  
CMRs with a small intrinsic colour scatter ($\sigma_{(U-V)}=0.076$). 
 
\subsubsection{The CMR in Antlia} 
\label{s.CMRA}

The CMR of spectroscopically confirmed Antlia members displays the most 
common features found for this relation in other nearby and distant  
clusters of galaxies. That is, it spans almost 10 magnitudes with no  
perceptible change of slope or increase in its scatter towards faint  
luminosities. When unconfirmed members and confirmed background galaxies  
are introduced in the relation, it displays a larger colour dispersion, not  
only in its faint end but also at intermediate luminosities. In fig.\,2b of 
Paper\,I we included in the same colour--magnitude diagram FS90 galaxies 
with spiral and irregular morphologies, some of them with spectroscopically 
confirmed membership. These objects introduced considerable dispersion towards 
the blue side of the relation, in agreement with the results of \citet{T01}. 
 
The revised slope of the relation for all definite members is steeper than  
that found in Paper\,I. However, this value is in agreement 
with the one found in that paper for the bright end of the relation. 
Therefore,  
the bright galaxies seem to define the slope of the relation and when  
confirmed early-type dwarf members are included in the analysis, the faint  
galaxies follow a relation with a similar slope and scatter. This steeper  
value for the slope of the CMR of Antlia is still consistent with those  
reported for other galaxy clusters.
 
The computed intrinsic scatter of the relation is  
$\sigma_{(C-T_1)}\sim0.08$ when only confirmed members are considered.  
Following \citet[][see their eq.\,1]{T01}, we calculate the intrinsic 
scatter as $\sigma_{intr}=\sqrt{\sigma_{obs}^2-\langle\epsilon\rangle^2}$, 
where $\langle\epsilon\rangle$ is the mean colour error of the sample.
If we translate our intrinsic dispersion to $(B-R)$ colours through  
$\sigma_{(B-R)}=0.704~\sigma_{(C-T_1)}$ \citep*[see eq.\,8 in][]{FFG07}, we  
obtain $\sigma_{(B-R)}\sim0.06$ which is similar to the mean value obtained by 
\citet{LC04} for clusters with $z<0.04$, and consistent with the value  
obtained by \citet{PC08} for the CMR of Perseus at the bright end  
($\sigma_{(B-R)}\sim0.05$).

If we take into account that $T_1-R\simeq0.02$  
\citep{gei96}, we can use the relation  
$\sigma_{(V-T_1)}=\sigma_{(V-R)}=0.256~\sigma_{(C-T_1)}$  
\citep[see eq.\,1 in][]{HH02b} and we obtain $\sigma_{(V-R)}\sim0.02$ which  
is a lower value than that obtained by \citet{An06} in Abell\,1185  
($\sigma_{(V-R)}=0.036$). In the Centaurus cluster, \citet{M09} have found 
intrinsic scatters of $\sigma_{(V-I)}=$0.06, 0.09 and 0.14 in three different 
magnitude intervals. Using $\sigma_{(V-I)}=0.49~\sigma_{(C-T_1)}$  
\citep[see the first equation in][]{FF01}, these values translate in 
$\sigma_{(C-T_1)}=$0.12, 0.18 and 0.28, much higher than our common  
value for the whole Antlia CMR. For the Hydra cluster, \citet{M08} have 
found a mean scatter for the relation of $\sigma_{(V-I)}=0.12$ which is 
equivalent to $\sigma_{(C-T_1)}=0.24$.  

\subsubsection{cE's location in the CMR}

Regarding the positions in the colour-magnitude diagram of the confirmed  
Antlia cEs, they can be compared with those displayed by the cE galaxies  
recently identified in Centaurus \citep{M09} and Coma (\citealp{P09}, see 
also \citealp{Chi10}). Centaurus' cEs are located within the CMR on its red  
side, in a similar manner to Antlia's cEs \citep[see fig.\,3 in][]{M09}.  
Despite all confident Coma's cEs (namely, CcGV9a, CcGV19a and CcGV19b) are  
found on the red side of Coma's mean CMR, only one of them shows a similar  
location to those found in Antlia and Centaurus (that is, within or near 
the general trend), while the other two display  
a considerable offset from the mean relation towards redder colours 
and/or fainter brightnesses \citep[see fig.\,2 in][]{P09}. 

Coma's cEs have age and abundances estimations. All three galaxies are coeval
within the age estimation errors. CcGV9a and CcGV19b present similar 
$(B-I)$ colours, and similar [Fe/H] and [Mg/Fe] abundances, while CcGV19a is
the reddest cE galaxy in the association, with the highest [Fe/H] and [Mg/Fe]
values. Therefore, the different locations of these three cEs in the  
colour--magnitude diagram can be explained through abundance differences 
and/or luminosity fading.

In this context it is interesting to point out that \citet{Faber73} noticed 
that {\it a small number of peculiar dwarf ellipticals are known that are 
quite red and have unusually high surface brightness} and that these objects 
{\it clearly do not fit into the sequence of normal ellipticals}. Moreover, 
Faber considered that these systems {\it are in fact remnants of tidal
encounters with more massive companion galaxies}.
 
\subsubsection{The scatter of the CMR} 
 
Given the universality of the CMR in clusters of galaxies, if the increase
in the scatter at low luminosities has a physical origin, it should be
observed at the same absolute magnitude in all cases. \citet{S97} have
obtained a CMR for the Coma cluster with a scatter that increases faintwards
from $R\simeq 19.5$ mag, corresponding to $M_R=-15.5$\,mag with a distance
modulus $(m-M)=35$. In the Perseus cluster, the scatter increases
significantly at $M_B=-15$ mag \citep[see fig. 3 in][]{PC08}. By means of the
transformations given by \citet{F95}, this is equivalent to $M_R=-16.6$ mag. 
In Fornax, \citet{H03} and \citet{M07} have found an increase of the CMR
scatter at $M_V\simeq-12.5$ mag or $M_R\simeq-13.1$ mag. In Hydra, the 
increase in the scatter is significant from $M_V\simeq-14$ mag 
\citep[fig. 10 in][]{M08} or $M_R\simeq-14.6$ mag. In Centaurus, from 
$M_V\simeq-13$ mag or $M_R\simeq-13.6$ mag \citep[fig. 3 in][]{M09}. In Virgo, 
the peak of the dispersion is found at $M_r=-17$ mag which is equivalent to 
$M_R\sim-17.3$ mag. In our case, the faintest Antlia confirmed members and 
candidates introduce a detectable dispersion increase starting from 
$M_R\sim-15$ mag.
 
Therefore, in all these clusters, the increase of the scatter starts at  
different absolute magnitudes, which tend to be brighter for more  
distant clusters, if Virgo is excluded from the sample. This trend could be  
interpreted as a diminishing of the precision of the photometric data and a 
more uncertain morphological/membership classification as we go to more  
distant clusters. 
 
From the analysis performed in \ref{s.CMRA}, it might be inferred that,
once photometric errors are discounted, different CMRs obtained using 
metallicity sensitive colours, like $(C-T_1)$ and $(B-R)$, display fairly low 
intrinsic dispersions ($\sigma_{(C-T_1)}<0.08$\,mag).

Several processes have been proposed as responsible for the intrinsic scatter 
of the relation (e.g., differences in ages, \citealp{TK99}, \citealp{F03}, 
\citealp{R08}; minor mergers, \citealp{K09}; accretion of small amounts of gas 
at high redshifts, \citealp{PM06}). Whatever the process dominating the 
intrinsic scatter is, from our comparison we could say that it arises  
with similar strength in different clusters. In this sense, \citet{J11} have  
found a colour scatter with no evolution with redshift or correlation with  
the velocity dispersions of the clusters included in their analysis.

\subsection{Luminosity--$\langle\mu_{\rm eff}\rangle$ Relation} 
\label{discusion_mueff_lum}

\subsubsection{Antlia's relation} 
 
As it was shown in Section\,\ref{mueff}, in a plot of $T_1$ magnitude versus  
$\langle\mu_{\rm eff}\rangle$, FS90 early-type dwarf members of  
Antlia follow the locus of constant effective radius of 1 kpc down to 
$T_1 \sim 18$ mag. Fainter FS90 members tend to deviate from this 
locus towards smaller radii. All of the new dwarf galaxy members, which 
have luminosities fainter than $T_1=18$ mag, show the same behaviour. 
The mean effective radius of the confirmed members with $T_1>18$ mag 
is $\langle r_{\rm eff}\rangle=0.58$ (rms 0.26) kpc. Therefore, there 
seems to be a faint limit in magnitude for the constant effective radius 
relation followed by dEs. 
   
We have already mentioned that this could be due to the isophotal limit of  
our photometry. It is possible that we are not reaching the most external 
regions of the galaxies as they are embedded in the background. Many of these 
faint objects are located near brighter galaxies as well. In this way, our 
effective radius could be underestimated for galaxies fainter than $T_1\sim18$ 
mag ($M_V\sim-14.1$ mag, \citealp{F95}), leading to the apparent discontinuity 
or ``break'' in  this relation.  
 
However, \citet*{Chi09} have found that newly discovered dwarf galaxies in
the M81 Group ($D=3.6$ Mpc), have effective radii in agreement to those
found for the new dwarf galaxies in Antlia, i.e. $r_{\rm eff}\lesssim 0.5$
kpc (see their table 3). Local Group dSphs have $r_{\rm eff}\lesssim 0.5$
kpc, as well (\citealp*{Za06}, and references therein). \citet{CB05} as well
as \citet{deRij09} found a similar trend to our faint break, in their
respective samples of dwarf galaxies.  Therefore, we cannot rule out the
possibility that the luminosity vs. $\langle\mu_{\rm eff}\rangle$ relation,
corresponding to a mean constant effective radius of $\sim 1$ kpc, presents
two physical breaks: one at magnitudes brighter than $M_R\sim-20$ mag, and
another at its faint end, at magnitudes fainter than $M_R\sim-14$ mag. We 
recall that the correlation between luminosity and 
$\langle\mu_{\rm eff}\rangle$ for magnitudes brighter than $M_R\sim-20$ mag
corresponds to the \citet{K77b} scaling relation defined by bright Es and 
bulges of S galaxies on the $\rm r_{eff}$ vs. $\mu_{\rm eff}$ diagram. This is 
a projection of the Fundamental Plane of E galaxies.
 
As mentioned above in this section, most of the galaxies fainter than 
the low luminosity break are located near brighter ones. The smaller 
$r_{\rm eff}$ found in our faint candidates might then be related with tidal 
effects. The newly  
confirmed cE galaxies deviate from the locus followed by the rest of early-type 
dwarfs of similar luminosities towards smaller effective radii, as well. They 
are companions of the giant ellipticals that dominate the central region of  
the cluster. In Paper\,II we showed that one of them, FS90\,110, displays a  
low surface brightness bridge that links it with NGC\,3258, confirming the 
existence of an interaction between both galaxies. Surface brightness  
profiles displaying a break, as is the case of FS90\,110 (see Paper\,II)  
would favour this interpretation \citep{choi02,MJ06}. 

\subsubsection{Dichotomy between E and dE galaxies}
 
\citet{Graham03} have argued that, with the exception of the very 
bright galaxies ($M_B\lesssim-20.5$, $M_V\lesssim-21.5$, i.e. ``core'' E),  
the break at the bright end of the luminosity versus  
$\langle\mu_{\rm eff}\rangle$ relation is not due to different formation  
mechanisms among dwarf and bright E galaxies. They emphasize that both type  
of galaxies display continuous trends in the central surface brightness  
($\mu_0$) versus luminosity plot, and in the $n$ S\'ersic index  
\citep{S68} versus luminosity diagram. This would show that dEs are the  
low-luminosity extension of E galaxies. Graham \& Guzm\'an explain the  
different behaviour of Es and dEs in the  
luminosity--$\langle\mu_{\rm eff}\rangle$ space, as due to an increasing  
difference between $\mu_0$ and $\langle\mu_{\rm eff}\rangle$ values with  
increasing $n$ as we go from dE to E (without core) systems (see their  
figures 11 and 12).   
 
On the contrary, \citet{Kormendy09} sustain the existence of a dichotomy  
between dE ({\it spheroidal} galaxies in their paper) and E galaxies pointing  
to distinct origins. The fact that E and dE follow similar correlations  
would indicate that the parameters involved in such relations are 
not sensitive to the physics that makes them different. Therefore, these 
authors consider that to distinguish galaxy types it is necessary to use all  
parameter correlations and find out which ones are sensitive to formation  
mechanisms. From an extensive analysis, Kormendy et al. conclude that E  
galaxies form via mergers being cE galaxies their low luminosity
counterparts, while dE are late-type galaxies transformed by  
environmental effects and by energy feedback from supernovae. 

In this context it is interesting to point out the top panel of figure 
1 in the work by \citet{Chilli09}. There the authors show the location in the 
$\langle\mu_{\rm eff}\rangle$--luminosity diagram of all cEs detected by them 
with the Virtual Observatory (21 objects), along other confirmed cEs and 
early-type systems. It can be seen that M\,32 seems to extend the locus 
defined by bright early-type galaxies towards brighter values of 
$\langle\mu_{\rm eff}\rangle$ (or smaller $r_{\rm eff}$). But the rest of the 
cEs included in the plot seem to define a parallel sequence to that of dEs 
and dS0s, towards brighter $\langle\mu_{\rm eff}\rangle$. Simulating the 
interaction of a disk galaxy with a galaxy cluster potential, \citet{Chilli09}
also found that tidal stripping can reduce the stellar mass of the galaxy and 
speculate that low-luminosity objects could be the progenitors of cEs. 
Considering that most (if not all) cEs are companions of brighter galaxies, it 
might be interpreted that the offset of different cE galaxies from the 
dE/dS0 sequence depends on the degree of interaction experienced by the 
system and/or on the morphology of the progenitor. In this sense, 
cE galaxies would not be thought as the natural extension of the family of 
giant ellipticals towards low luminosities, as the origins of both 
populations would be quite different. It is worth noting that our two cE 
galaxies also seem to lie along a sequence parallel to that of dEs 
rather than to follow the trend of bright ellipticals. 

It is clear that the interpretation of the behaviour of bright and dwarf  
early-type galaxies in the luminosity--$\langle\mu_{\rm eff}\rangle$ diagram  
is an open question. Our present results for the Antlia cluster do not show  
any discontinuity between bright and dwarf ones in this diagram, but  
it should be taken into account that bright early-type galaxies in Antlia  
are mainly lenticulars, not ellipticals. In addition, we should enlarge the 
photometric sample at the low-luminosity end before arriving at sensible  
conclusions. 

At the moment, it is not possible to discern if   
$\langle\mu_{\rm eff}\rangle$ keeps almost constant from the faint break
towards fainter magnitudes, or if the effective surface brightness increases 
with decreasing luminosity, showing a similar trend to that displayed by E  
galaxies. However, it is remarkable the good match showed by the galaxies  
belonging to poor groups to the locus of constant effective radius defined  
by Antlia's systems. This fact is pointing to a relation that arises  
with similar characteristics in very different environments.  
 
\subsection{Projected Galaxy Distribution} 
 
X-ray observations of the Antlia cluster \citep*{Pedersen97,Nakazawa00}, 
have shown that the X-ray diffuse emission is elongated in the direction  
defined by a line joining both dominant galaxies. Studies of the globular  
cluster systems of NGC\,3258 and NGC\,3268 \citep{dir03,Bass08} have  
revealed that these systems display an elongated distribution in the same  
direction. These results may be the consequence of tidal forces between  
both galaxies, which may be ultimately pointing to an ongoing merger  
in the central region of Antlia.  
Thus, the two structures found in this region (Fig.\,\ref{posiciones})  
might be two groups in interaction: one with NGC\,3258 as dominant galaxy,  
and other with NGC\,3268 as the central galaxy, both of them containing a  
cE galaxy. More radial velocities from galaxies in the outer regions of  
Antlia are needed to confirm this hypothesis.  
 
We have shown in Section\,\ref{distribution} that there is no 
conclusive evidence of the existence of a colour--density relation in Antlia
in the sense reported by \citet{Barazza09} for Abell 901/902. These 
authors propose that the existence of this   
relation in the projected radial distribution is an indication  
that the formation of dEs is closely related to the processes that  
affect the cluster itself, like ram-pressure stripping and harassment. The
ram-pressure stripping would remove more efficiently (and faster) the 
interstellar medium of galaxies located closer to cluster centres. That may 
affect star formation in a way that these galaxies would have, on the average,  
an older stellar population and thus be redder than galaxies located  
further away from the cluster centres. 
 
The harassment scenario predicts a colour gradient in terms of effective 
radius \citep{Mastro05}, in the sense that bluer (outskirts) galaxies display  
larger $\rm r_{eff}$ than the redder (central) ones. In agreement  
with Barazza et al. results, we do not find a relation between  
colour and effective radius. In this context, as the early-type dwarf  
galaxy populations of Antlia and Abell 901/902 do not show such a  
gradient, the harassment process seems not to be a dominant one  
in these associations. \citet{Bos08} have also found that  
the lack of correlations with clustercentric distances in the  
scaling relations favours a ram-pressure stripping scenario for  
the evolution of the galaxies.

\subsection{dSph candidates in Antlia} 
 
Our selection of the low luminosity dSph candidates in Antlia was mainly
based on their limiting magnitude ($M_V\gtrsim -13$ mag, \citealp{Kalirai10})
and morphology (diffuse appearance and no nucleus).  Local Group counterparts
of these objects are also devoid of gas and dominated by old and
intermediate-age stars. Their extended sizes, low luminosity and measured
velocity dispersions, as well as their large mass-to-light ratios, point to
dSph being objects with large amounts of dark matter \citep[e.g.][]{M98}.
 
Our new dSph candidates have integrated magnitudes in the range $T_1 = 20.2
- 22.3$ mag, which corresponds to $-10.4 \gtrsim M_R\gtrsim -12.5$. In 
agreement with the tidal origin proposed for dSph galaxies 
\citep[e.g.][]{Penarrubia09}
they appear to cluster tightly around massive galaxies. In the Local Group,
most dSphs are located within a radius of 300 kpc from the Milky Way and
M\,31 \citep{Gallagher94,G05}.  Though our dSph candidates sample in Antlia
is in fact small, their positions displayed in Fig.\,\ref{posiciones} and
identified by open pentagons, place them close to the two giant dominant
galaxies or to bright lenticulars, all of them confirmed cluster members.
Their average projected distance to the closest bright galaxy (with $T_1 <
13$ mag) is $6.4 \pm 1.3$ arcmin, which at the Antlia distance correponds
to about $\approx$ 65 kpc.

Regarding their sizes, the study of dSphs in M\,31 by \citet{MI06b} has
shown that their tidal radii ($\approx 1.2 - 7$\,kpc) are about twice as
extended as their counterparts in the Milky Way. The classical dSphs (i.e.,
not including the newly discovered ultra faint dwarfs) surrounding the Milky
Way have tidal radii in the range $\sim 0.5$ kpc to 3 kpc \citep{Irwin95}. Even
their spatial distribution is different, being the Milky Way satellites more
clustered around their host than those of M\,31 \citep{MI06a}.
 
Globally, we can consider that tidal radii of Local Group dSphs range from
$\sim$ 0.5 to 7 kpc, which correspond to $\approx$ 3 - 45 arcsec at the
Antlia distance. From Table\,\ref{candidates}, we can see that the Antlia
dSph candidates display total radii in the range 5 - 11 arcsec, that is
within the expected values if we compare to the Local Group ones, though in
the range of the smaller sizes.  This can be understood like a selection
effect, as it would be even more difficult to detect larger, but more
diffuse, galaxies at the Antlia distance.

\section{Summary and Conclusions} 
\label{conclusions} 
 
In this paper we have deepened our study of the galaxy populations in the
central region of the Antlia cluster, paying particular attention to the
faint systems composed by dwarf ellipticals and dwarf spheroidals. We have
obtained radial velocities for a total of 28 faint galaxies ($M_V \gtrsim
-18$ mag) located in this region through GEMINI-GMOS and MAGELLAN--MIKE
spectra. With these new information, we have improved the colour-magnitude 
and surface brightness vs. luminosity relations originally studied in Paper\,I, 
and we have analyzed the projected spatial distribution of the whole galaxy sample,
presenting Washington photometry for 5 new confirmed members from the FS90
catalogue as well as for 16 new low surface brightness candidates that have
never been catalogued before. Total and/or effective radii have been
measured for the 5 new members and the 16 new candidates except one. In
addition, images in the $R$ band are shown for all the galaxies with new
radial velocities and the 16 new candidates. Our main conclusions are as
follows:\\
 
\noindent - Early-type galaxies in the Antlia cluster follow a 
tight colour-magnitude relation that extends, with no perceptible 
change of slope, from the giant ellipticals down to the dSphs 
spanning almost 12 magnitudes in luminosity. When only confirmed 
early-type members are considered, we find no significant increase 
in the scatter towards the faint end of the relation. When the 
intrinsic scatter of the CMR of Antlia is compared with those 
of other groups and clusters' CMRs, it is found that Antlia's CMR is 
one of the tightest relations reported up to now. We suggest that
this finding is related with the use of a metallicity sensitive 
colour index like $(C-T_1)$. \\ 
 
\noindent - The five previously unknown early-type Antlia members and the 16
new candidates follow the same photometric relations as the FS90 early-type
members and candidates, although introducing some dispersion. In particular,
the new dwarf candidates increase the scatter of the CMR towards its faint
end. These new dwarf members and candidates also depart, towards smaller
effective radii, from the surface brightness vs. luminosity relation traced
by FS90 dE galaxies with $14 < T_1 < 18$ that follow the locus of constant
effective radius of $\approx 1$\,kpc. This latter effect could be due to the
isophotal limit of our photometry. However, similar sizes ($r_{\rm eff} <
1$\,kpc) have been reported for faint members of the Local Group, M\,81 and
NGC\,5044 groups. Therefore, there seems to exist a physical brightness
limit for the relation corresponding to constant effective radius, a
relation that seems to arise in different environments as it has been found
in clusters like Virgo and Coma (see fig. 5 in Paper\,I) as well as in poor
groups. \\
 
\noindent - With regard to the possible existence of a discontinuity between
bright and faint ellipticals in the surface brightness vs. luminosity
relation \citep[e.g.][]{Graham03, Kormendy09}, our results for the Antlia
cluster do not show any gap in this diagram. It should be taken into account
that the bright early-type galaxies in Antlia are mostly lenticular galaxies
and few ellipticals. \\
 
\noindent - The dE galaxies with $13 < T_{1_0} < 18$ present an almost constant 
effective radius around $\langle r_{\rm eff} \rangle \approx 0.9$\,kpc, 
irrespective if their colours are bluer or redder than that of the mean CMR for 
the same luminosity. The almost constant effective radii of dEs are in  
agreement with what has been found for other clusters and groups  
(Paper\,I and references therein). For galaxies fainter than $T_{1_0} = 18$ mag
the mean value decreases to $\langle r_{\rm eff} \rangle \approx 0.6$\,kpc, which
is consistent with what is found for the faintest galaxies in the Local Group,  
M81 and NGC\,5044 groups. \\

\noindent - We have not found a colour--density or a colour--size relation among
our Antlia dwarf galaxy sample. This behaviour might favour a ram-pressure 
stripping scenario for the evolution of dwarf galaxies in this cluster. However, 
we should confirm these trends by enlarging our dwarf galaxy sample considering 
dwarf galaxies present in the outer regions of the cluster.\\

\noindent - We have confirmed the existence of two cE galaxies in Antlia.  
Each of them is the neighbour of one of the two giant ellipticals that  
dominate the central region of Antlia. For their luminosities, they are  
both the reddest and most compact objects of the early-type Antlia galaxies.  
If we accept that their colours are mainly driven by metallicity, as  
likely happens with galaxies on the CMR, both of them would be the  
richest galaxies for their luminosities in our sample.  
The cEs found in the Centaurus and Coma clusters are also located on the  
red side of their corresponding mean CMRs.   
However, the cE galaxies found in Coma present [Fe/H] abundances comparable  
to those of some dEs. Therefore, at the moment it is difficult to establish if 
cEs should be treated as a separate class of objects, or should be considered  
as part of the general early-type dwarf galaxy family experiencing, perhaps,  
particular dynamical processes.   \\ 
 
\noindent - We have found a small sample of 7 unstudied dSph candidates,  
whose membership cannot be confirmed with our spectra due to their low  
surface brightness. They cover a range in integrated magnitudes  
$-10.4 \gtrsim M_R\gtrsim -12.5$ and total radii between 5 and 11 arcsec,  
which at the Antlia distance correspond to approximately 850 pc -- 2 kpc.  
They are found close to bright galaxies ($T_1< 13$\,mag), being their  
average projected distance from them of $\approx 65$\,kpc. \\ 
 
\noindent - The projected spatial distribution of the whole sample  
points to a complex structure for Antlia. At least two groups are present  
in the central region, with some evidence of an ongoing merger. We will extend  
our study adding new fields adjoining the central one, and increasing the  
spectroscopic data set, in order to add kinematic information to the spatial 
distribution.
 
\section*{Acknowledgements} 
We are grateful to the anonymous referee for his/her comments and 
corrections which helped to improve the content of this paper. 
A.S.C. and G.R. warmly thank to Nidia Morrell for her help with the
reduction of echelle spectra.
A.S.C. and F.F. acknowdlege finantial support from Agencia de
Promoci\'on Cient\'ifica y Tecnol\'ogica of Argentina (BID AR PICT 
2010-0410 and BID AR PICT 2007-885).  
T.R. acknowledges financial support from the Chilean Center for Astrophysics,
FONDAP Nr. 15010003, from FONDECYT project Nr. 1100620, and
from the BASAL Centro de Astrof\'isica y Tecnolog\'ias
Afines (CATA) PFB-06/2007. 
G.R. was supported by ALMA/Conicyt (grant 31070021) and ESO/comite mixto.
This work was also funded with grants from Consejo Nacional de  
Investigaciones Cient\'{\i}ficas y T\'ecnicas de la Rep\'ublica  
Argentina, Agencia Nacional de Promoci\'on Cient\'{\i}fica y Tecnol\'ogica 
and Universidad Nacional de La Plata (Argentina). 
Part of the data used in this paper were obtained under 
GEMINI programs GS-2008A-Q-56, GS-2009A-Q-25 and GS-2010A-Q-21.

\clearpage
 
\appendix 
 
\section{FS90 galaxies with new radial velocities} 
\label{FS90_vr} 

\begin{figure*} 
\begin{center} 
\includegraphics[scale=0.16]{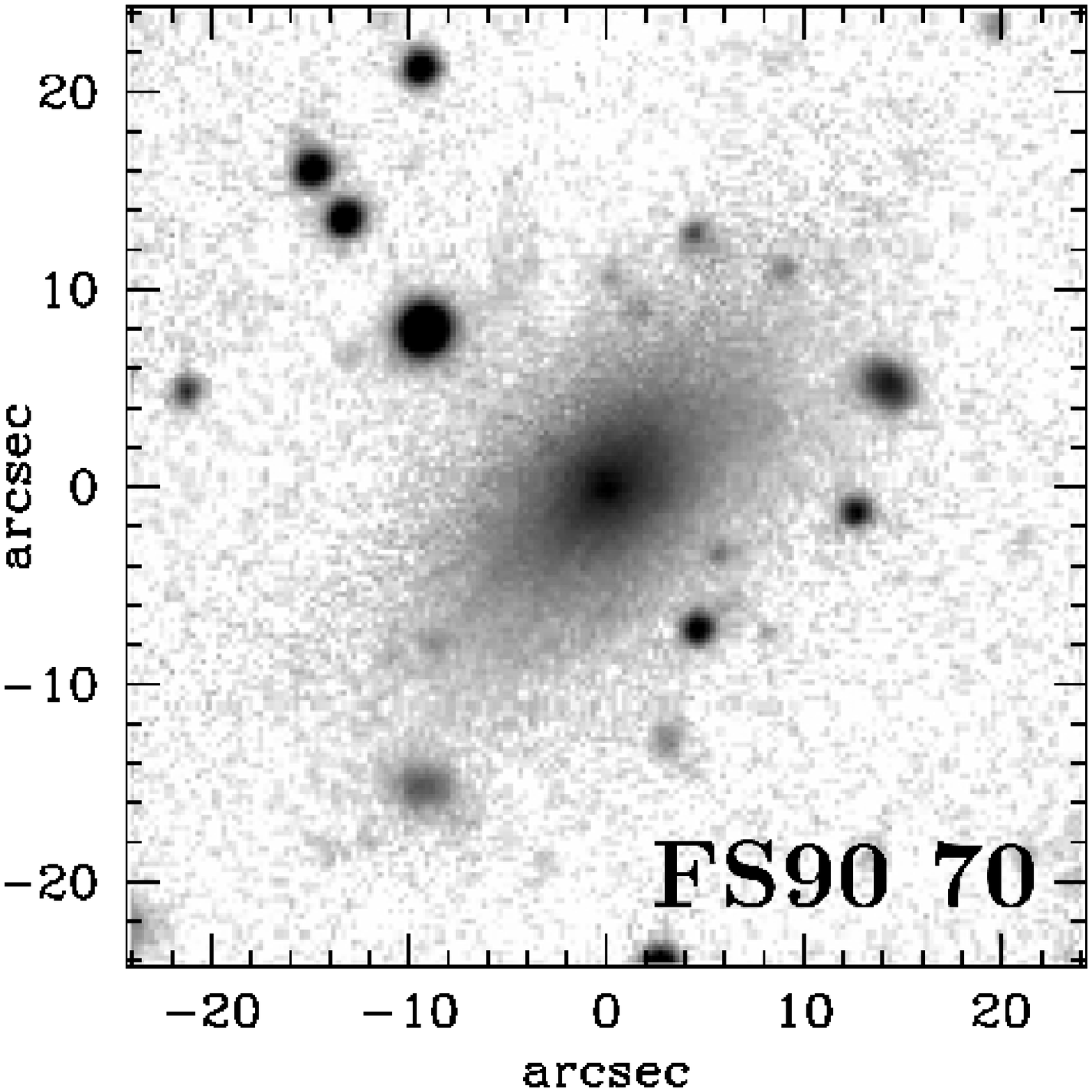}   
\includegraphics[scale=0.16]{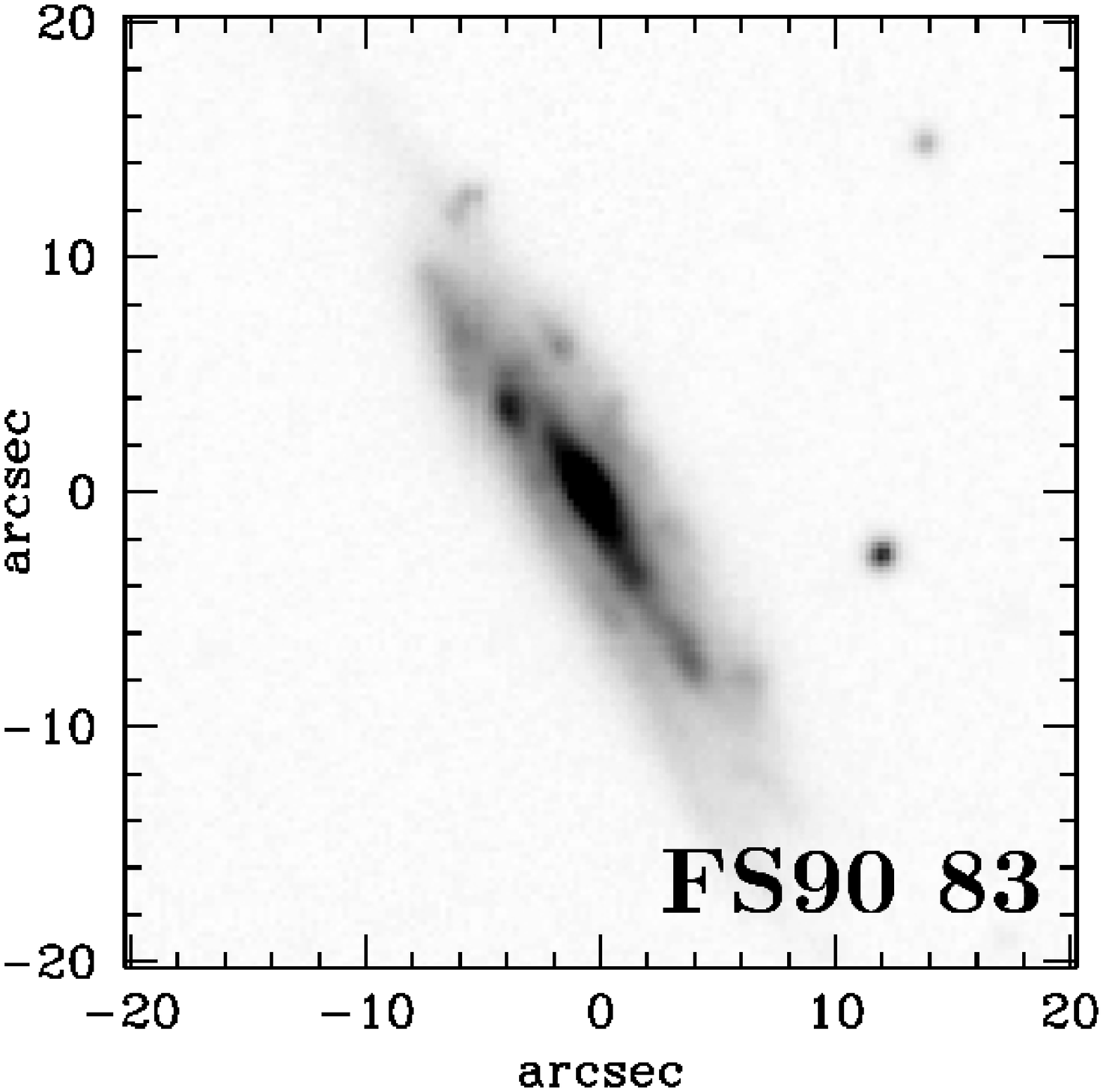}   
\includegraphics[scale=0.16]{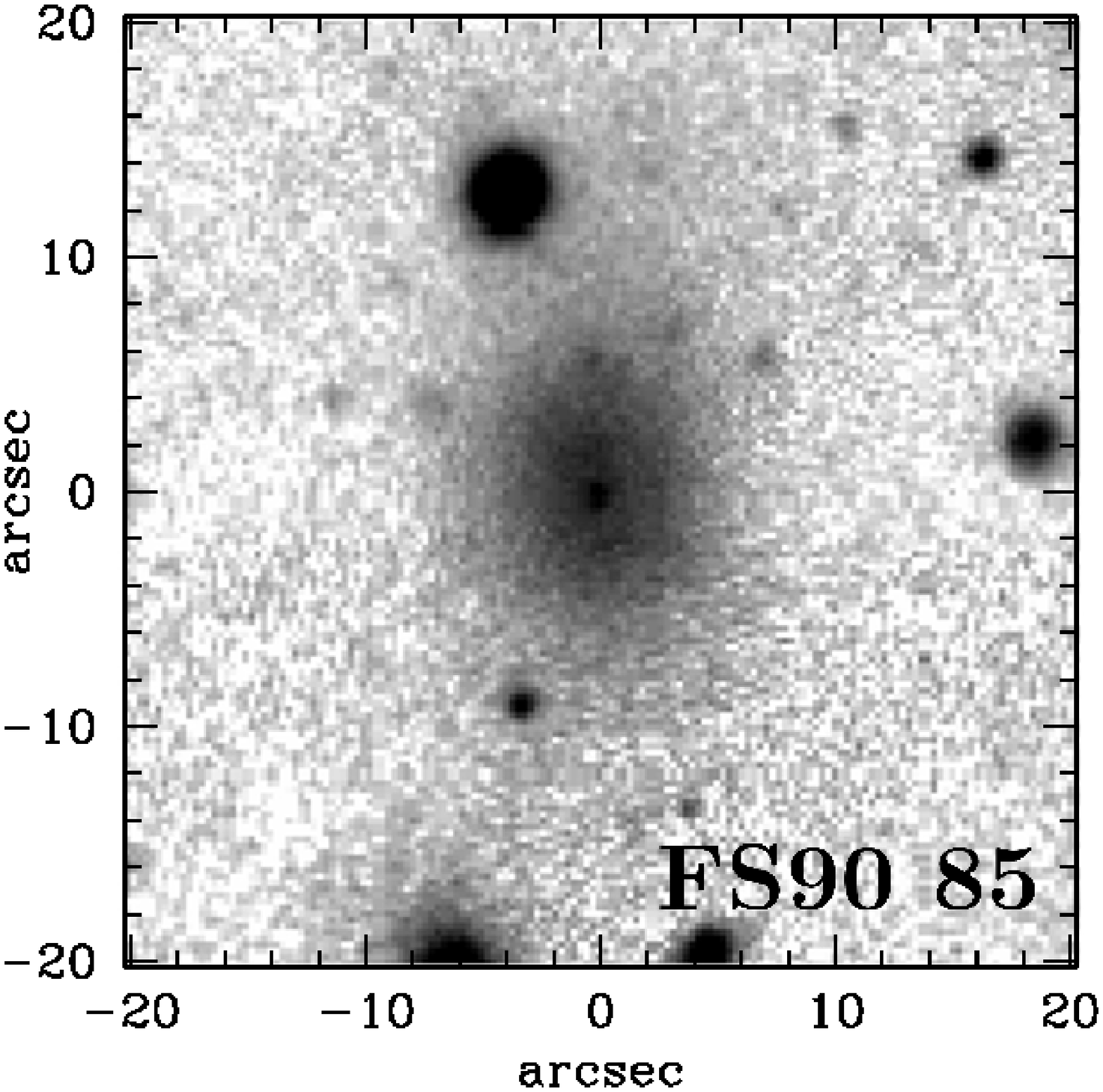}   
\includegraphics[scale=0.16]{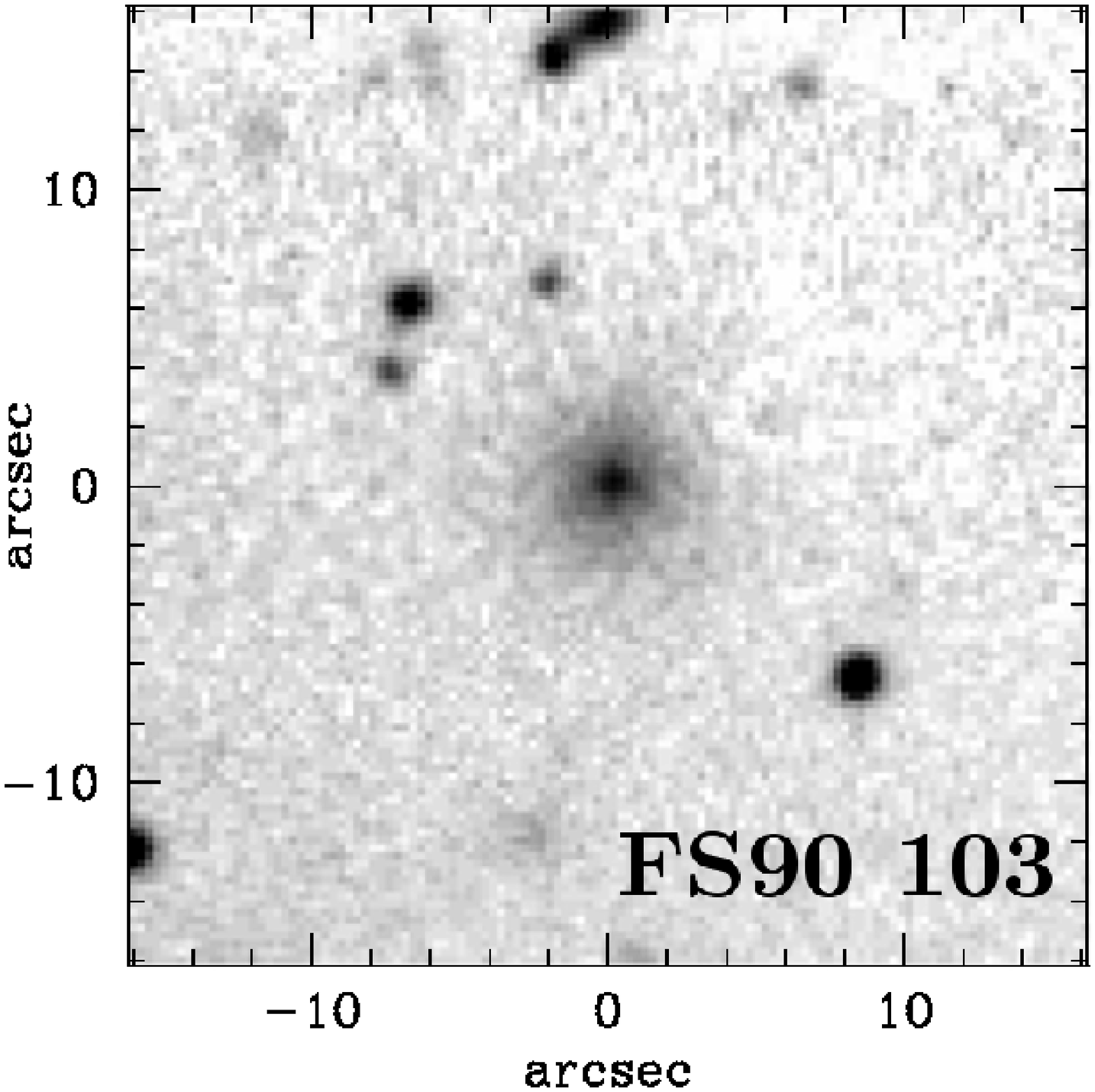}  
\includegraphics[scale=0.16]{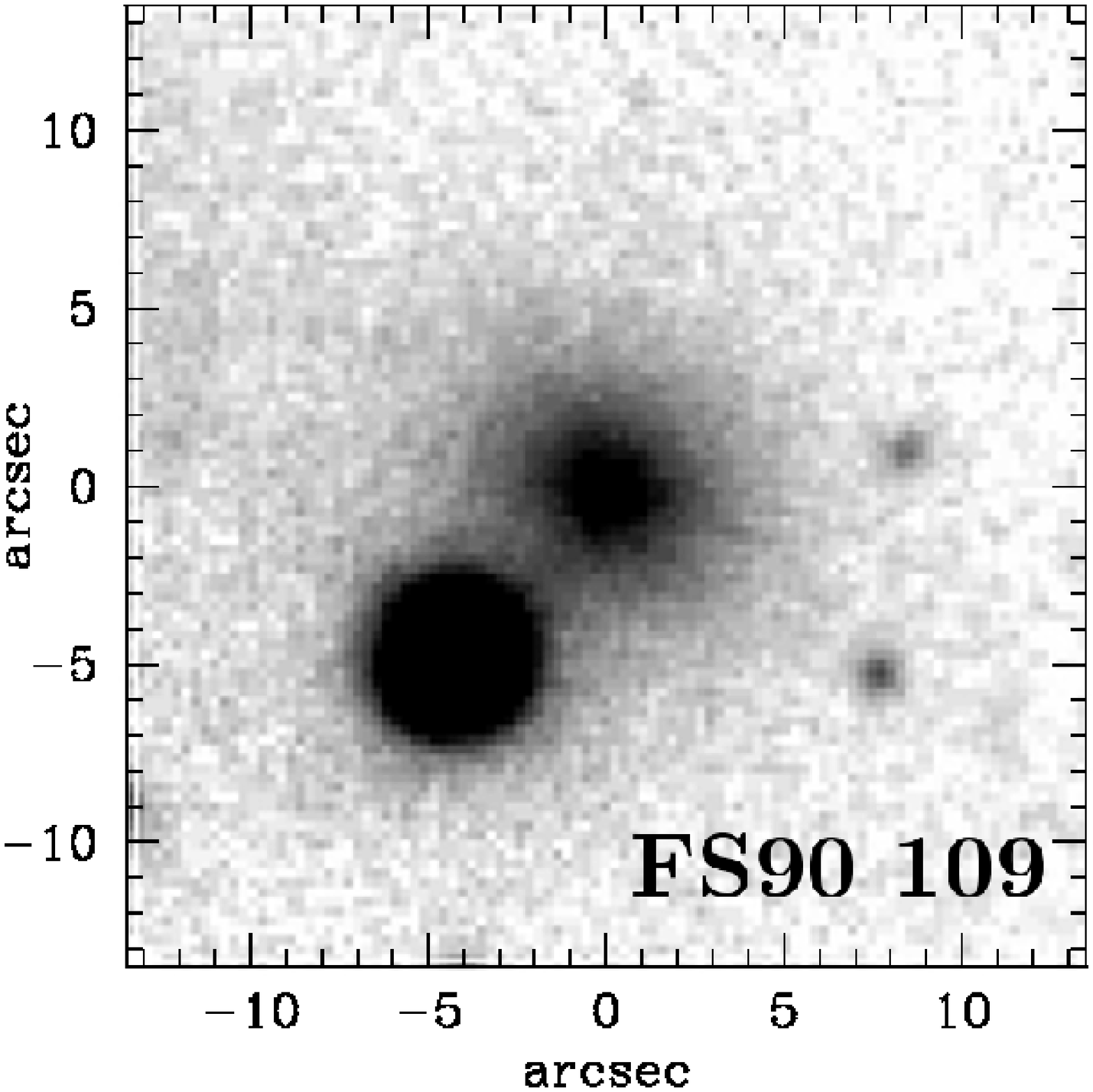}  
\includegraphics[scale=0.16]{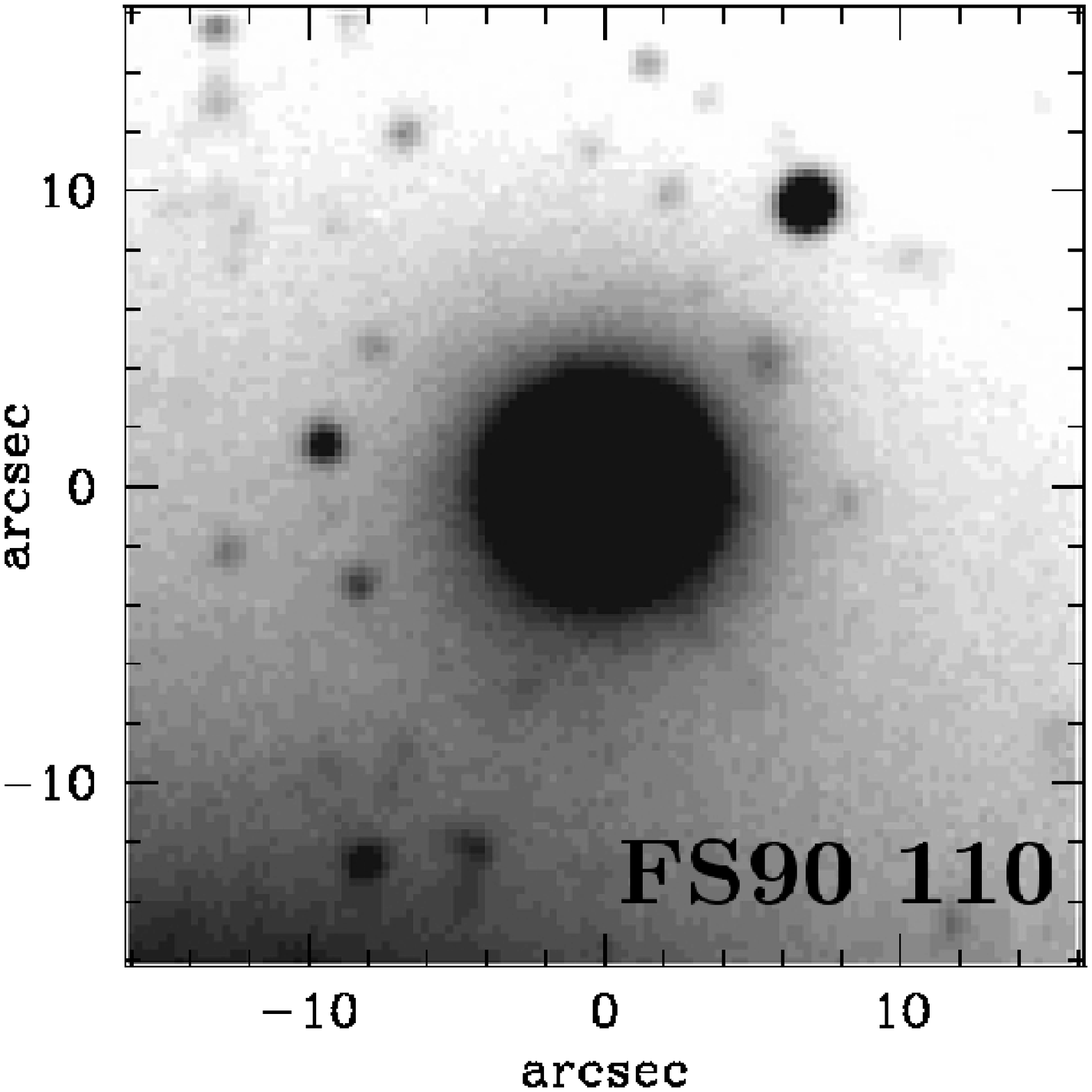}  
\includegraphics[scale=0.16]{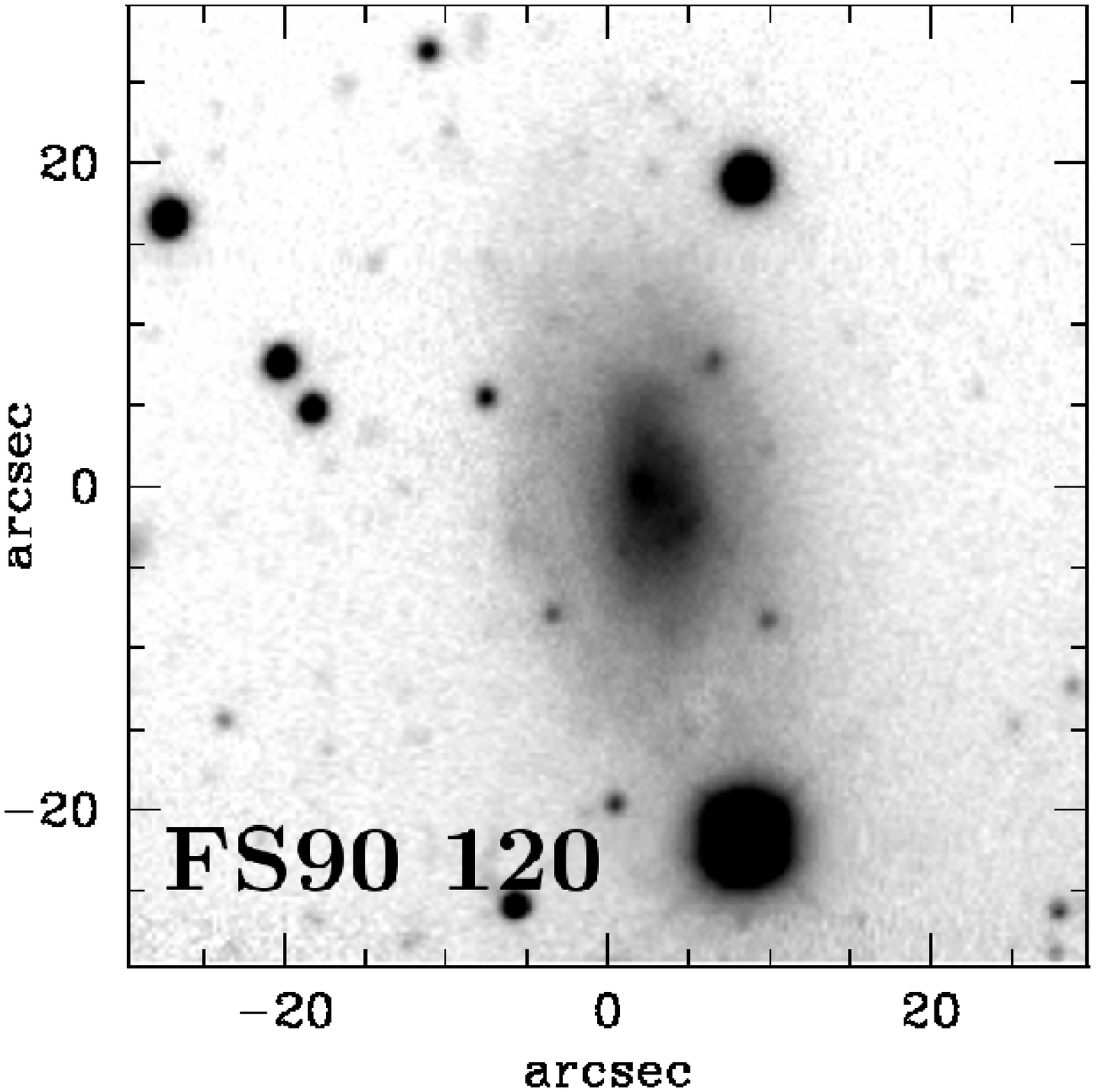}  
\includegraphics[scale=0.16]{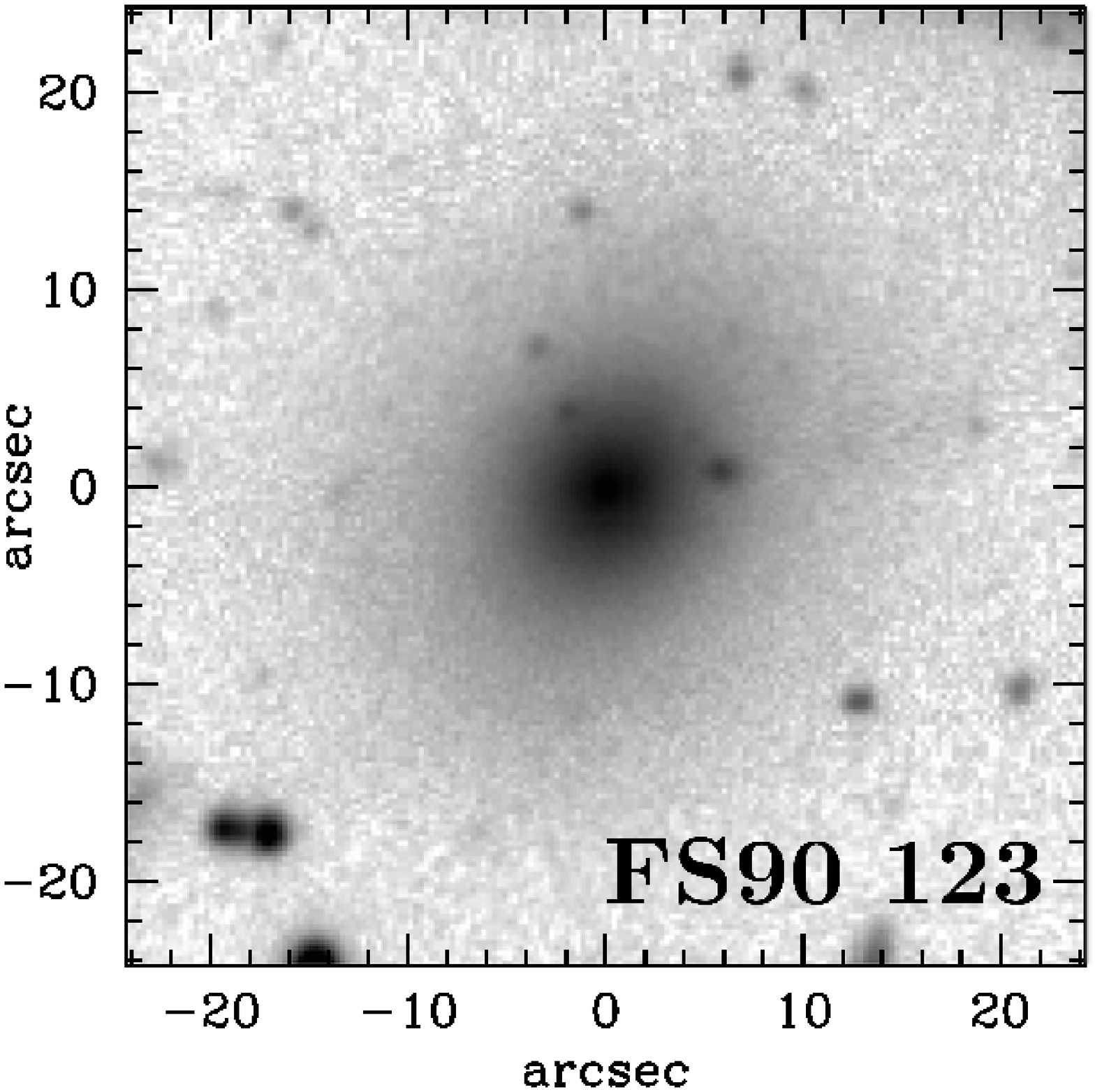}  
\includegraphics[scale=0.16]{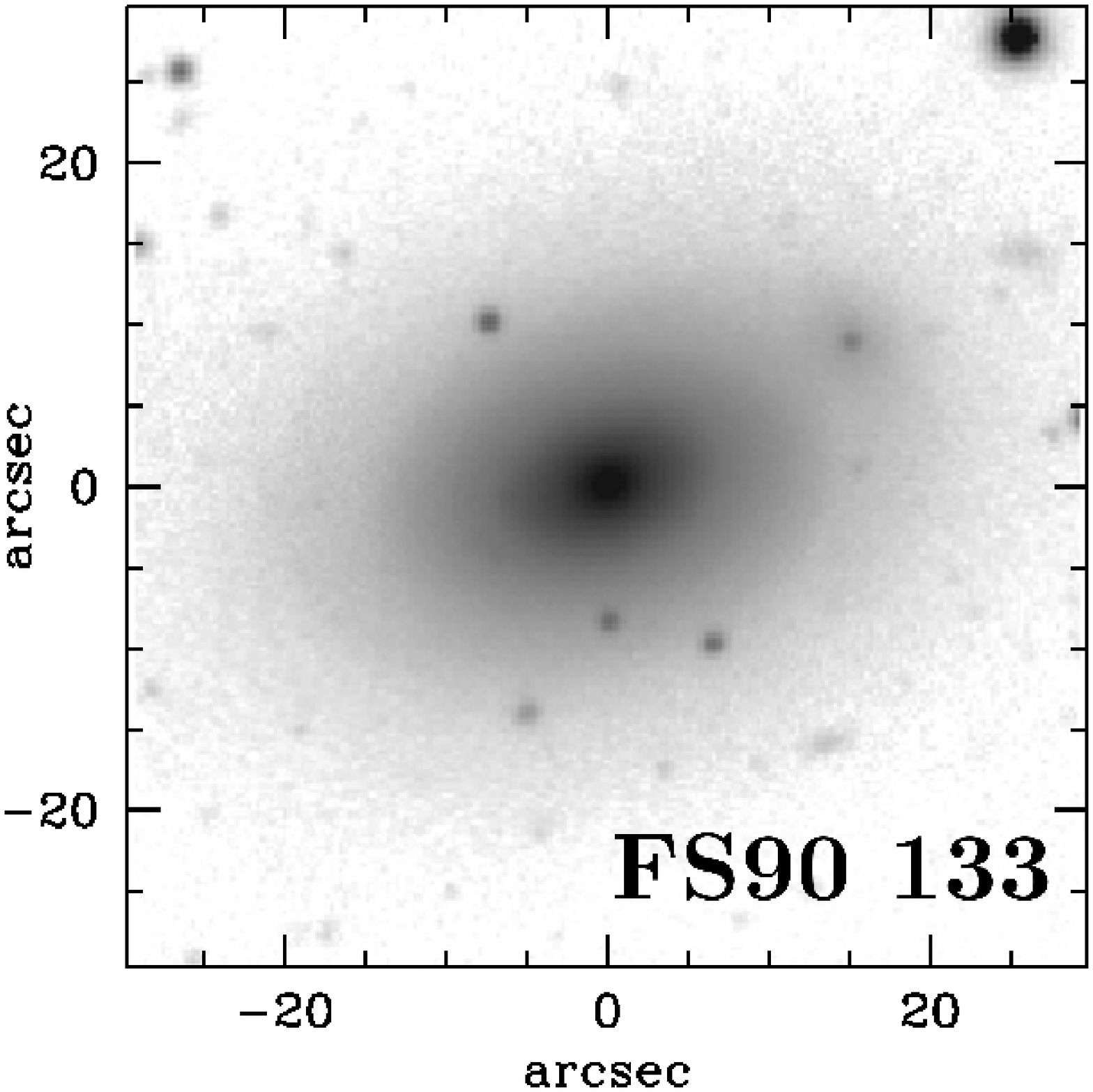}  
\includegraphics[scale=0.16]{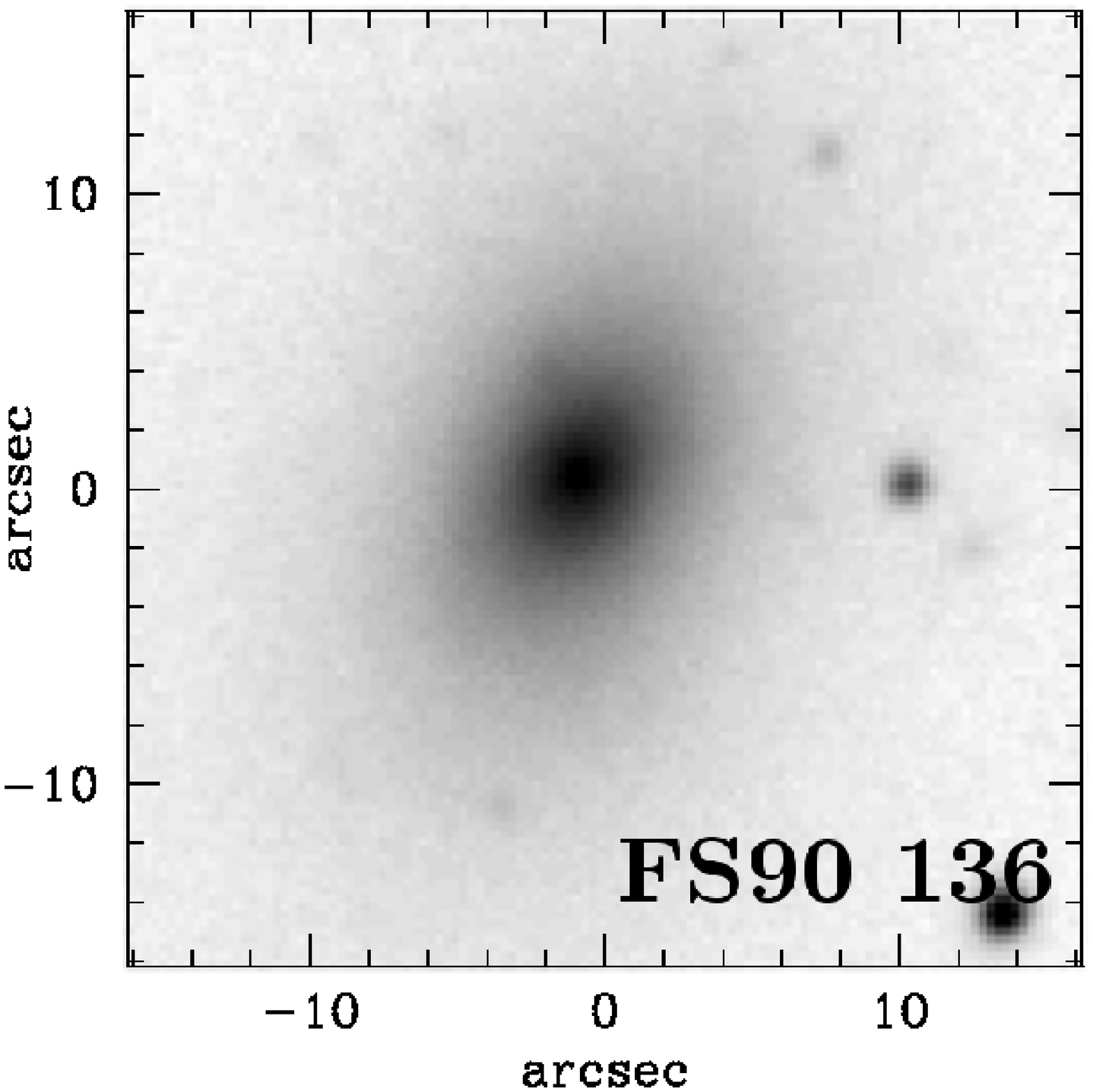}  
\includegraphics[scale=0.16]{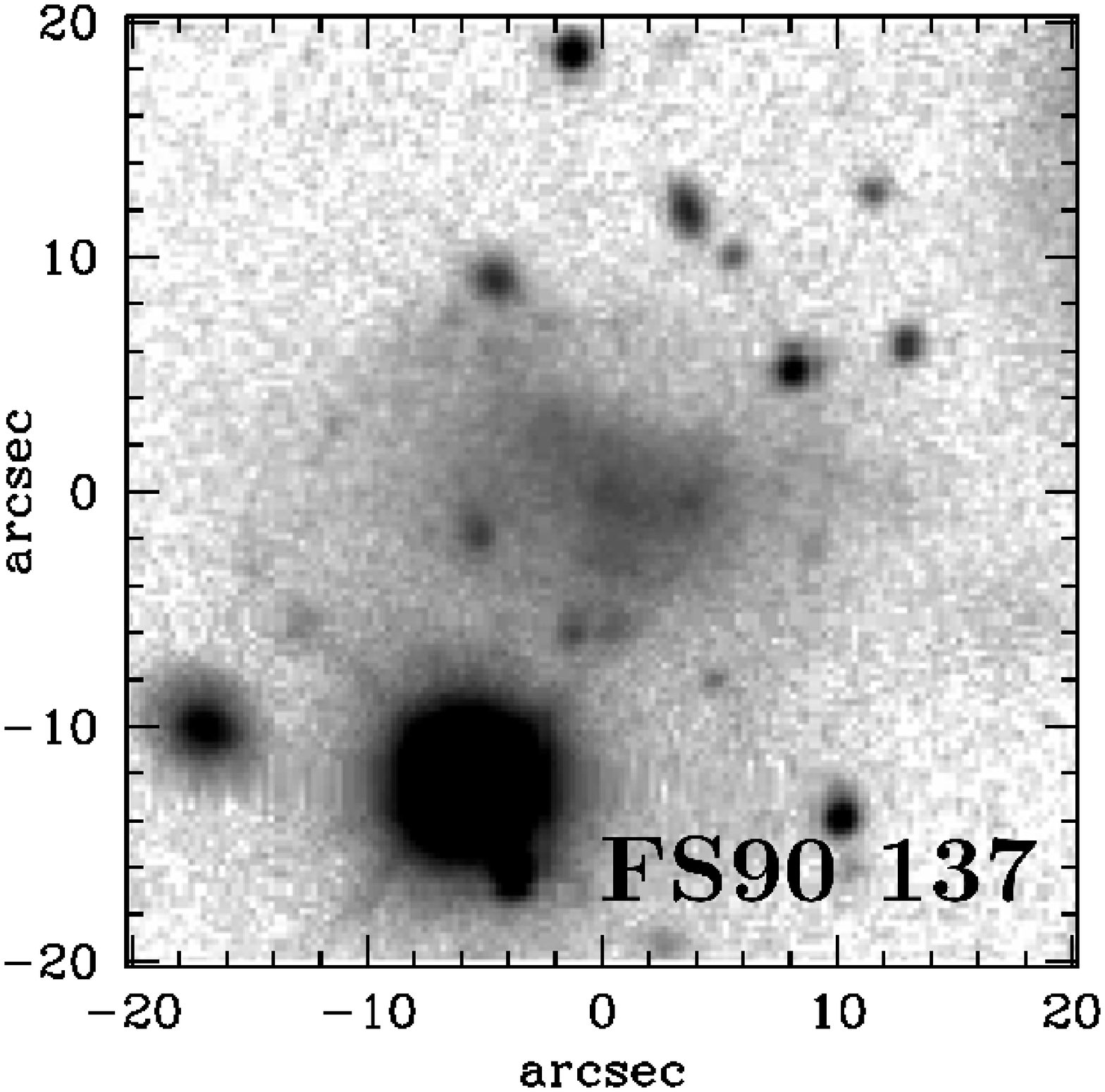}  
\includegraphics[scale=0.16]{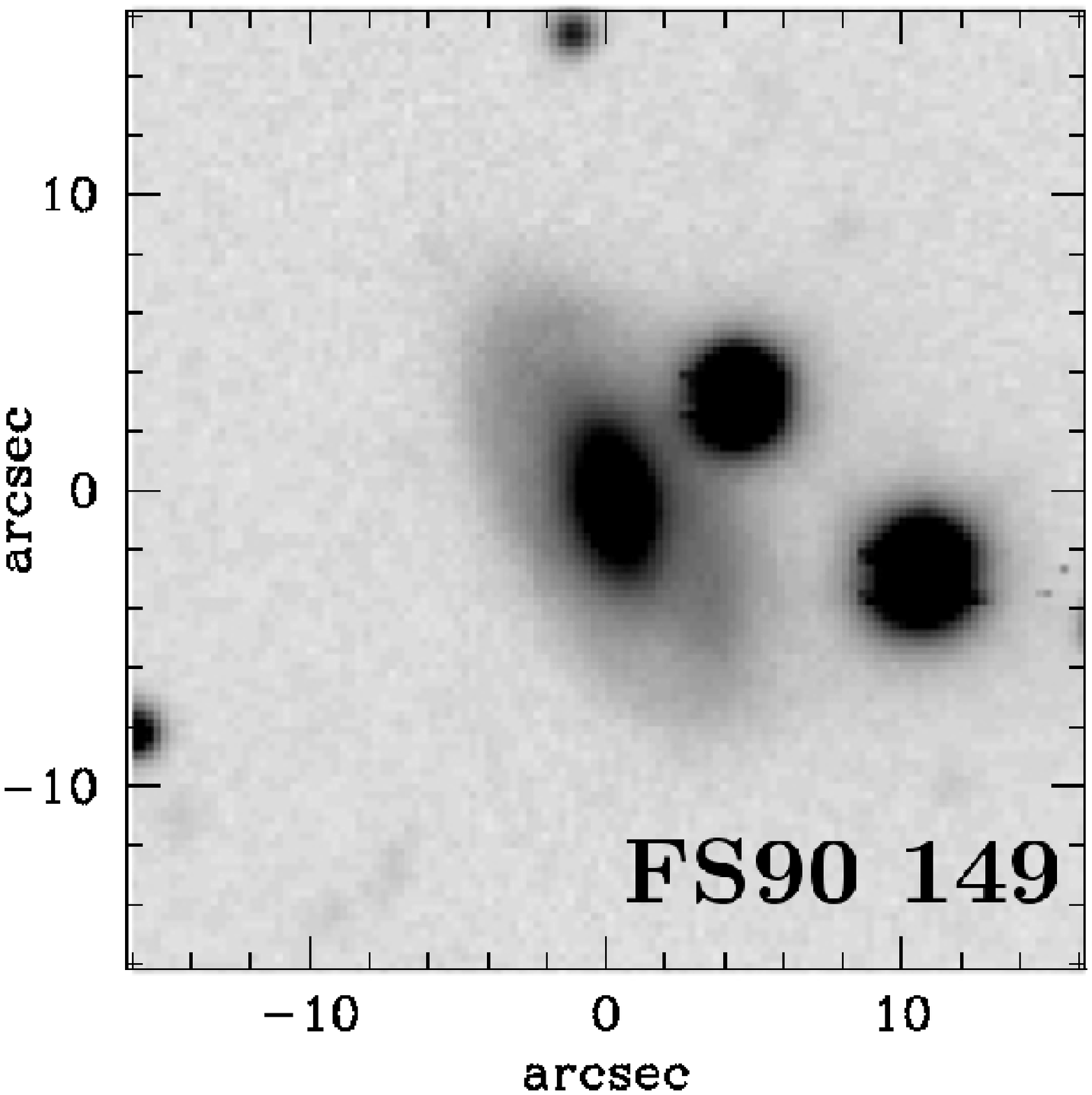}  
\includegraphics[scale=0.16]{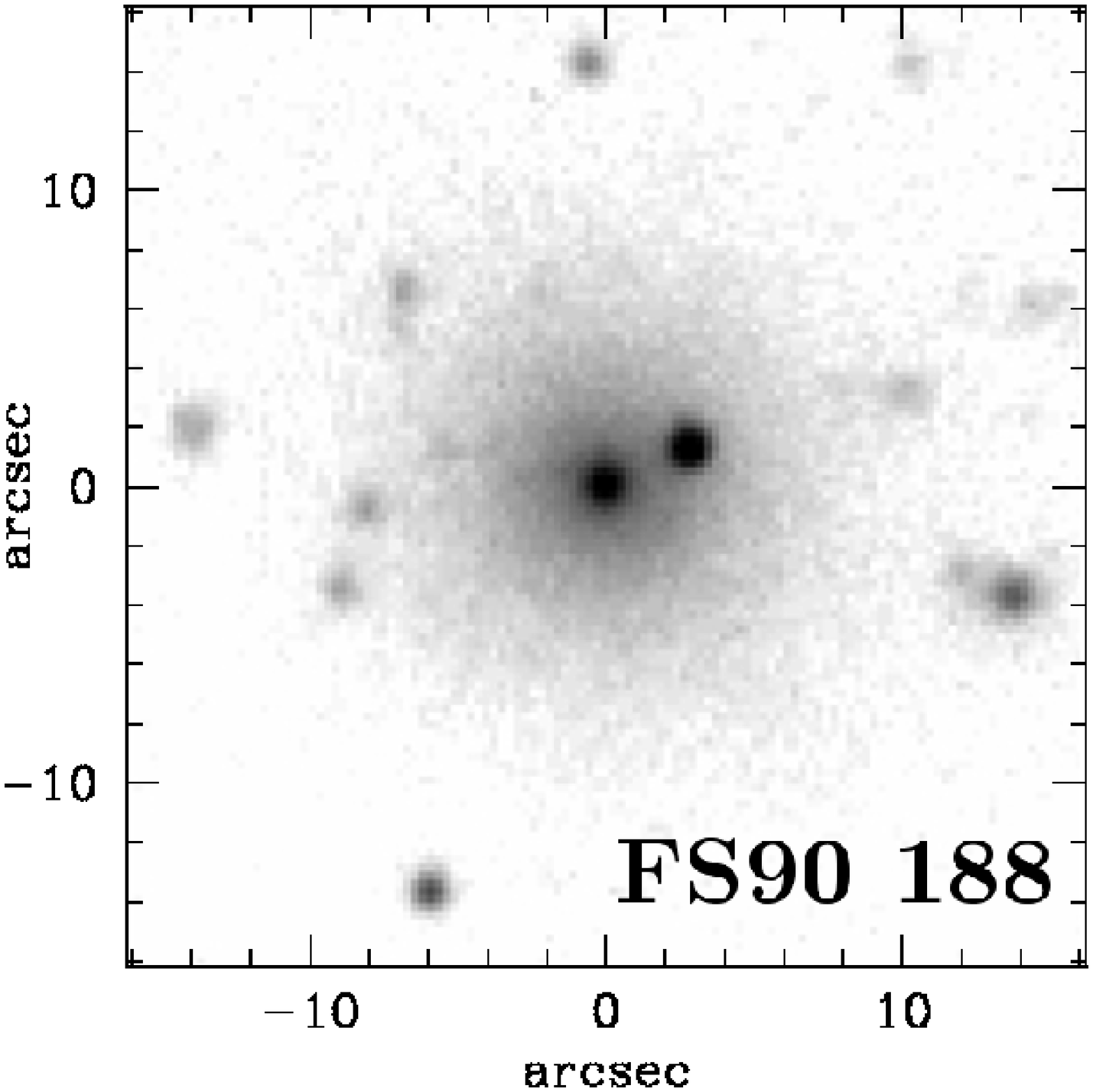}  
\includegraphics[scale=0.16]{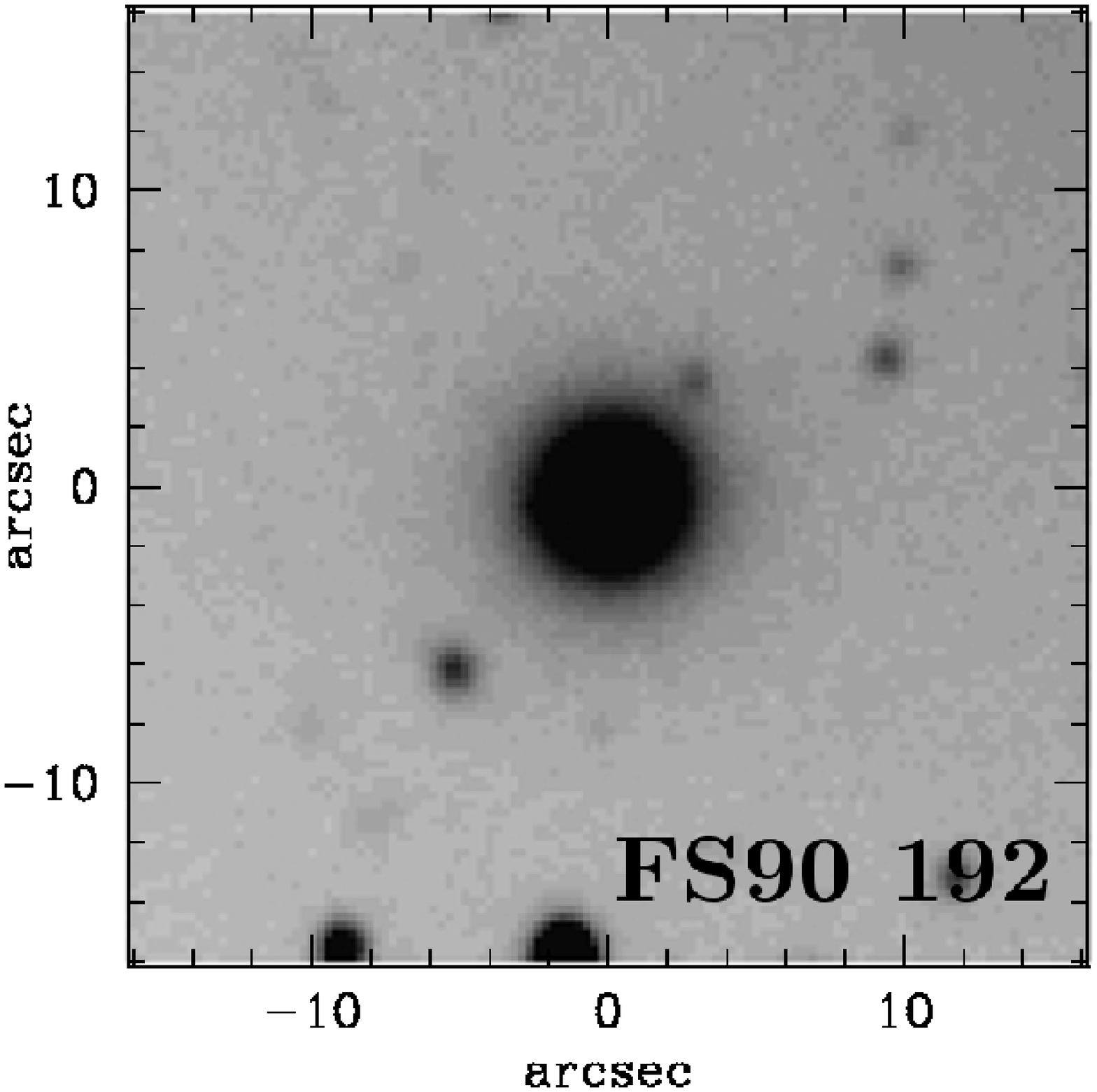}  
\includegraphics[scale=0.16]{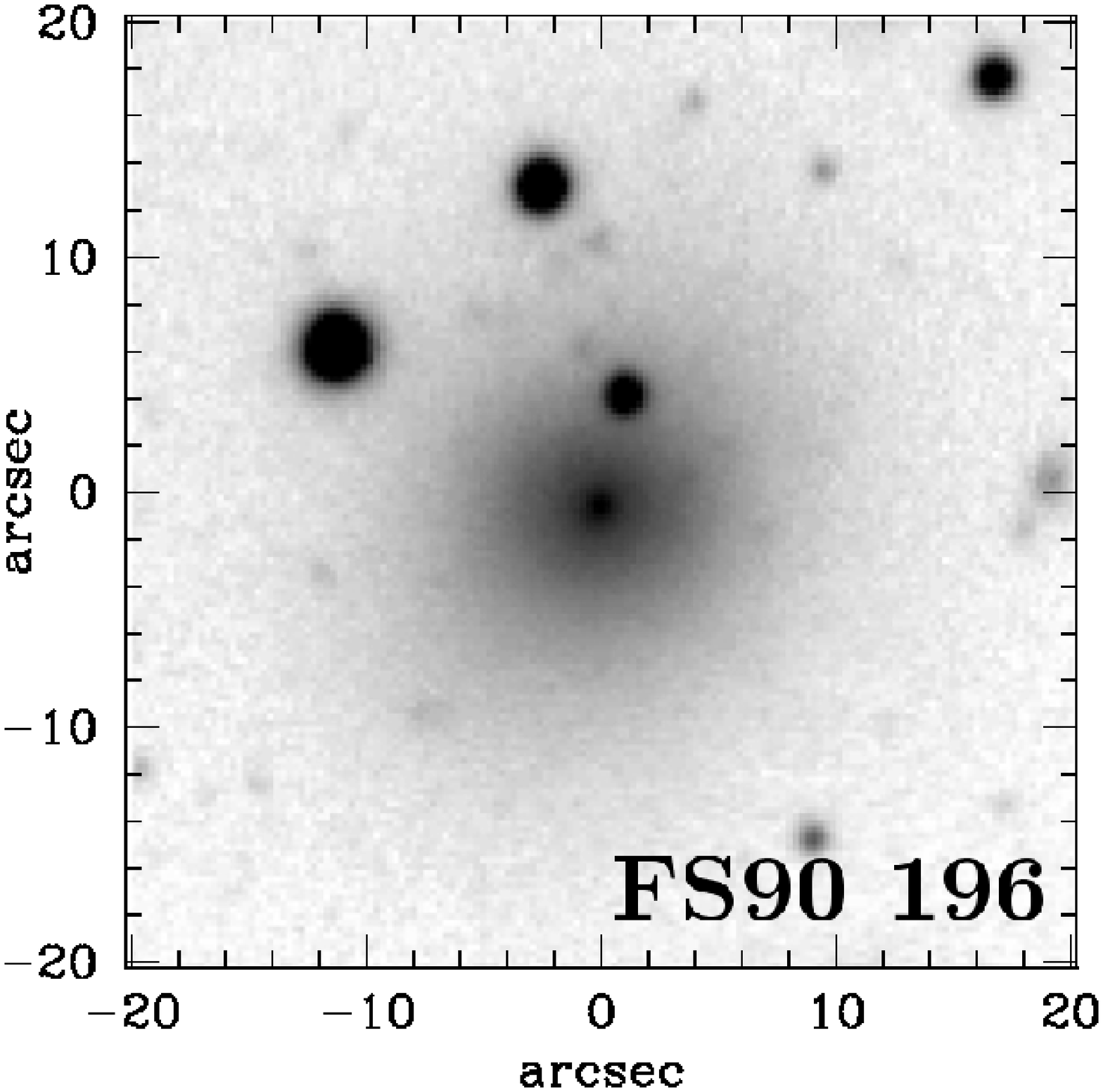}  
\includegraphics[scale=0.16]{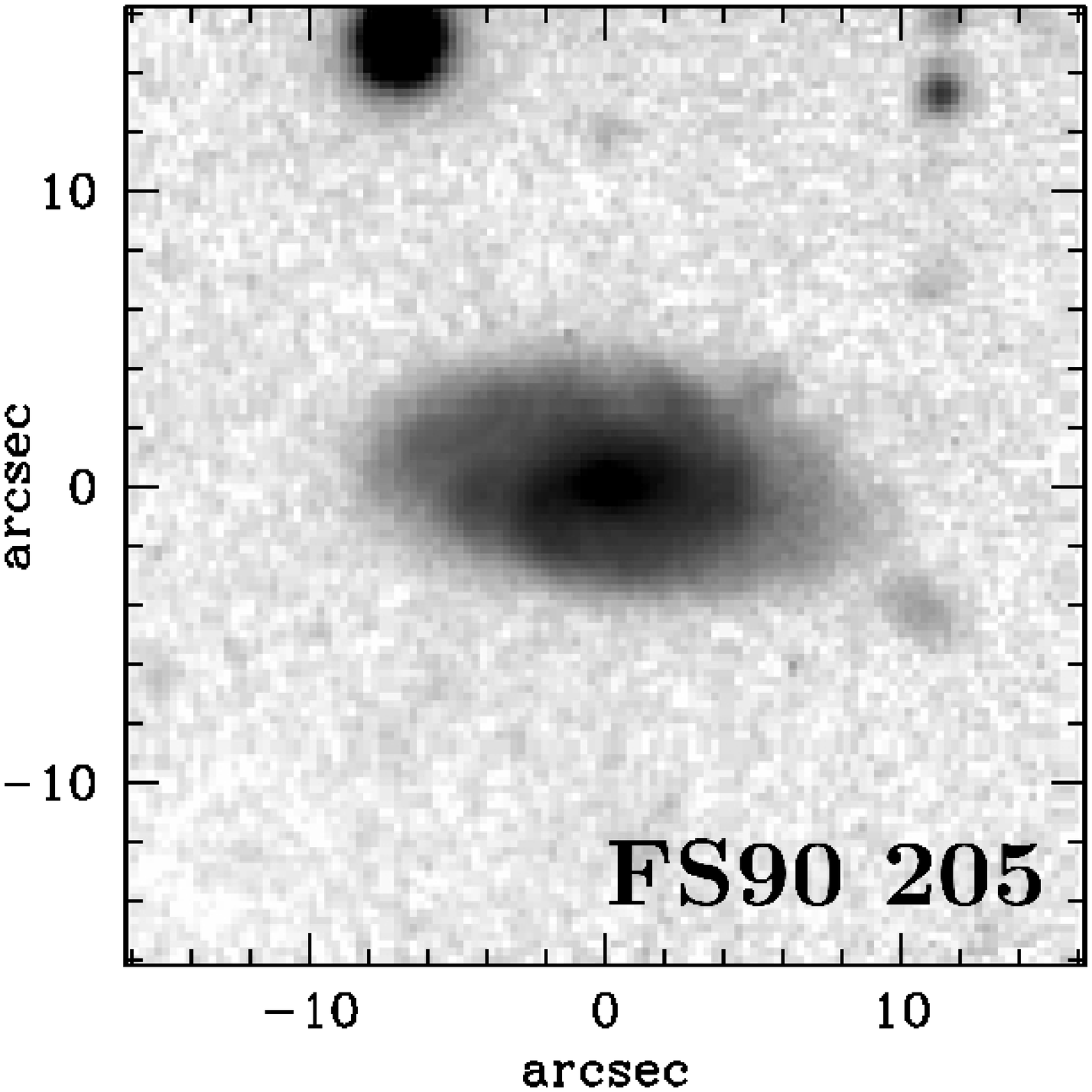}  
\includegraphics[scale=0.16]{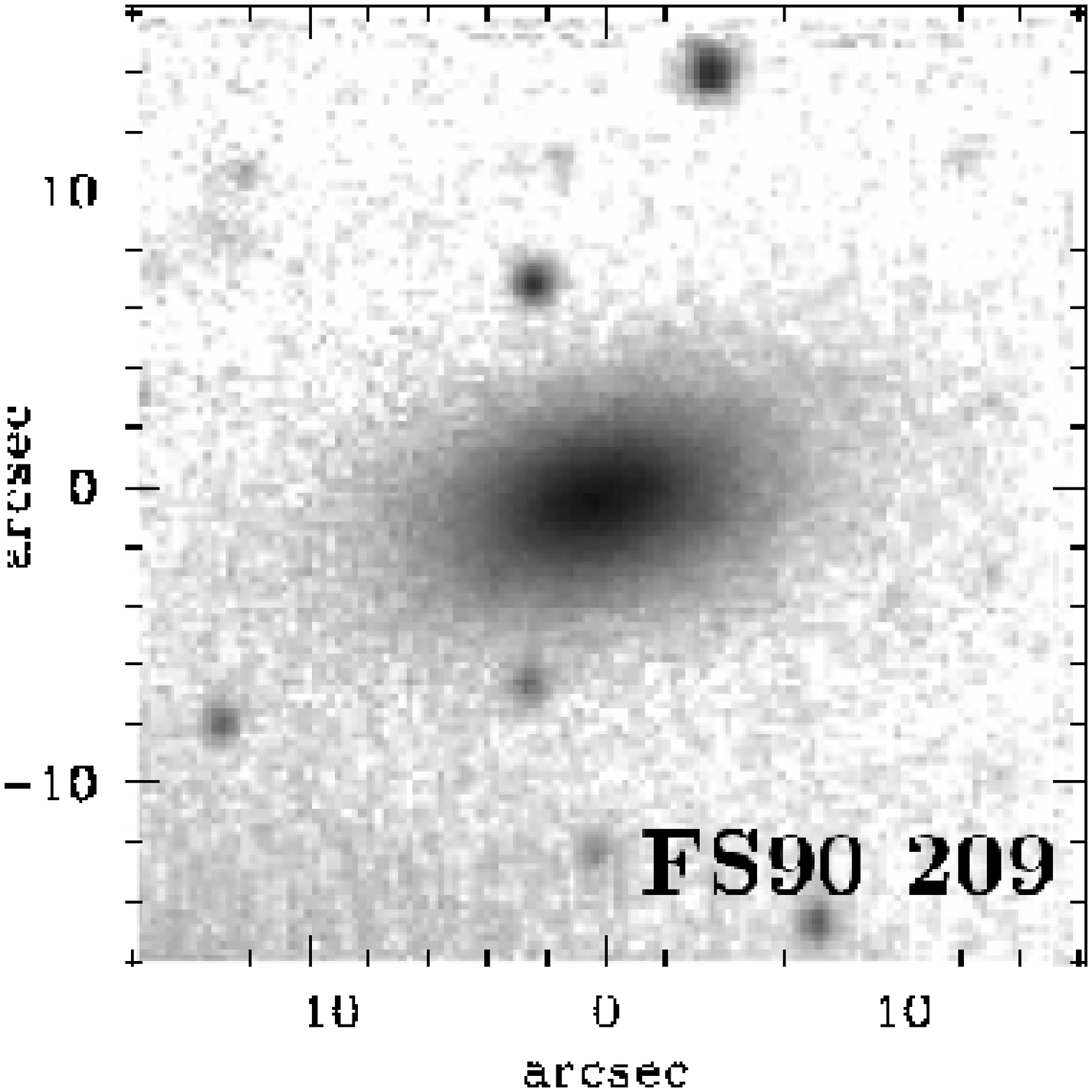}  
\includegraphics[scale=0.16]{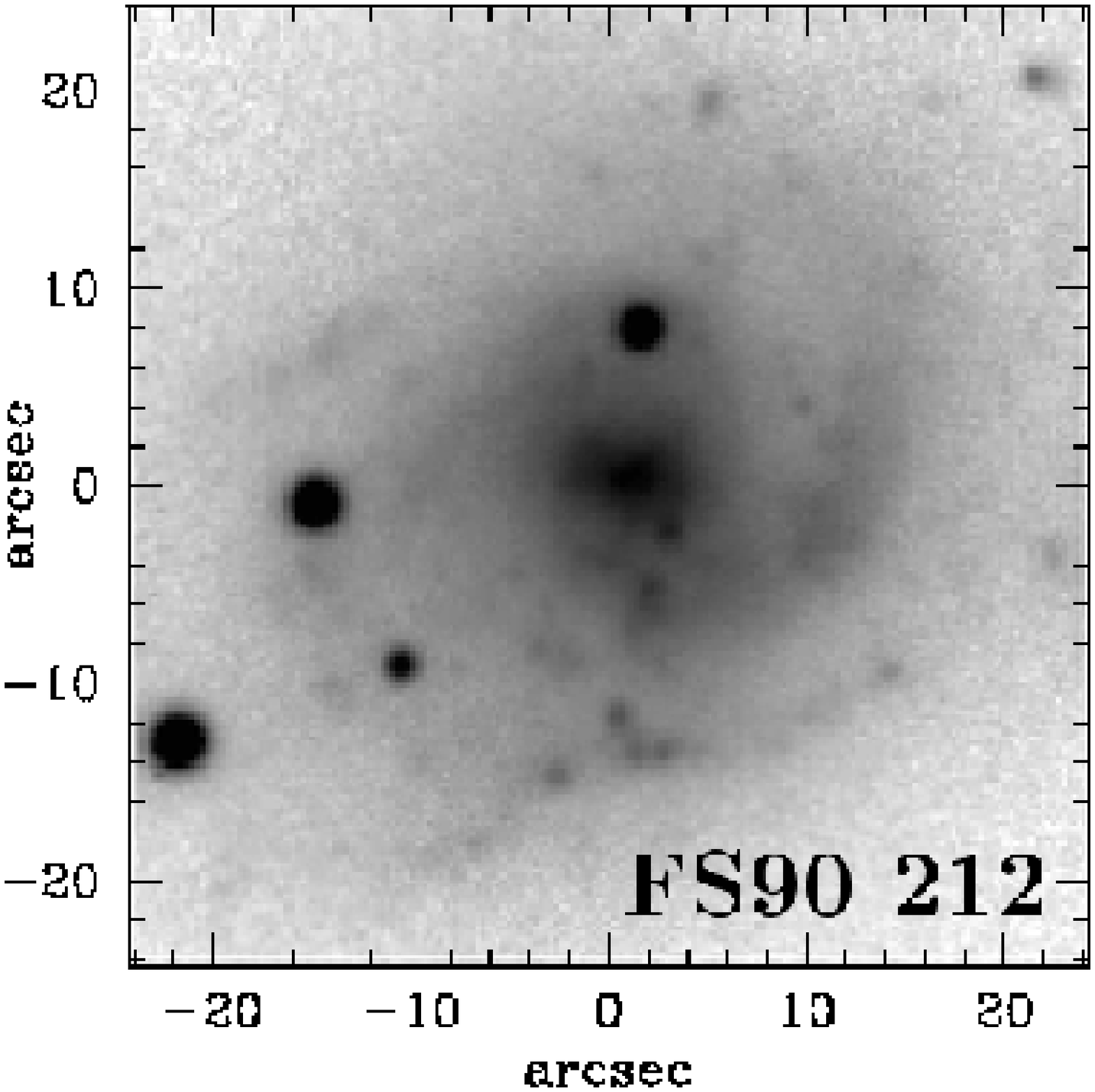}  
\includegraphics[scale=0.16]{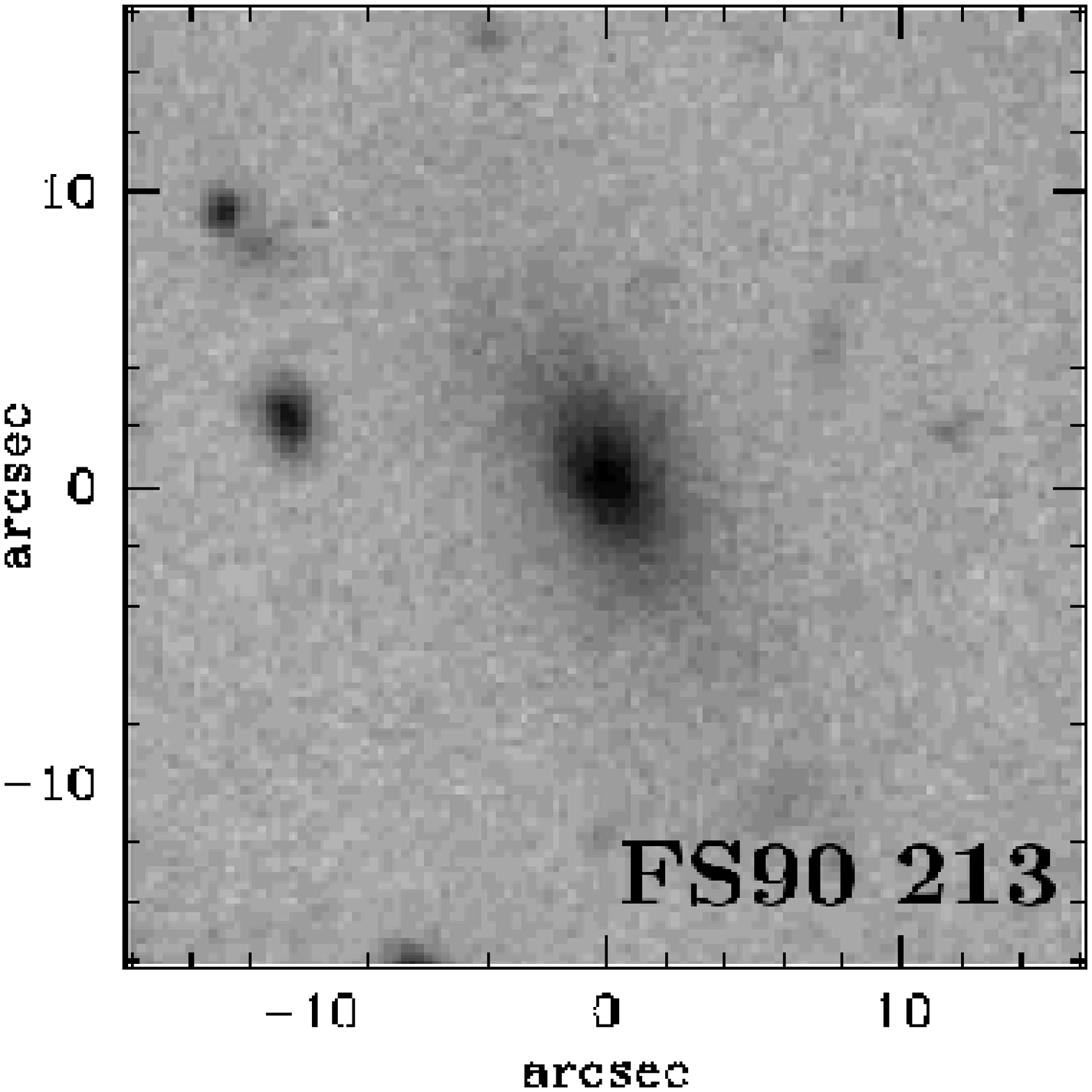}  
\includegraphics[scale=0.16]{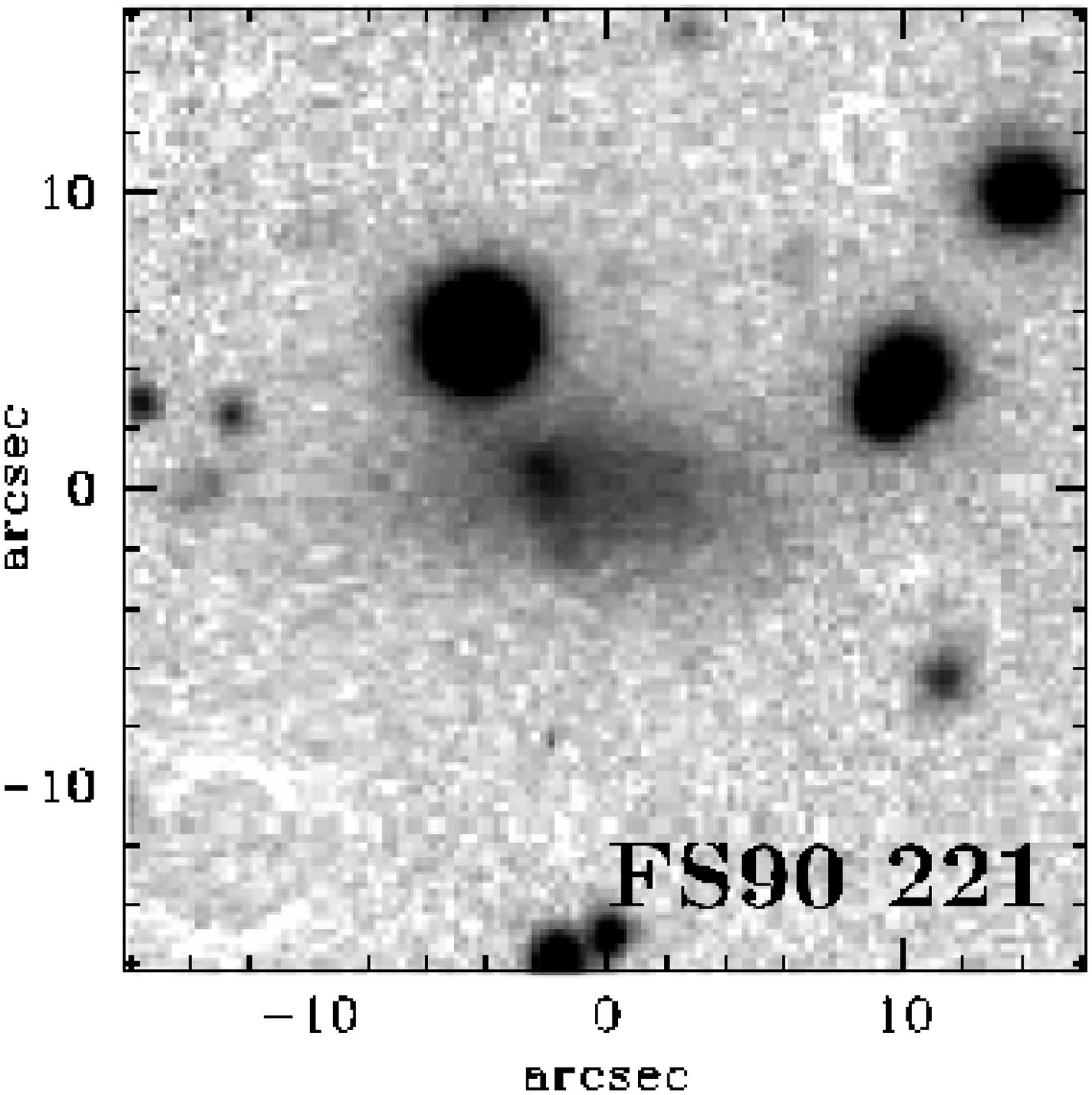}  
\includegraphics[scale=0.16]{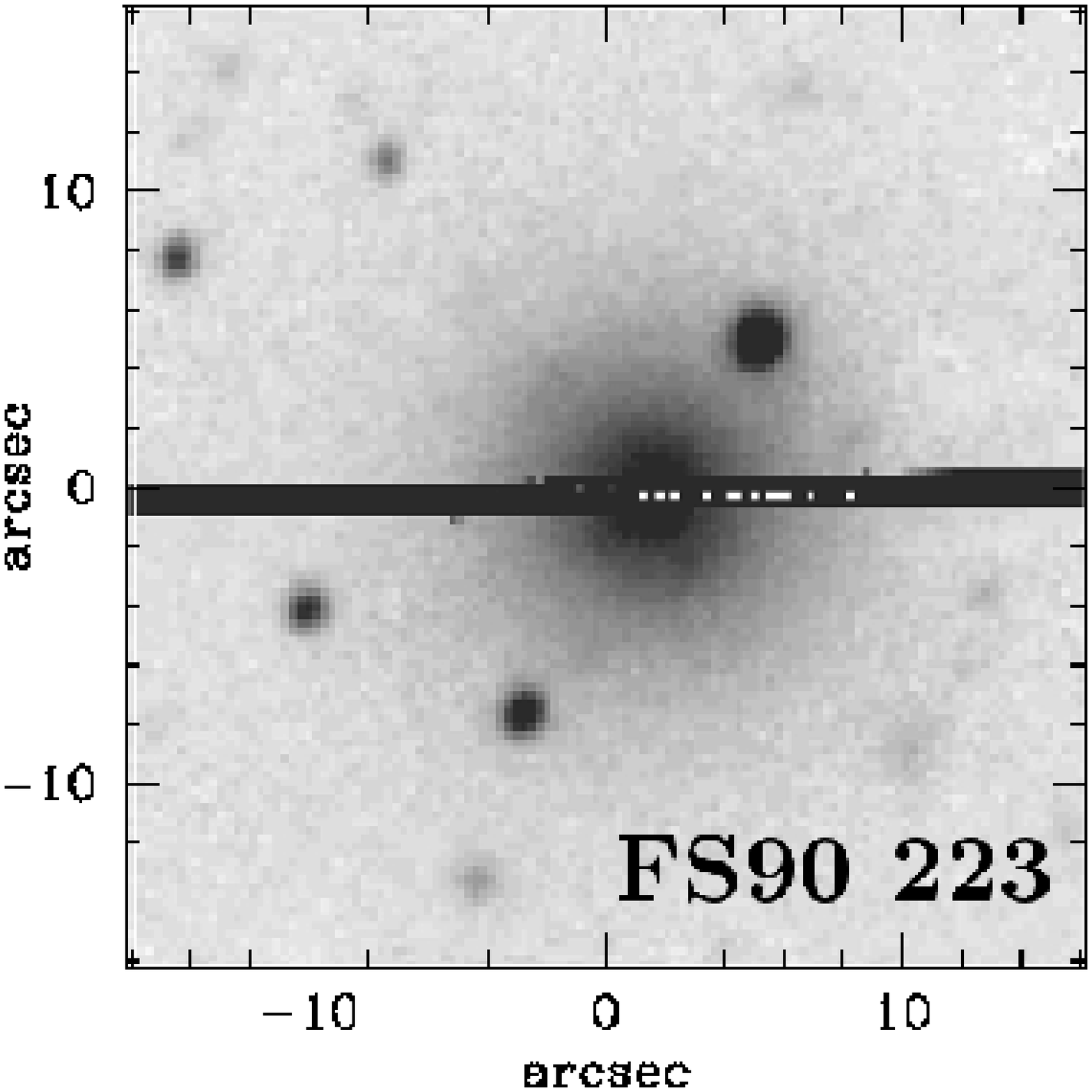}  
\includegraphics[scale=0.16]{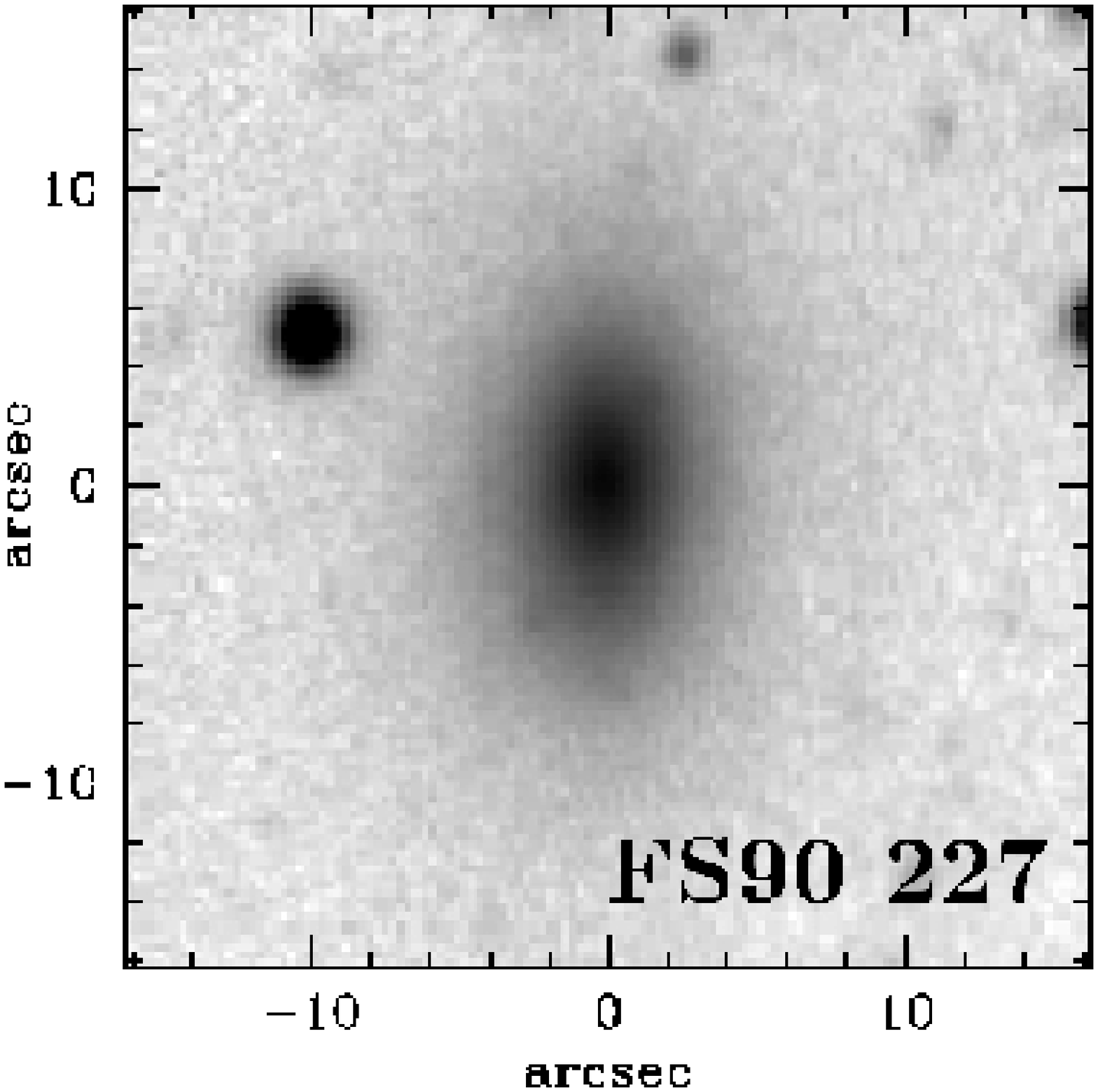}  
\includegraphics[scale=0.16]{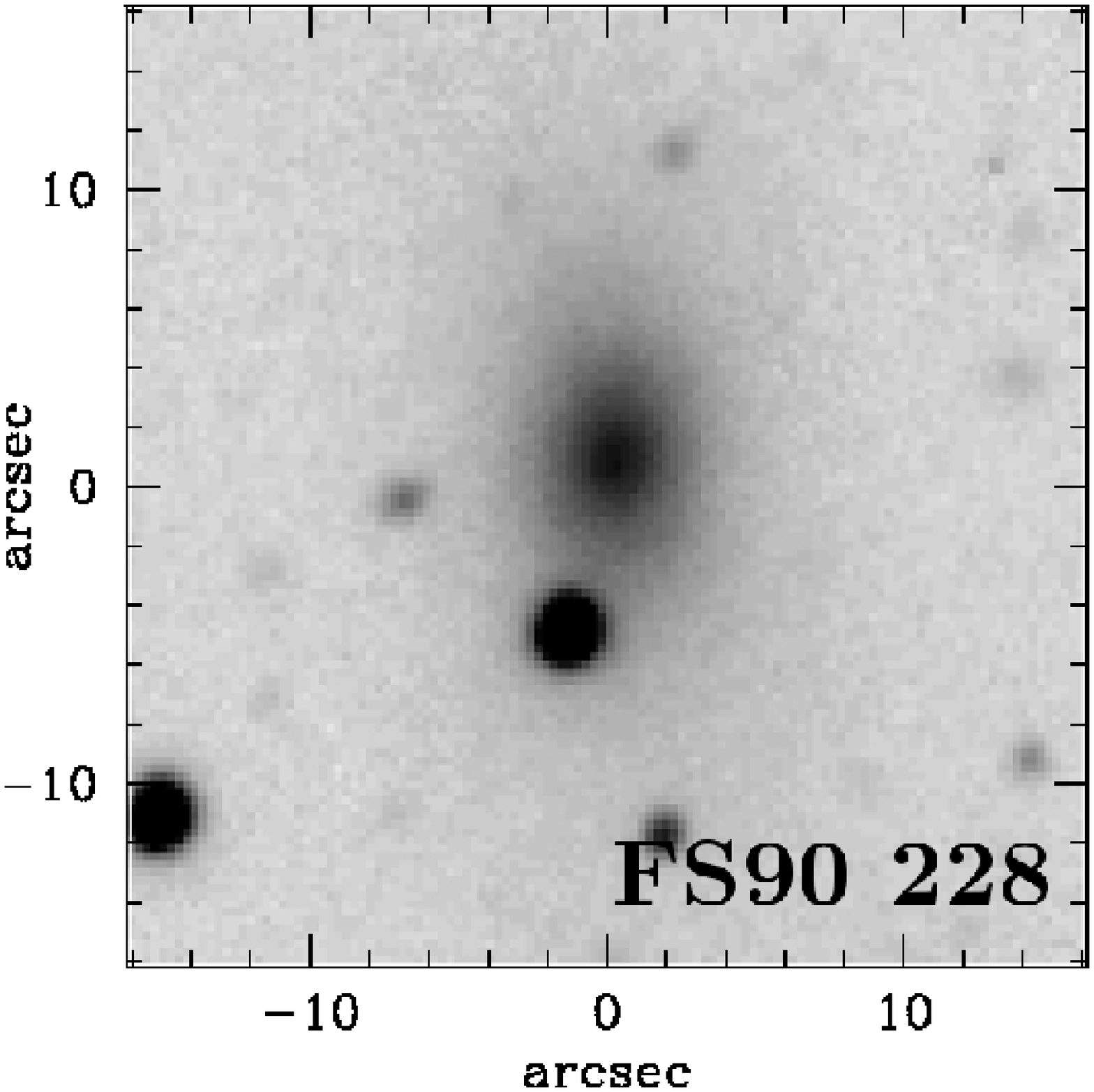}  
\caption{Logarithmic scale $R$ images of FS90 galaxies with new radial 
velocities. North is up and east to the left. Background galaxies are FS90\,83,
FS90\,149 and FS90\,205. The rest are all Antlia members.} 
\label{fig1_ch4} 
\end{center} 
\end{figure*}

In this appendix we present the new measured radial velocities 
of 23 FS90 galaxies (Table\,\ref{velocidades}), as well as 
their logarithmic scale $R$ images (Fig.\,\ref{fig1_ch4}). \\ 
 
\begin{table} 
\caption{FS90 galaxies with new radial velocities.}
\label{velocidades} 
\begin{tabular}{@{}lcccc@{}c@{}r@{}c@{}l@{}} 
\hline 
\\ 
\multicolumn{1}{c}{ } & \multicolumn{1}{c}{FS90} 
& \multicolumn{1}{c} {$\alpha$} &
\multicolumn{1}{c} {$\delta$} & \multicolumn{1}{c}{FS90} &
\multicolumn{1}{c}{FS90} & \multicolumn{3}{c}{v$_r$} \\
\multicolumn{1}{c}{ } & \multicolumn{1}{c}{ID} &\multicolumn{1}{c} {(J2000)}&\multicolumn{1}{c}
            {(J2000)} & \multicolumn{1}{c}{mor.} &
            \multicolumn{1}{c}{status} & \multicolumn{3}{c}{km s$^{-1}$} \\
\hline 
\\ 
1 & FS90\,  70  & 10:28:06.9 & -35:35:20 & dE 	    & 1 &  2864&$\pm$&70  \\ 
2 & FS90\,  83  & 10:28:23.0 & -35:30:57 & S or Sm  & 3 & 19685&$\pm$&33  \\ 
3 & FS90\,  85  & 10:28:24.0 & -35:34:22 & dE 	    & 1 &  2000&$\pm$&200 \\ 
4 & FS90\,103 & 10:28:45.1 & -35:34:40 & dE         & 3 &  2054&$\pm$&29  \\ 
5 & FS90\,109 & 10:28:53.0 & -35:32:52 & dE         & 2 &  1618&$\pm$&24  \\ 
6 & FS90\,110 & 10:28:53.0 & -35:35:34 & E(M32?)    & 3 &  2911&$\pm$&7   \\ 
7 & FS90\,120 & 10:29:02.1 & -35:34:04 & ImV        & 1 &  2634&$\pm$&13  \\ 
8 & FS90\,123 & 10:29:03.1 & -35:40:30 & dE,N       & 2 &  1865&$\pm$&25  \\ 
9 & FS90\,133 & 10:29:12.0 & -35:39:28 & d:E,N      & 1 &  2205&$\pm$&24  \\ 
10 & FS90\,136 & 10:29:15.3 & -35:25:58 & dE,N      & 1 &  2989&$\pm$&10  \\ 
11 & FS90\,137 & 10:29:15.1 & -35:41:34 & ImV       & 2 &  3987&$\pm$&36  \\ 
12 & FS90\,149 & 10:29:27.3 & -35:27:10 & S0 or dS0 & 3 & 46175&$\pm$&13  \\ 
13 & FS90\,188 & 10:30:02.4 & -35:24:28 & dE        & 1 &  2673&$\pm$&17  \\ 
14 & FS90\,192 & 10:30:04.5 & -35:20:31 & E(M32?)   & 3 &  2526&$\pm$&4   \\ 
15 & FS90\,196 & 10:30:06.4 & -35:23:31 & dE        & 1 &  3593&$\pm$&9   \\ 
16 & FS90\,205 & 10:30:18.4 & -35:24:43 & dE        & 2 & 45909&$\pm$&18  \\ 
17 & FS90\,209 & 10:30:19.4 & -35:34:48 & dE        & 2 &  3065&$\pm$&13  \\ 
18 & FS90\,212 & 10:30:21.3 & -35:35:31 & SmIII     & 1 &  2364&$\pm$&27  \\ 
19 & FS90\,213 & 10:30:21.6 & -35:12:14 & dE        & 2 &  2185&$\pm$&21  \\ 
20 & FS90\,221 & 10:30:25.4 & -35:23:38 & dE        & 2 &  3556&$\pm$&130 \\ 
21 & FS90\,223 & 10:30:25.6 & -35:13:19 & dE        & 1 &  2661&$\pm$&9   \\ 
22 & FS90\,227 & 10:30:31.4 & -35:23:06 & dE?       & 2 &  2921&$\pm$&60  \\ 
23 & FS90\,228 & 10:30:31.6 & -35:14:38 & dE,N      & 1 &  2417&$\pm$&13  \\ 
\\ 
\hline 
\end{tabular} 
\label{tabla_FS90nrv} 
\end{table} 
 
\section{New dwarf galaxy members and candidates} 
\label{candidatas} 
\begin{figure*} 
\begin{center} 
\includegraphics[scale=0.23]{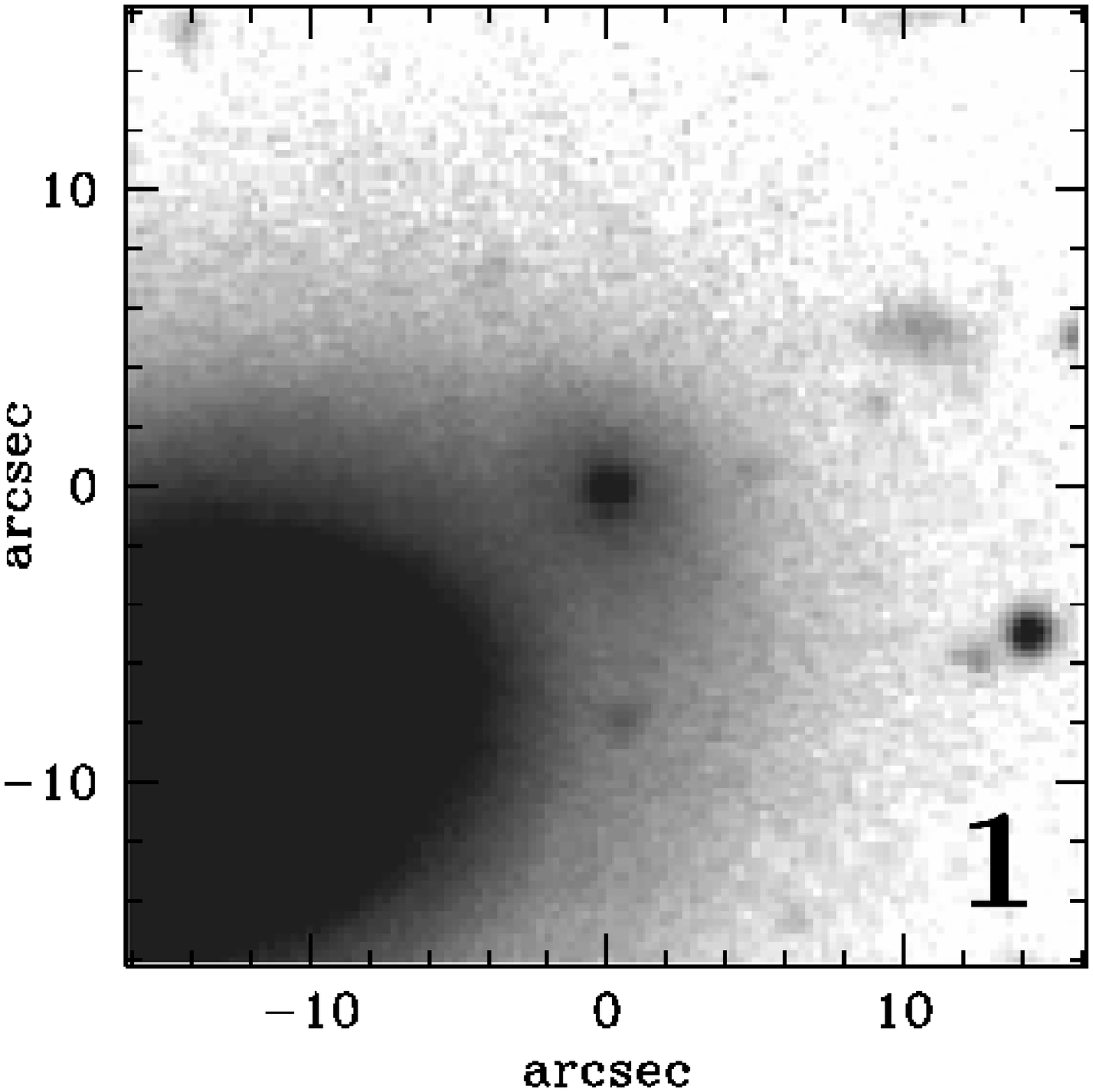} 
\includegraphics[scale=0.23]{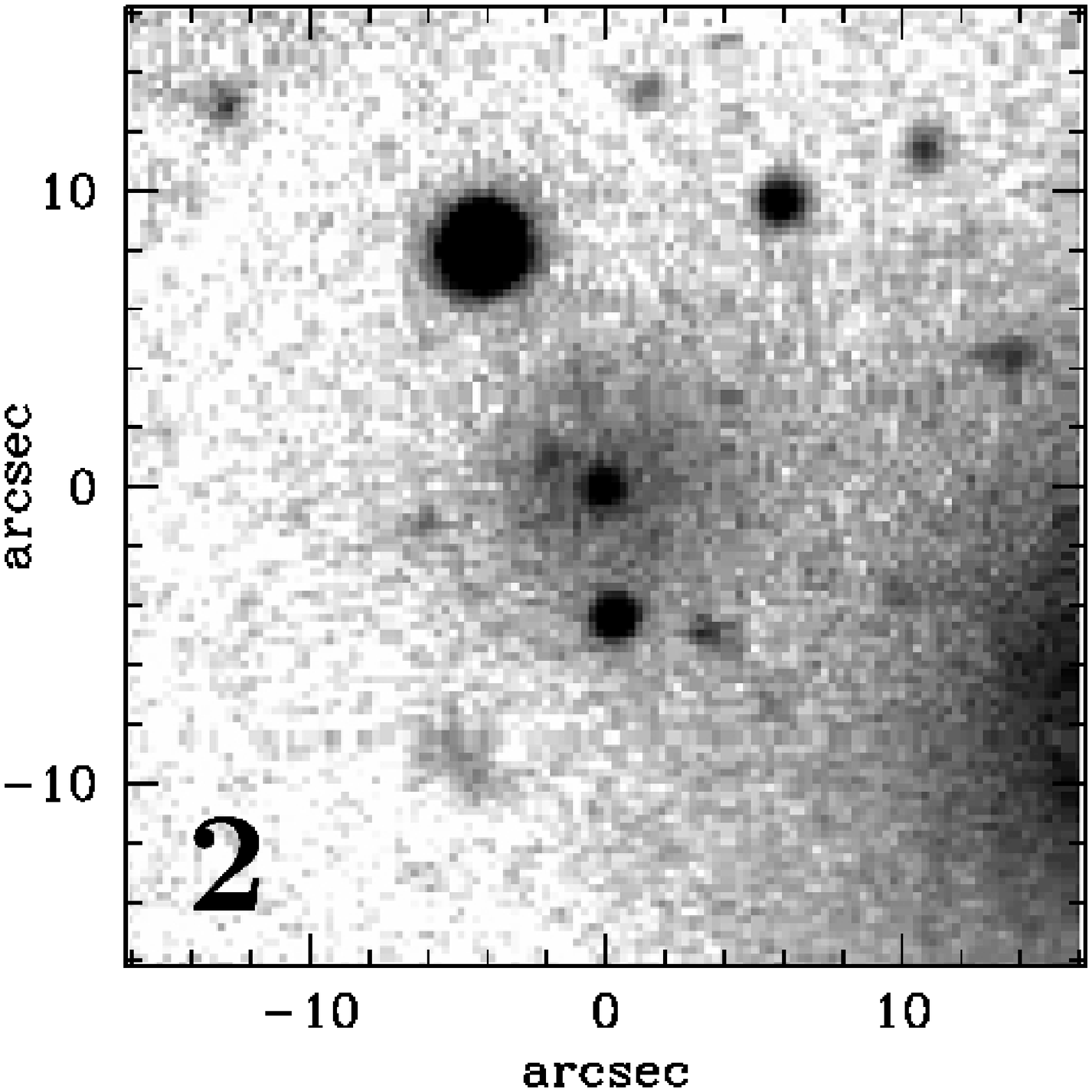} 
\includegraphics[scale=0.23]{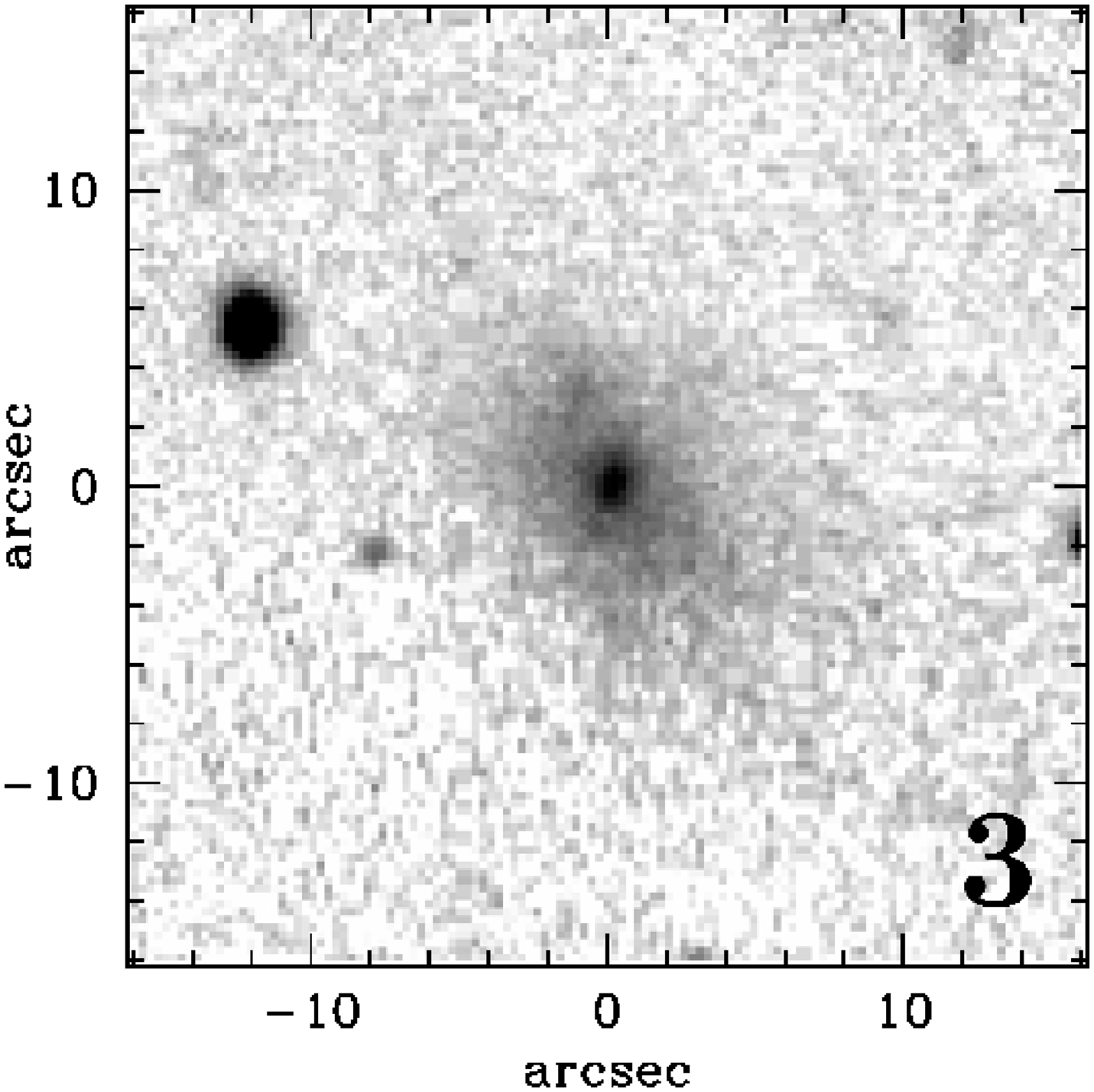} 
\includegraphics[scale=0.23]{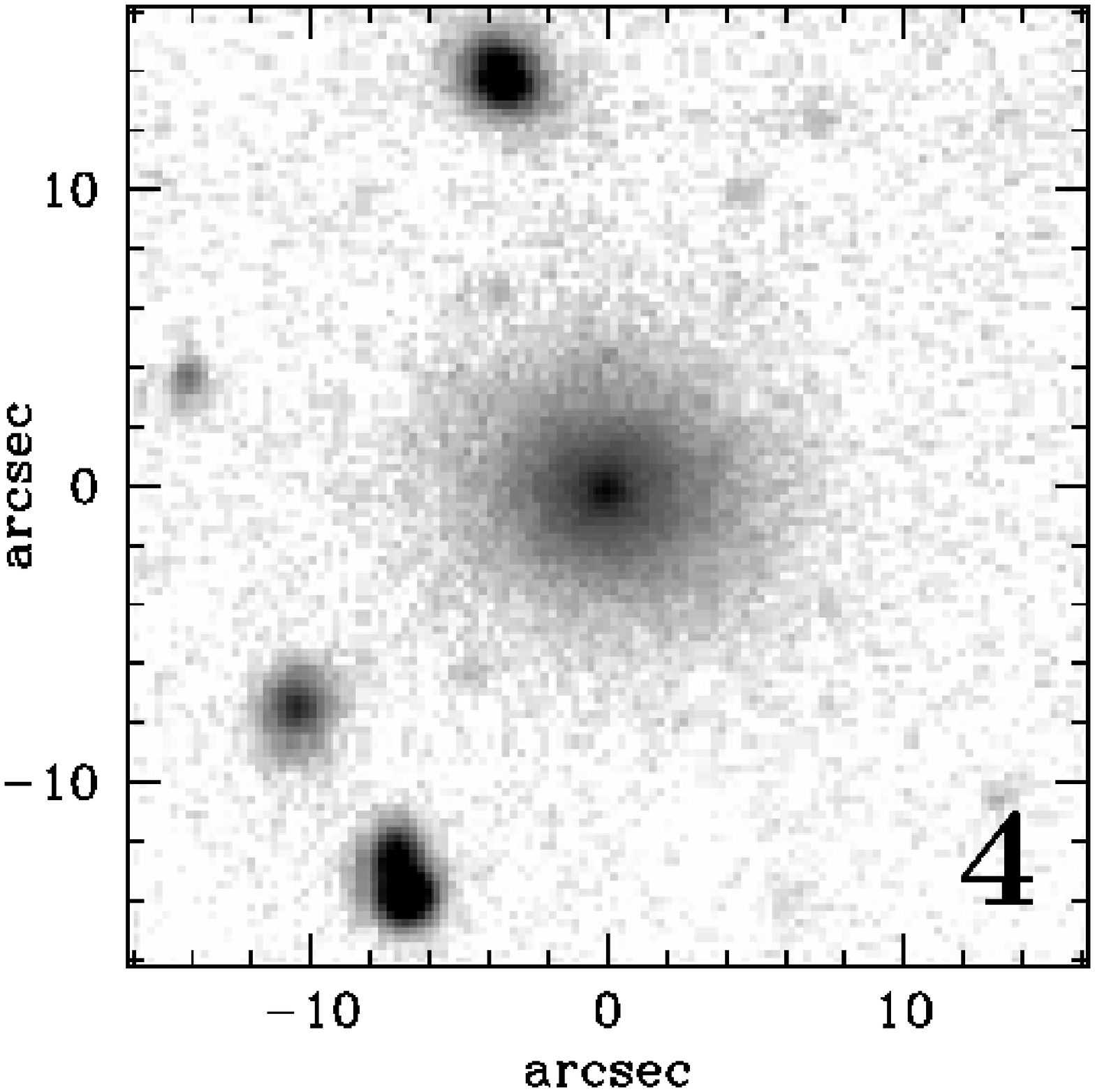} 
\includegraphics[scale=0.23]{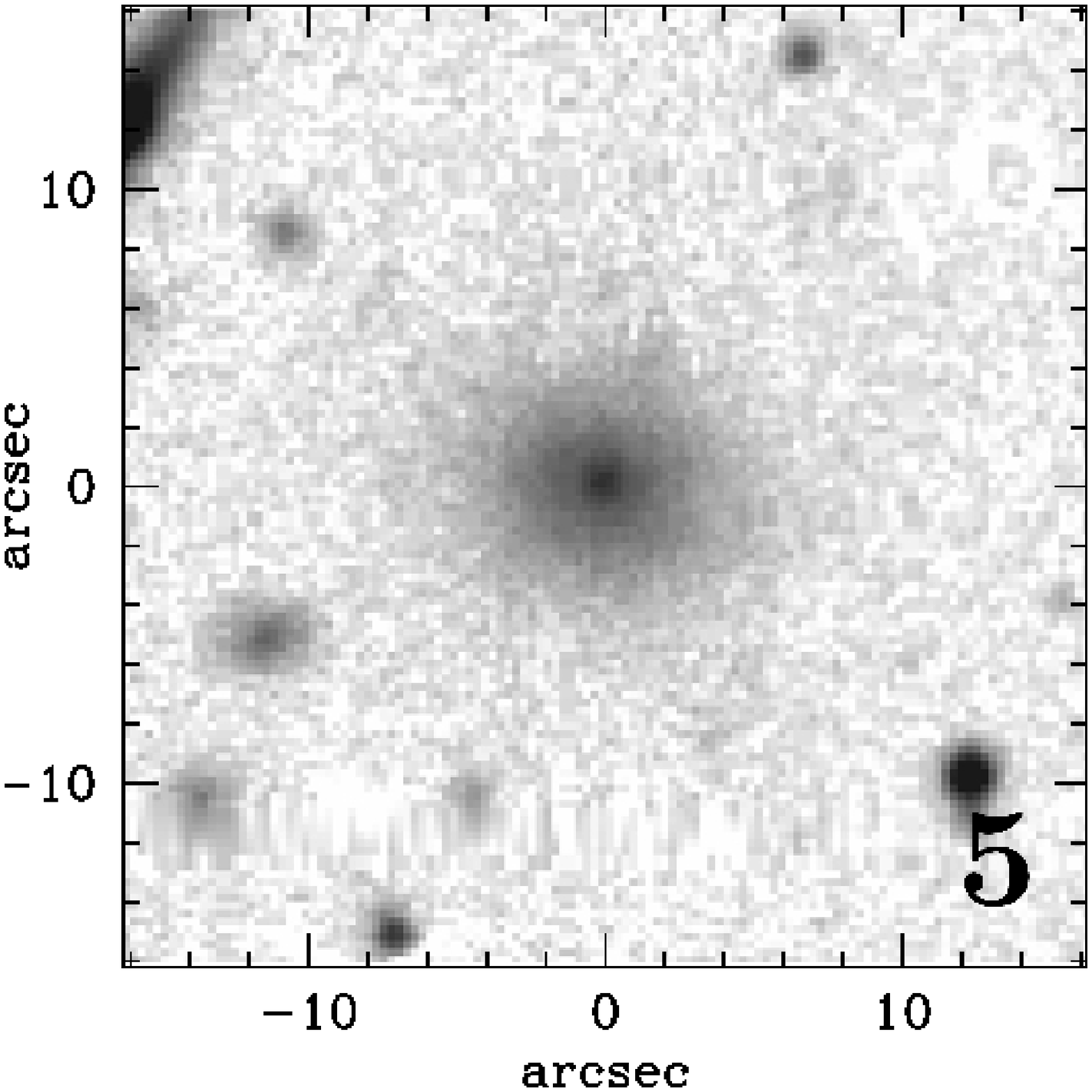} 
\caption{Logarithmic scale $R$ images of new confirmed dwarf members of  
Antlia. They are identified by a number at one of the lower corners, that  
corresponds to the order number in Table\,\ref{tabla_ndgm}. North is up and  
east to the left.} 
\label{fig2_ch4} 
\end{center} 
\end{figure*}

We present the coordinates, radial velocities and photometric parameters for
five new dwarf galaxy members identified from spectroscopic data obtained
with GEMINI-GMOS (see Table\,\ref{tabla_ndgm}). In Fig.\,\ref{fig2_ch4} we
show their logarithmic scale $R$ images. All of them  
display a bright central region that could be associated to a nucleus.
 
\begin{figure*} 
\includegraphics[scale=0.20]{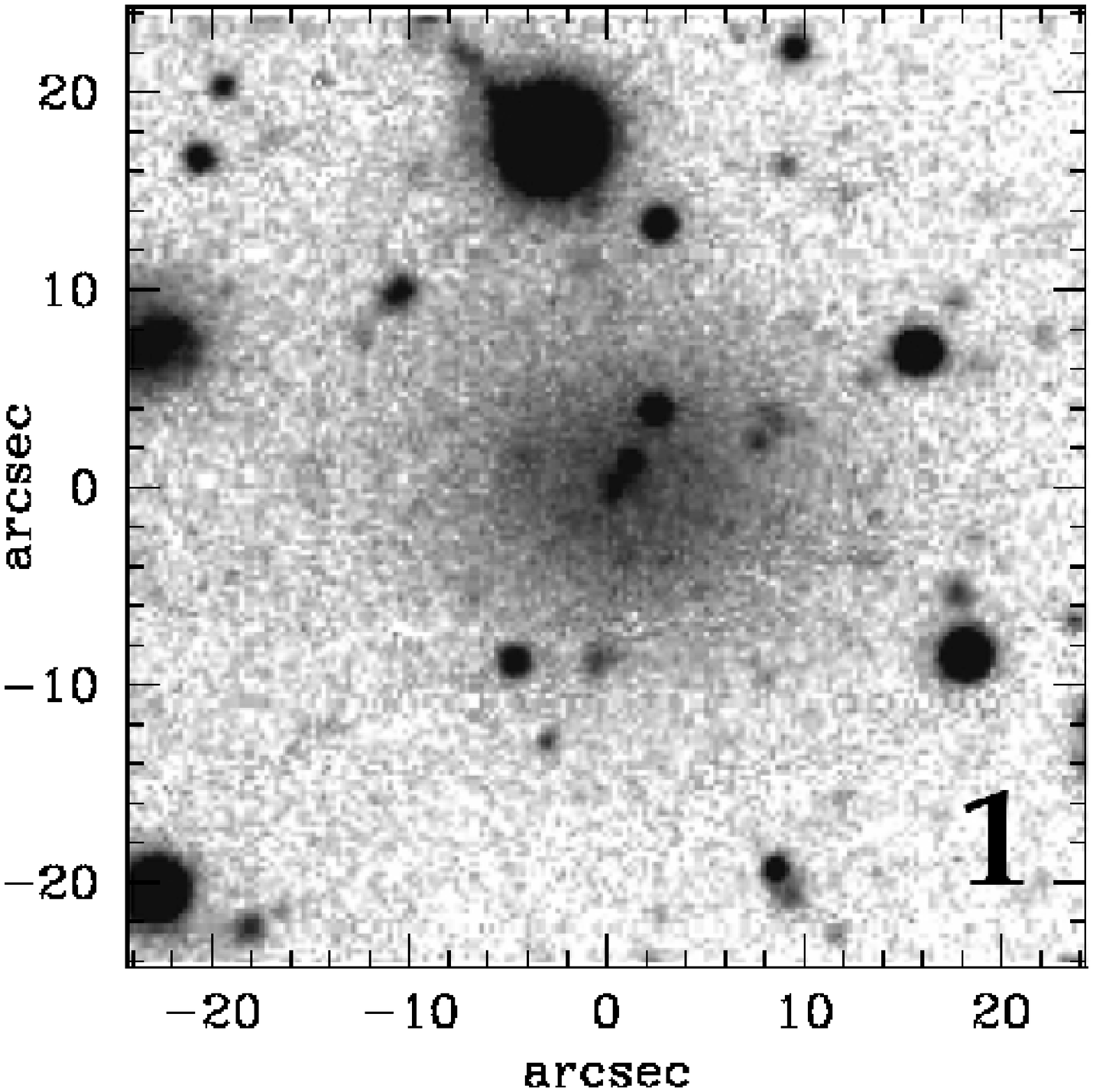} 
\includegraphics[scale=0.20]{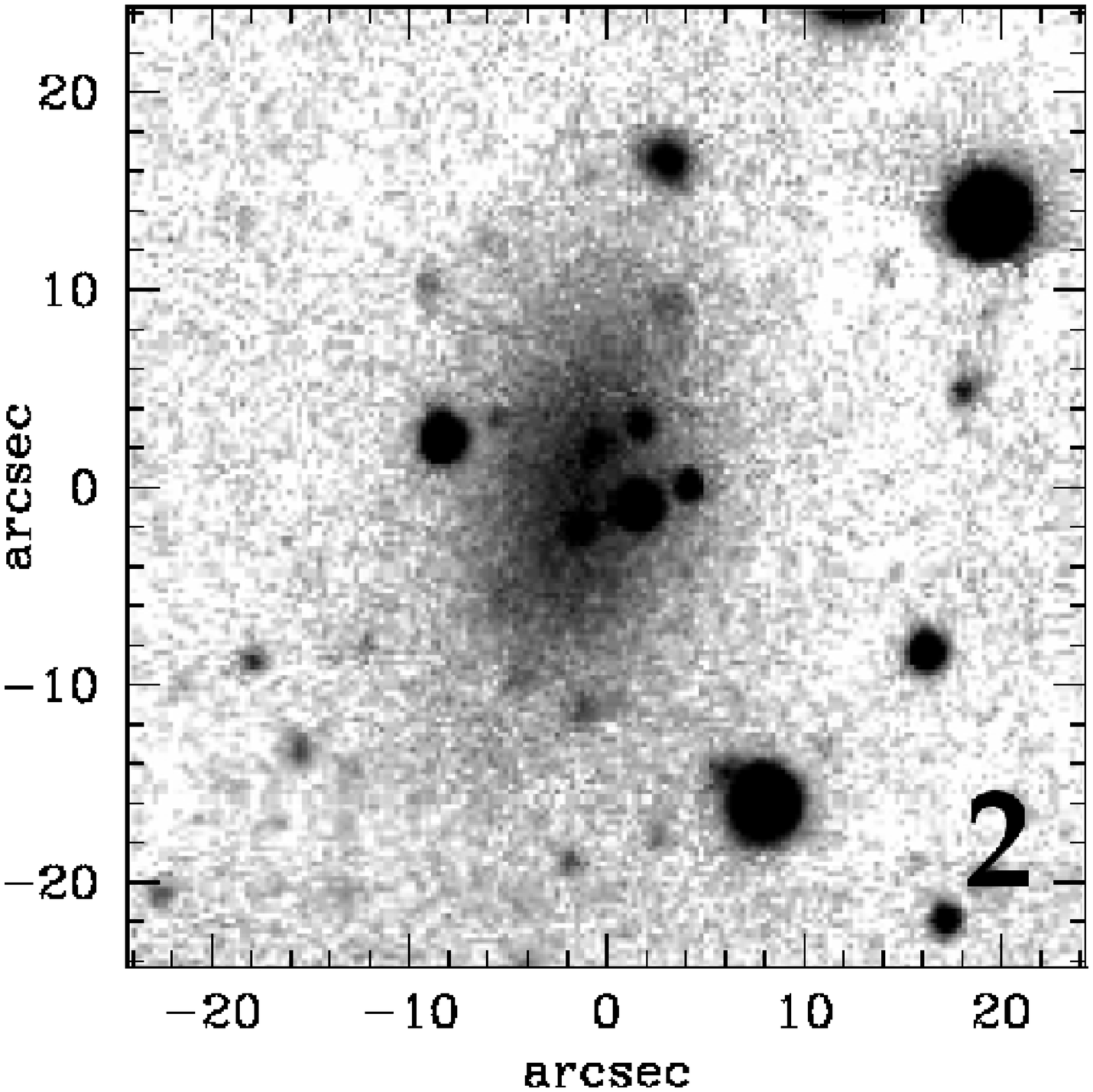} 
\includegraphics[scale=0.20]{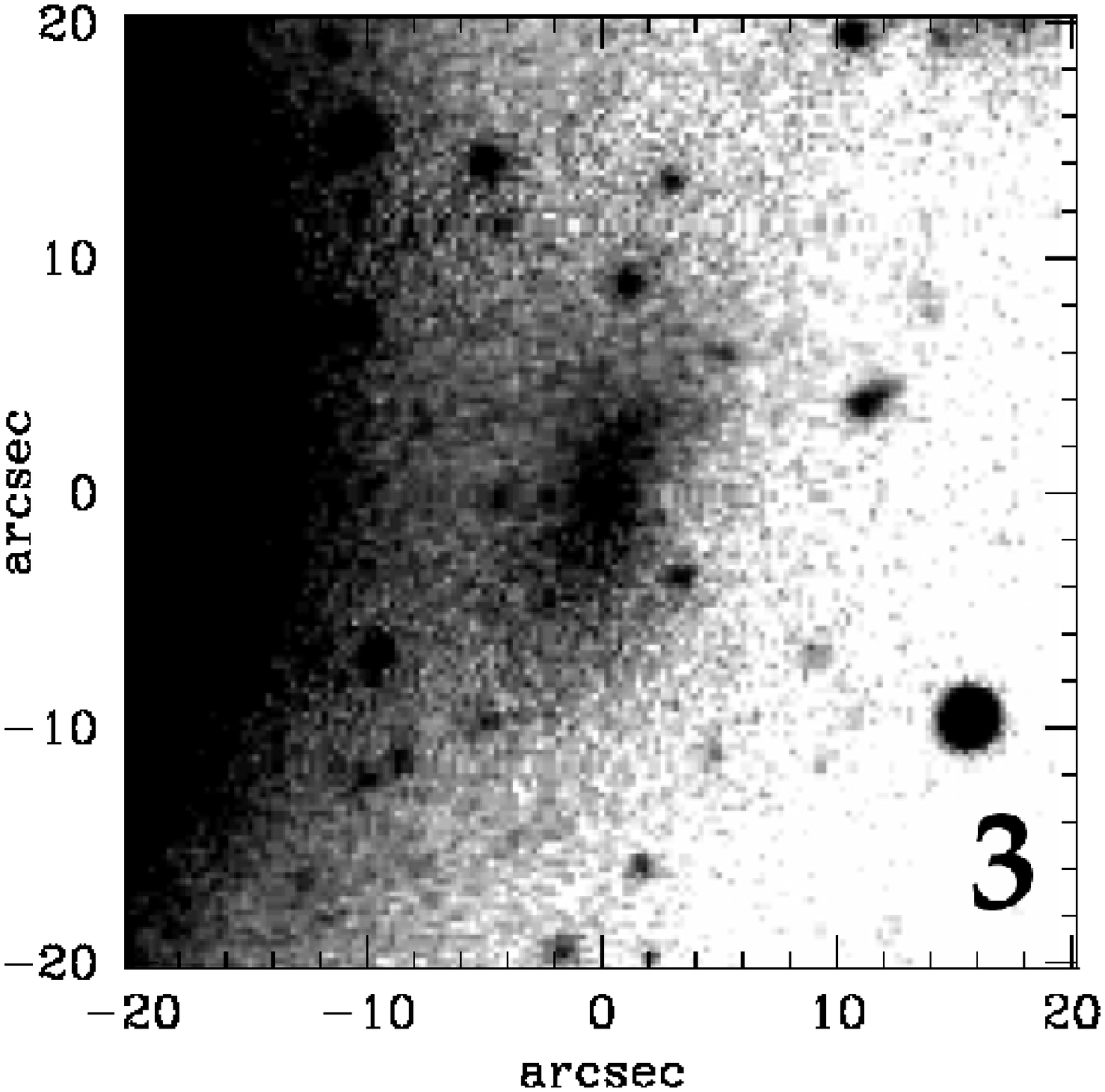} 
\includegraphics[scale=0.20]{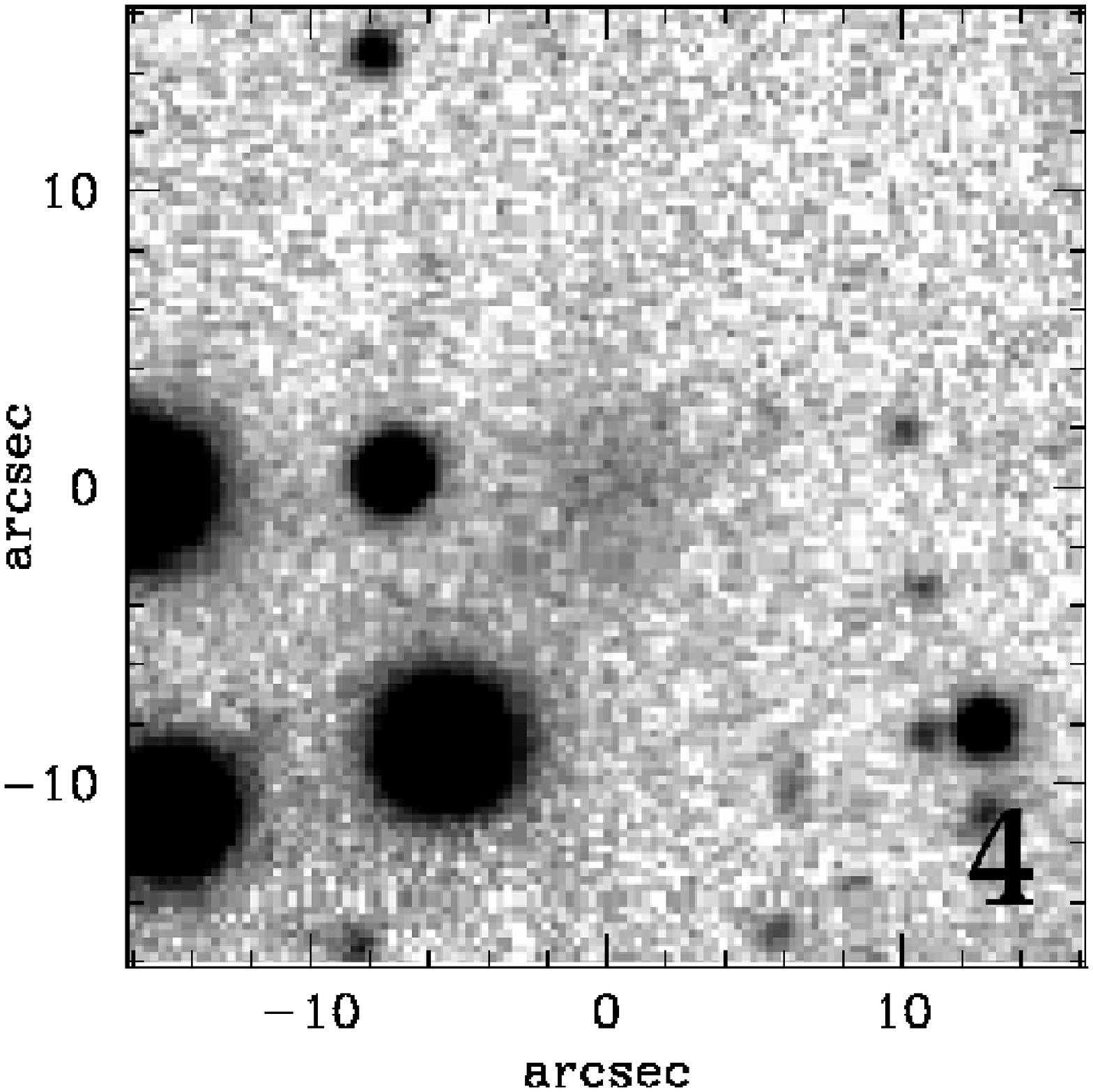} 
\includegraphics[scale=0.20]{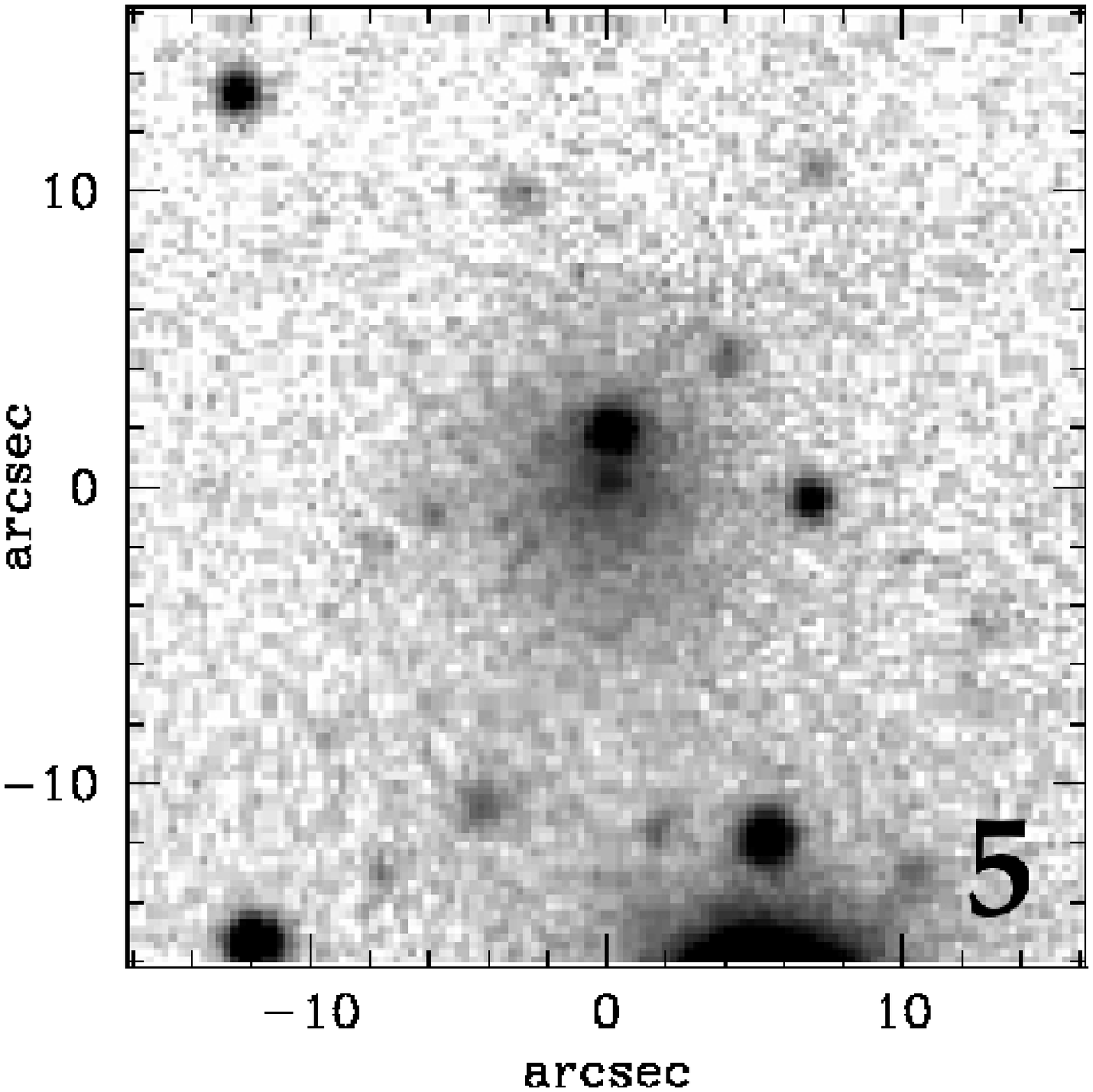} 
\includegraphics[scale=0.20]{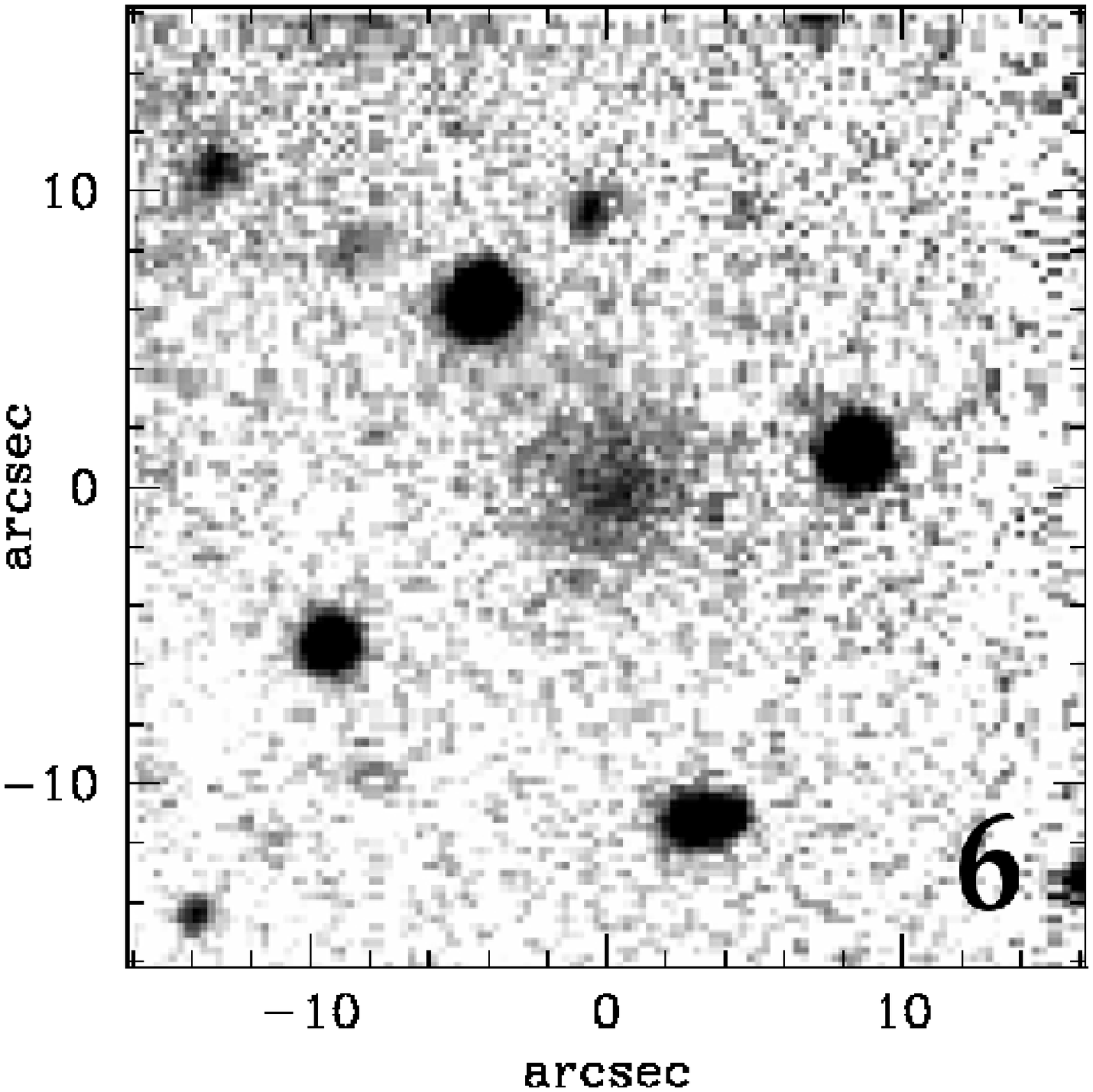} 
\includegraphics[scale=0.20]{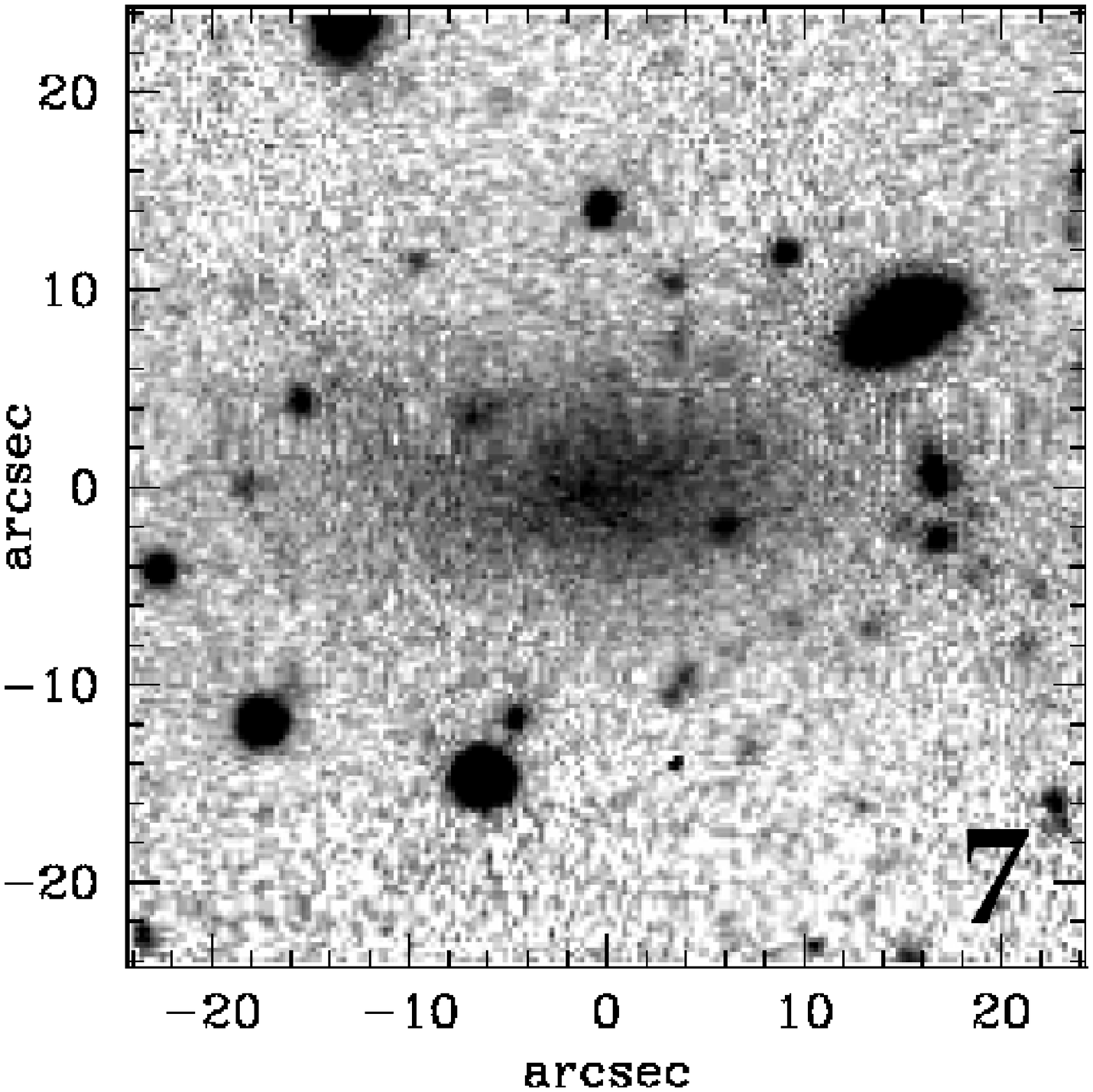} 
\includegraphics[scale=0.20]{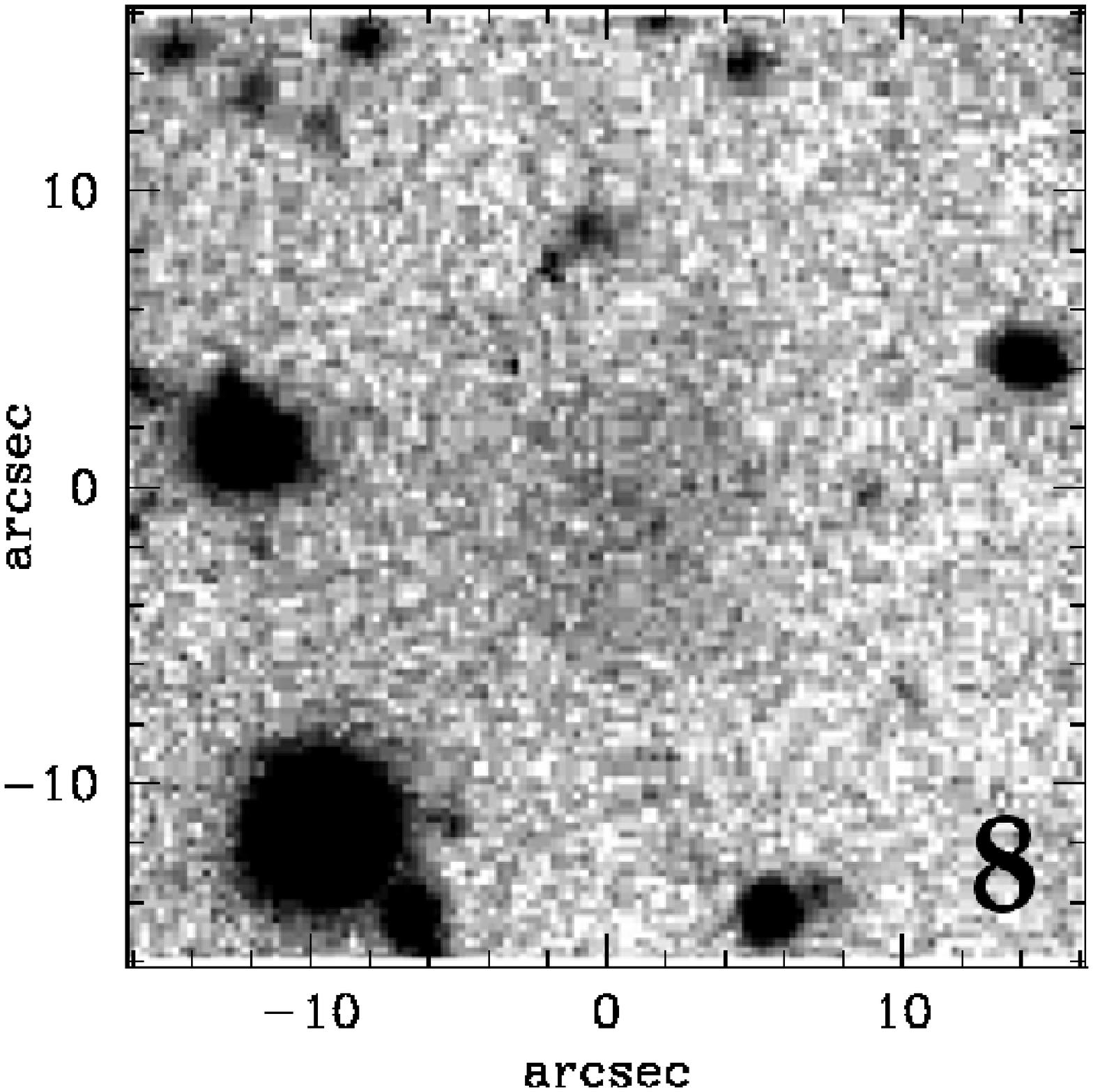} 
\includegraphics[scale=0.20]{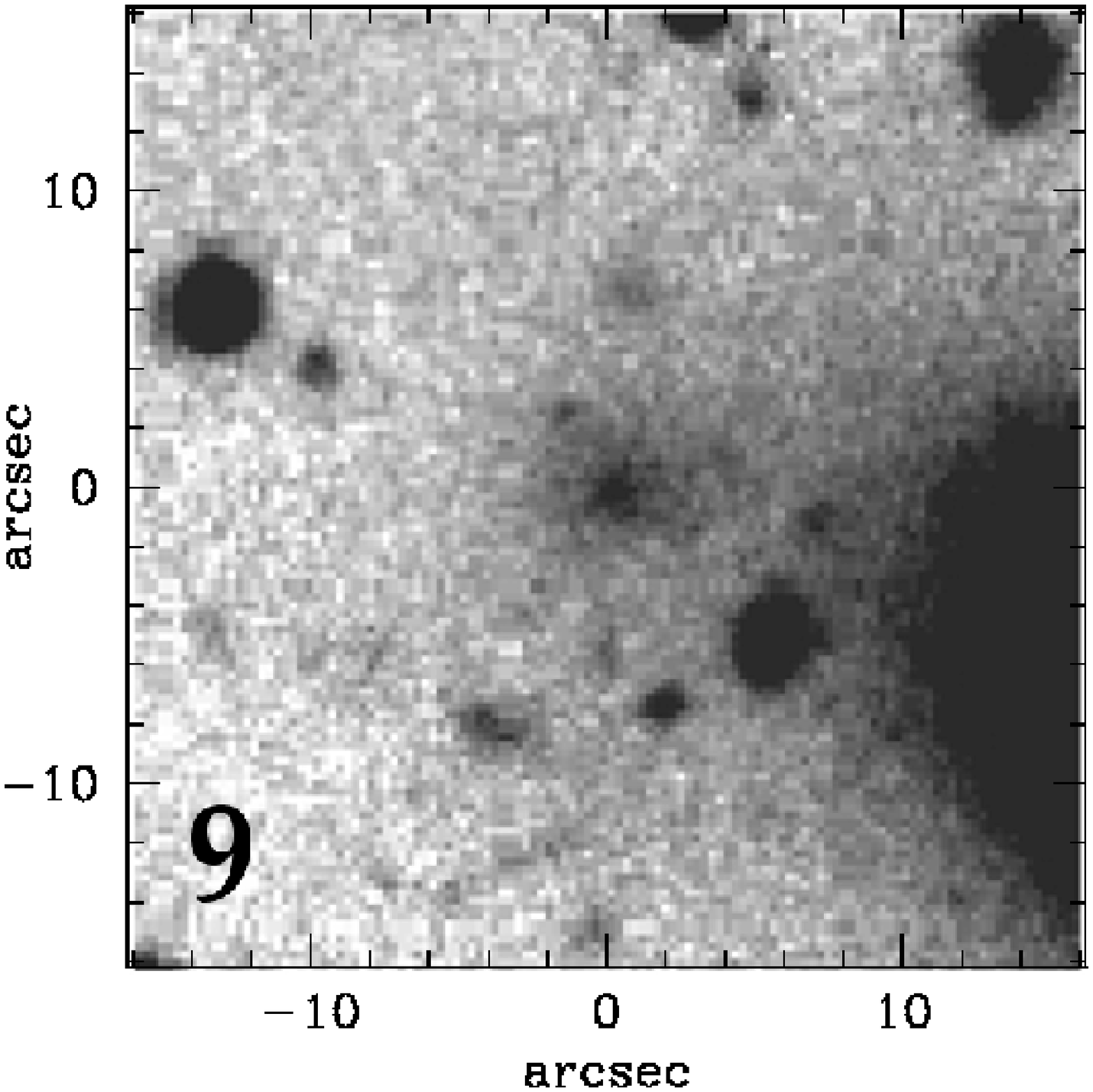} 
\includegraphics[scale=0.20]{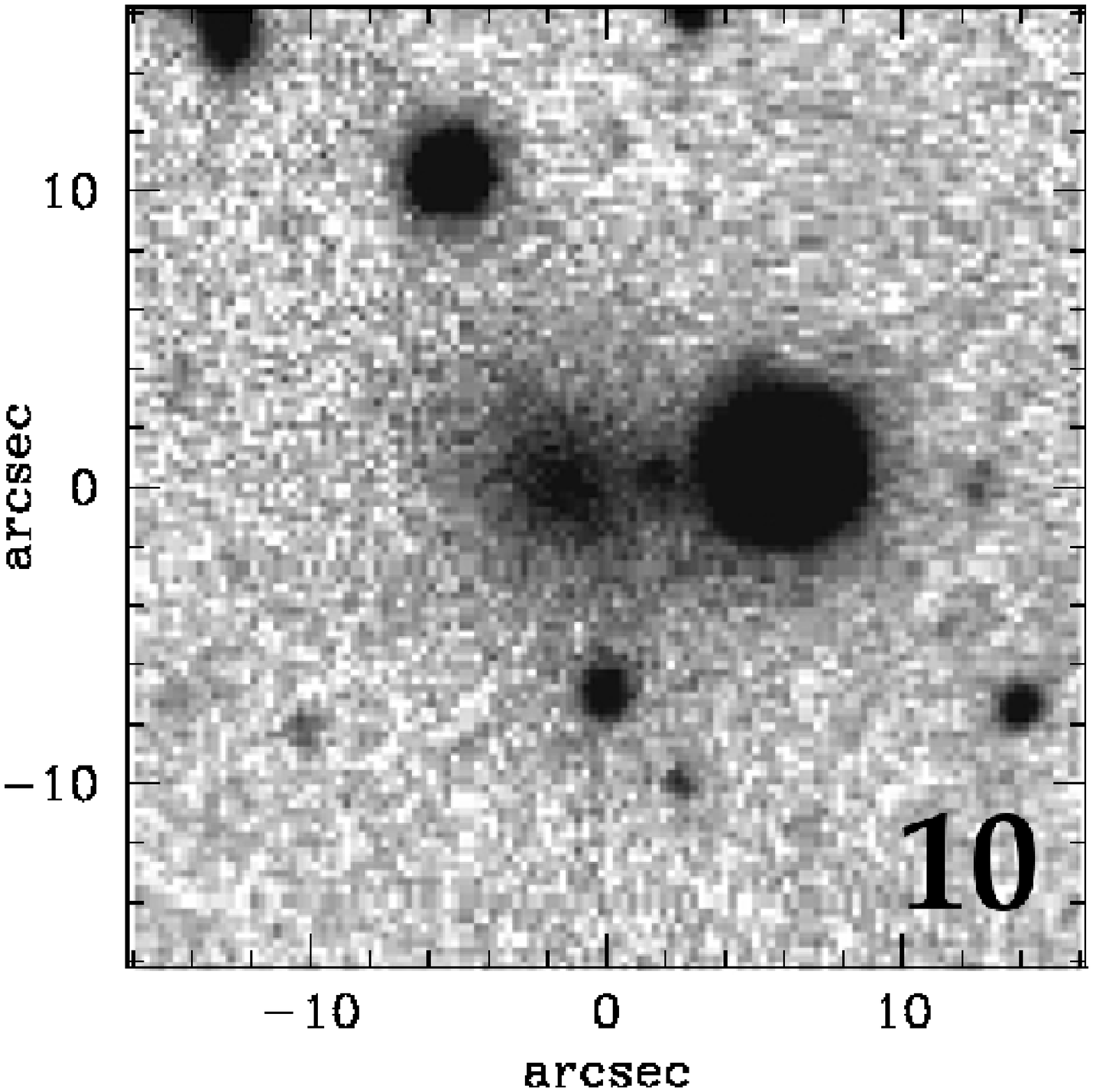} 
\includegraphics[scale=0.20]{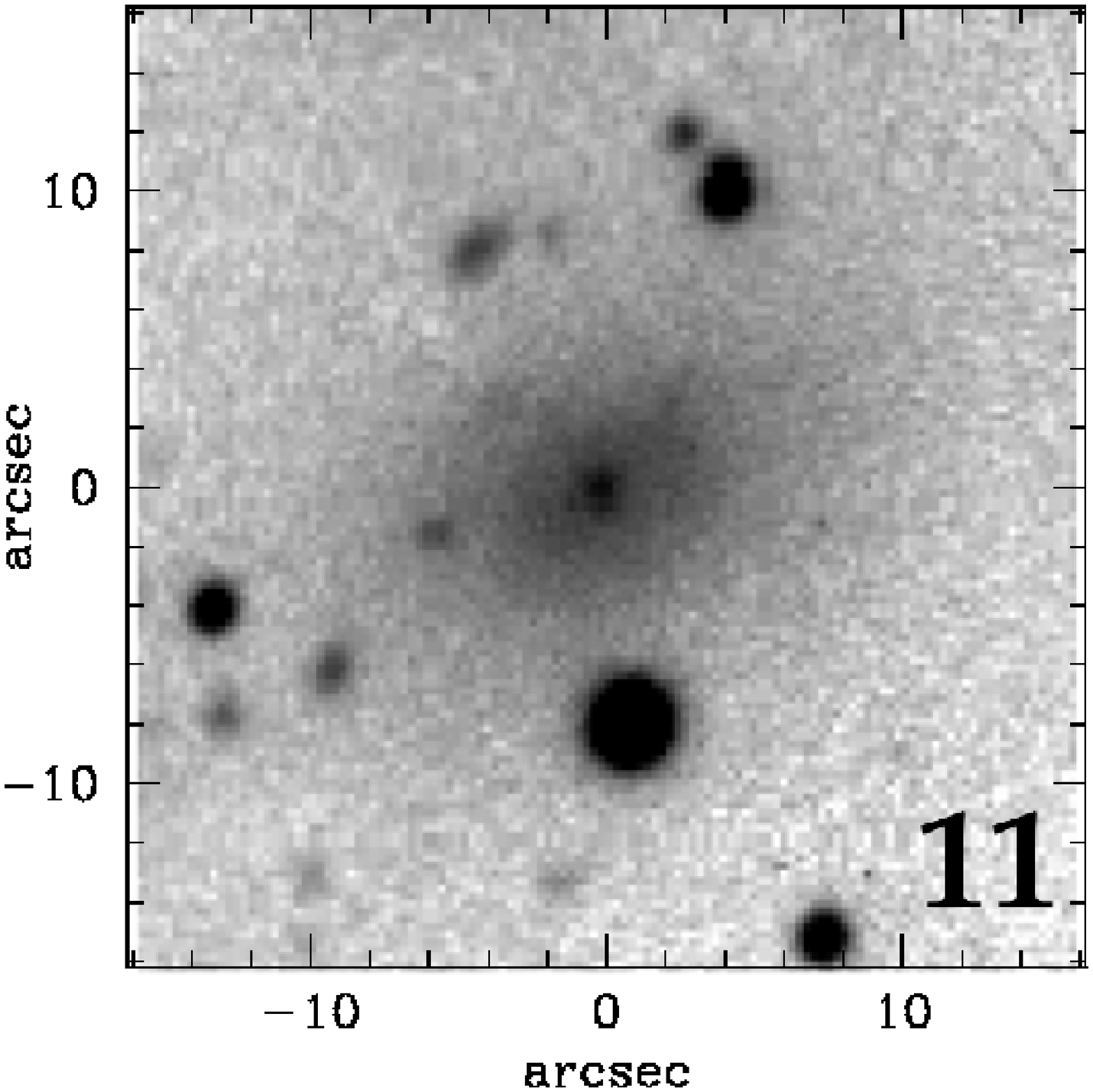} 
\includegraphics[scale=0.20]{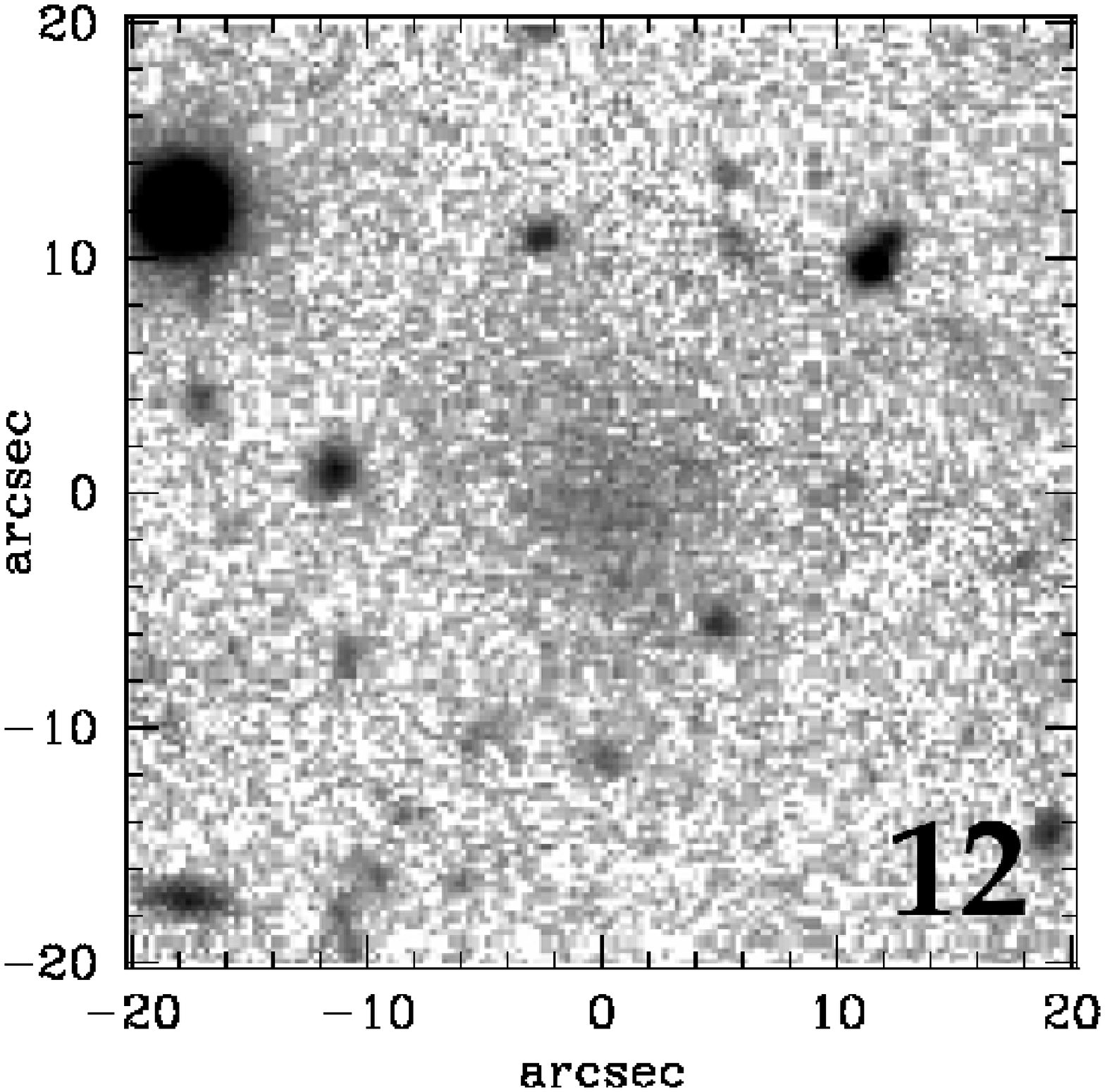} 
\includegraphics[scale=0.20]{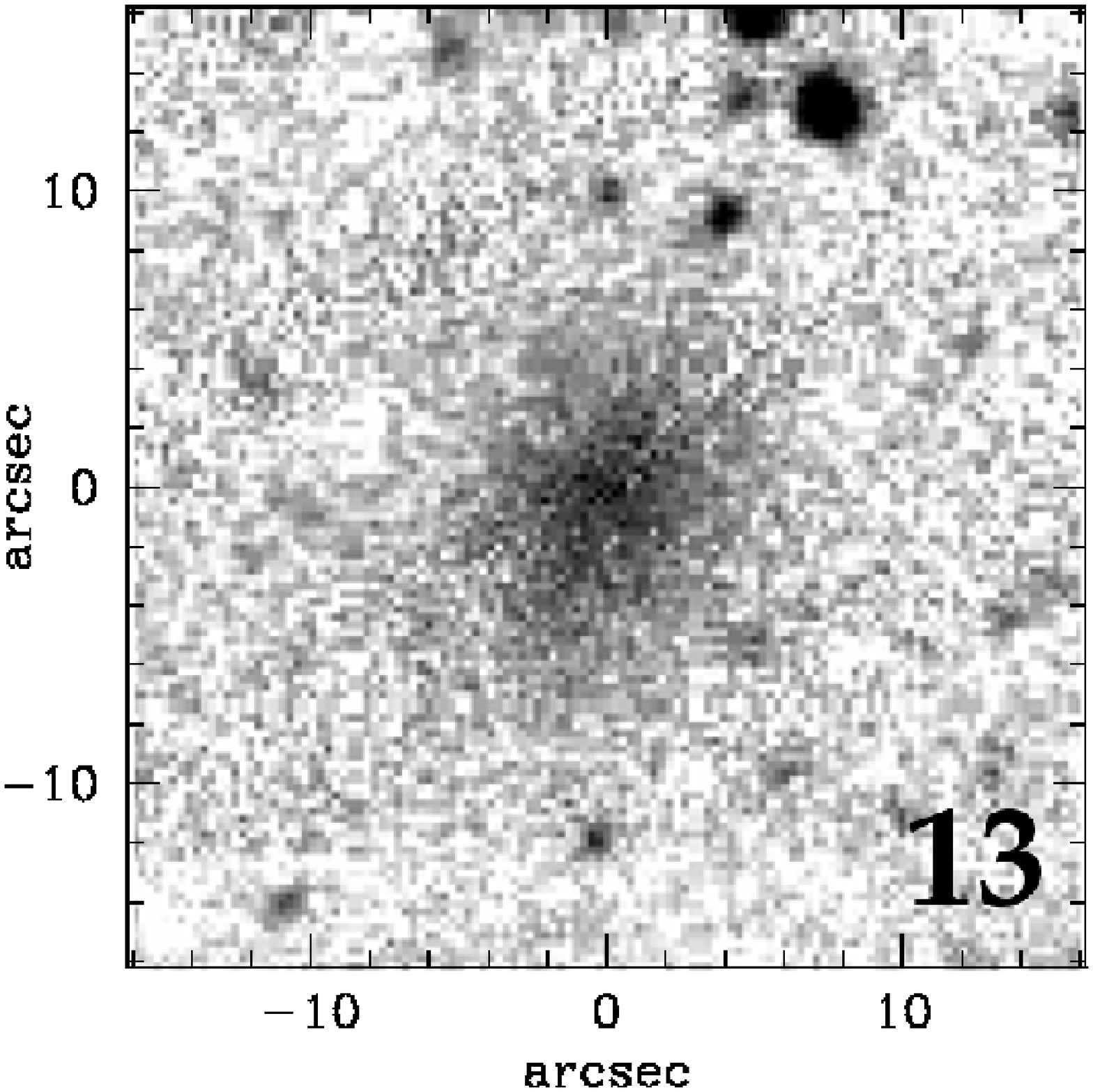} 
\includegraphics[scale=0.20]{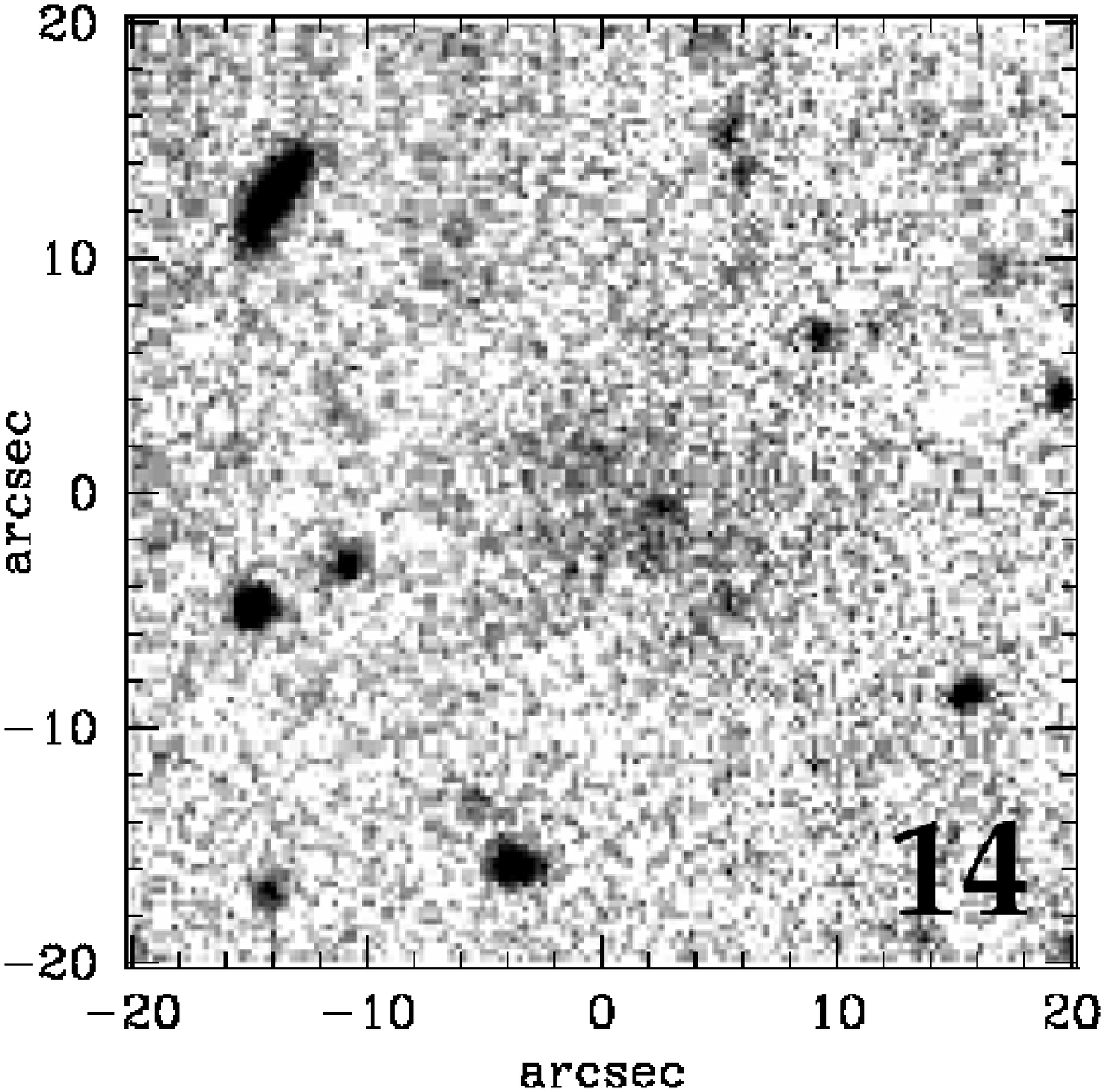} 
\includegraphics[scale=0.20]{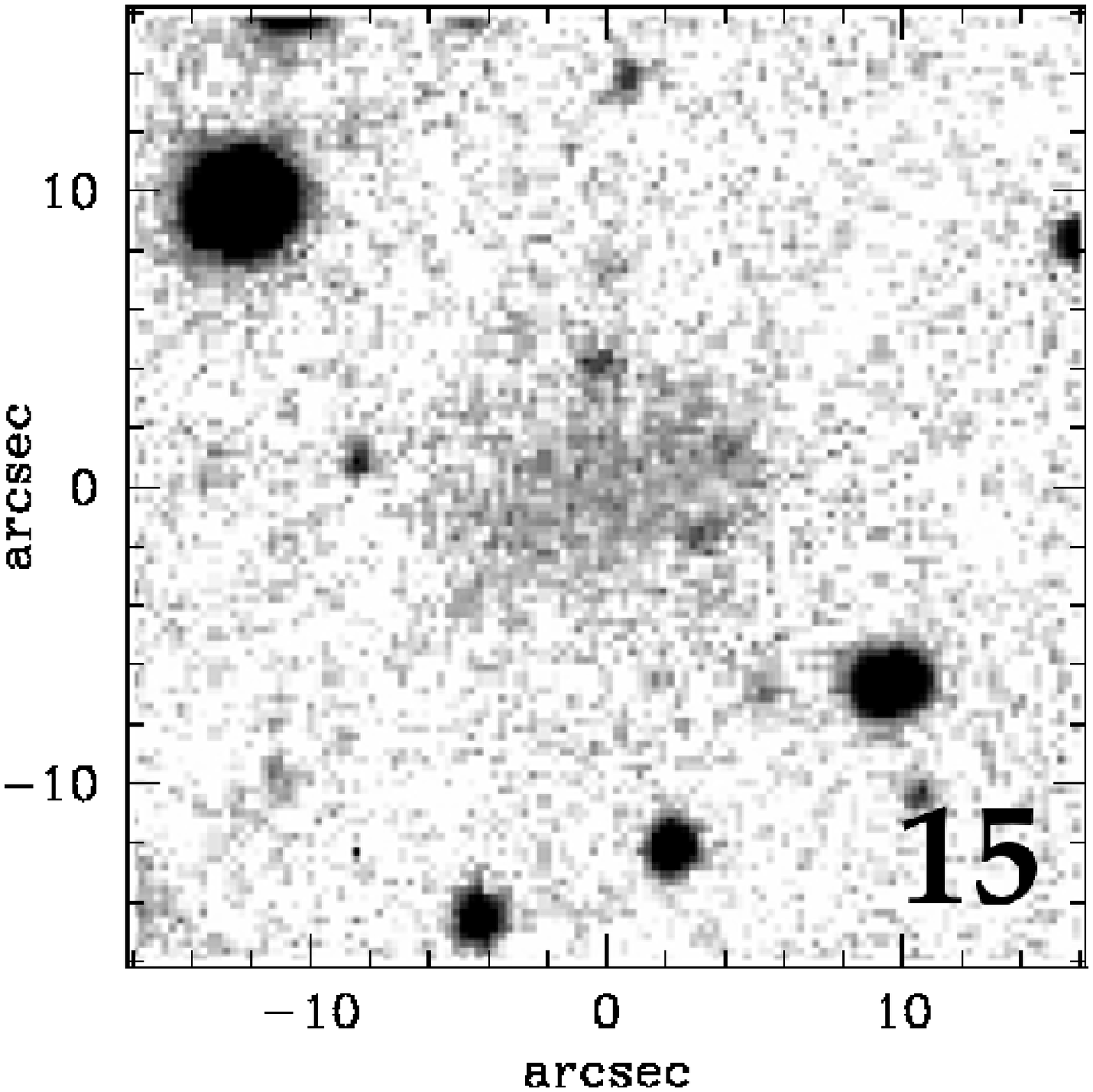} 
\includegraphics[scale=0.20]{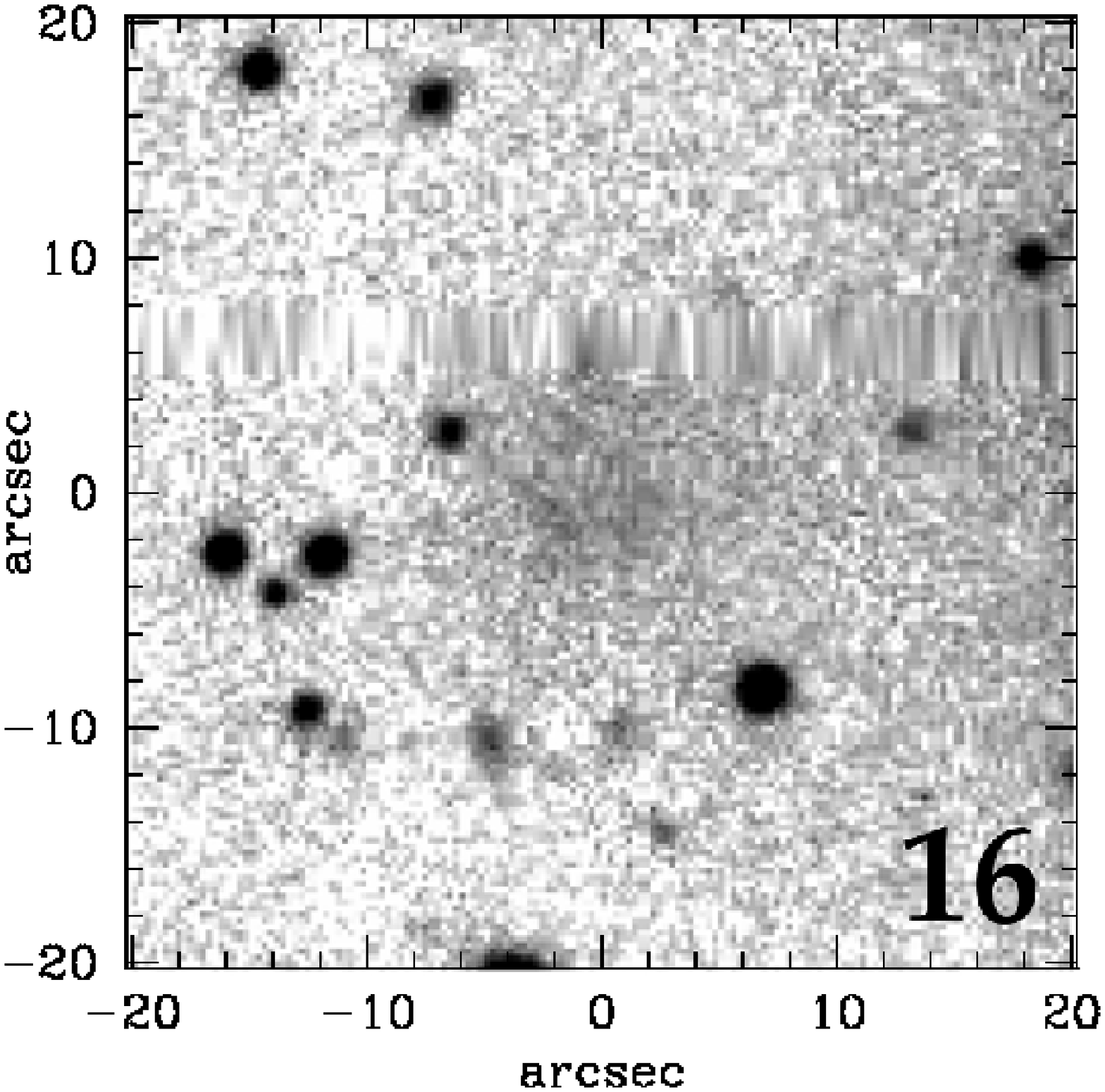} 
\caption{Logarithmic scale $R$ images of the new dwarf galaxy candidates  
identified in the central region of the Antlia cluster. The number at one of  
the lower corners corresponds to the number in Table\,\ref{candidates}. North  
is up and east to the left.} 
\label{fig_new_dwarfs} 
\end{figure*}

We also show the photometric parameters (when available) for 16 new dwarf  
galaxy candidates (see Table\,\ref{candidates}) and their images in  
Fig.\,\ref{fig_new_dwarfs}.\\  
 
\begin{table*} 
\caption{New dwarf galaxy members located in the central region of the Antlia  
cluster.}
\label{tabla_ndgm} 
\scriptsize 
\begin{tabular}{ccccccccccccc} 
\hline 
\\ 
\multicolumn{1}{c}{} & \multicolumn{1}{c}{ID} & \multicolumn{1}{c} {$\alpha$} & \multicolumn{1}{c} {$\delta$} & \multicolumn{1}{c} {$E(B-V)$} &  \multicolumn{1}{c} {$T_1$} &  \multicolumn{1}{c}{$(C-T_1)$}  &  \multicolumn{1}{c} {$\mu_{_{T_1}}$}&  \multicolumn{1}{c} {$r_{_{T_1}}$}  &  \multicolumn{1}{c} {$\langle\mu_{\rm eff}\rangle$} &  \multicolumn{1}{c} {$r_{\rm eff}$} & \multicolumn{1}{c}{v$_r$} & \multicolumn{1}{c}{Remarks}\\ 
\multicolumn{1}{c}{} & \multicolumn{1}{c}{} &\multicolumn{1}{c} {(J2000)}&\multicolumn{1}{c} {(J2000)} & \multicolumn{1}{c}{} & \multicolumn{1}{c}{\scriptsize (mag)}& \multicolumn{1}{c}{\scriptsize (mag)} & \multicolumn{1}{c}{\scriptsize (mag arcsec$^{-2}$)} & \multicolumn{1}{c}{\scriptsize (arcsec)} & \multicolumn{1}{c}{\scriptsize (mag arcsec$^{-2}$)} & \multicolumn{1}{c}{\scriptsize (arcsec)} & \multicolumn{1}{c}{km s$^{-1}$} & \multicolumn{1}{c}{}\\ 
\hline 
\\ 
1 & ANTL\,J102910-353920.1 & 10:29:10.3 & -35:39:20.1 & 0.083 & 19.76 (0.02) & 1.54 (0.03) & 26.4 & 4.8 & 23.4 & 2.1 & 1940$\pm$155  & {\tiny ELL} \\
2 & ANTL\,J102914-353923.6 & 10:29:14.4 & -35:39:23.6 & 0.083 & 20.25 (0.06) & 1.67 (0.08) & 26.3 & 5.6 & 24.6 & 3.0 & 4067$\pm$115  & {\tiny ELL} \\
3 & ANTL\,J103013-352458.3 & 10:30:13.8 & -35:24:58.3 & 0.104 & 19.49 (0.04) & 1.59 (0.06) & 25.8 & 7.6 & 25.7 & 3.7 & 2613$\pm$200  & {\tiny SE,NS} \\
4 & ANTL\,J103033-352638.6 & 10:30:33.3 & -35:26:38.6 & 0.104 & 19.07 (0.02) & 1.68 (0.04) & 26.6 & 6.6 & 23.4 & 2.9 & 2311$\pm$130  & {\tiny ELL} \\
5 & ANTL\,J103037-352708.8 & 10:30:37.5 & -35:27:08.8 & 0.103 & 19.16 (0.03) & 1.60 (0.04) & 25.8 & 6.8 & 25.7 & 3.1 & 2400$\pm$100  & {\tiny SE} \\
\\ 
\hline 
\end{tabular} 
\small 
\medskip
 
{\it Notes.-} $\mu_{_{T_1}}$ corresponds to the 
threshold above which SEXTRACTOR detects and measures the object  
(MU\_THRESHOLD), or to the surface brightness of the outermost isophote for 
ELLIPSE. $r_{_{T_1}}$ is the radius that contains 90 per cent of the light 
for SEXTRACTOR, or the equivalent radius 
($r=\sqrt{a\cdot b}=a\cdot \sqrt{1-\epsilon}$) of the most  
external isophote for ELLIPSE. In both cases, $\langle\mu_{\rm eff}\rangle$ 
is obtained from $r_{\rm eff}$, the radius that contains one-half of the light 
(see eq. 1 in Paper\,I) and is the output parameter FLUX\_RADIUS for 
SEXTRACTOR. Remarks refer to: SE= magnitudes and colours measured with 
SEXTRACTOR; ELL= magnitudes and colours obtained from ELLIPSE; NS= nearby a 
bright star.
\end{table*}

\begin{table*} 
\begin{minipage}{185mm} 
\caption{New dwarf galaxy candidates with early-type morphologies located in  
the central region of the Antlia cluster. }  
\setlength\tabcolsep{1.35mm} 
\label{candidates} 
\begin{tabular}{ccccccccr@{}c@{}lcr@{}c@{}l@{}c} 
\hline 
\multicolumn{1}{c}{} & \multicolumn{1}{c}{ID} & \multicolumn{1}{c} {$\alpha$} & \multicolumn{1}{c} {$\delta$} & \multicolumn{1}{c} {$E(B-V)$} &  \multicolumn{1}{c} {$T_1$} &  \multicolumn{1}{c}{$(C-T_1)$}  &  \multicolumn{1}{c} {$\mu_{_{T_1}}$}&  \multicolumn{3}{c} {$r_{_{T_1}}$}  &  \multicolumn{1}{c} {$\langle\mu_{\rm eff}\rangle$} &  \multicolumn{3}{c} {$r_{\rm eff}$} & \multicolumn{1}{c}{Remarks}\\ 
\multicolumn{1}{c}{} & \multicolumn{1}{c}{} &\multicolumn{1}{c} {(J2000)}&\multicolumn{1}{c} {(J2000)} & \multicolumn{1}{c}{} & \multicolumn{1}{c}{\scriptsize (mag)}& \multicolumn{1}{c}{\scriptsize (mag)} & \multicolumn{1}{c}{\scriptsize (mag arcsec$^{-2}$)} & \multicolumn{3}{c}{\scriptsize (arcsec)} & \multicolumn{1}{c}{\scriptsize (mag arcsec$^{-2}$)} & \multicolumn{3}{c}{\scriptsize (arcsec)}  & \multicolumn{1}{c}{}\\ 
\hline 
1  & ANTL\,J102828-354128 & 10:28:28.2 & -35:41:28.5 & 0.074 & 18.15 (0.05) & 1.63 (0.07) & 25.7 & 14&.&8 & 24.4 & 7&.&3 & {\tiny SE}        \\
2  & ANTL\,J102839-351053 & 10:28:39.0 & -35:10:53.5 & 0.096 & 18.57 (0.10) & 1.48 (0.13) & 25.8 & 17&.&6 & 24.7 & 6&.&8 & {\tiny SE}        \\
3  & ANTL\,J102843-353933 & 10:28:43.8 & -35:39:33.1 & 0.077 & 18.74 (0.07) & 1.76 (0.11) & 25.8 & 14&.&2 & 24.8 & 6&.&6 & {\tiny SE}        \\
4  & ANTL\,J102914-353855 & 10:29:14.7 & -35:38:55.2 & 0.084 & 21.35 (0.11) & 1.31 (0.14) & 25.6 &  5&.&1 & 25.7 & 2&.&9 & {\tiny SE,FG}     \\
5  & ANTL\,J102918-352900 & 10:29:18.8 & -35:29:00.2 & 0.090 & 19.74 (0.01) & 1.52 (0.02) &   -  &  9&.&4 &   -  &  &-&  & {\tiny NS,PHOT}   \\
6  & ANTL\,J102932-354216 & 10:29:32.2 & -35:42:16.4 & 0.084 & 22.35 (0.08) & 1.68 (0.11) & 26.5 &  2&.&6 & 25.4 & 1&.&6 & {\tiny ELL,PB}    \\
7  & ANTL\,J102936-352445 & 10:29:36.2 & -35:24:45.7 & 0.094 & 19.00 (0.08) & 1.50 (0.11) & 25.8 & 13&.&0 & 25.0 & 6&.&4 & {\tiny SE,BL}     \\
8  & ANTL\,J102943-352238 & 10:29:43.3 & -35:22:38.8 & 0.098 & 21.08 (0.04) & 1.17 (0.06) &   -  &  9&.&4 &   -  &  &-&  & {\tiny FG,NG}     \\
9  & ANTL\,J102948-352354 & 10:29:48.6 & -35:23:54.0 & 0.098 & 22.39 (0.02) & 1.90 (0.04) & 26.1 &  2&.&1 & 24.5 & 1&.&1 & {\tiny ELL,NG}    \\
10 & ANTL\,J102954-352744 & 10:29:54.2 & -35:27:44.6 & 0.094 &      -       &      -      &   -  &   &-&  &   -  &  &-&  & {\tiny FG,NS}     \\
11 & ANTL\,J102955-351517 & 10:29:55.9 & -35:15:17.6 & 0.103 & 18.42 (0.03) & 1.70 (0.05) & 26.0 &  9&.&7 & 25.3 & 4&.&0 & {\tiny ELL,NG}    \\
12 & ANTL\,J102959-354227 & 10:29:59.8 & -35:42:27.1 & 0.089 & 20.70 (0.03) & 1.93 (0.06) &   -  & 10&.&5 &   -  &  &-&  & {\tiny FG,PHOT}   \\
13 & ANTL\,J103027-350957 & 10:30:27.5 & -35:09:57.4 & 0.101 & 20.24 (0.04) & 1.38 (0.05) & 25.9 &  5&.&0 & 25.4 & 4&.&4 & {\tiny ELL}       \\
14 & ANTL\,J103027-352941 & 10:30:27.5 & -35:29:41.8 & 0.098 & 21.24 (0.05) & 1.82 (0.09) &   -  & 10&.&5 &   -  &  &-&  & {\tiny FG,PHOT}   \\
15 & ANTL\,J103042-351519 & 10:30:42.1 & -35:15:19.5 & 0.099 & 21.02 (0.04) & 0.99 (0.06) &   -  & 10&.&8 &   -  &  &-&  & {\tiny FG,PHOT}   \\
16 & ANTL\,J103047-354025 & 10:30:47.1 & -35:40:25.1 & 0.102 & 20.37 (0.03) & 1.02 (0.04) &   -  & 10&.&8 &   -  &  &-&  & {\tiny FG,BL,PHOT}\\
\\ 
\hline 
\end{tabular} 
\scriptsize
 
\medskip{\it Notes .-} $\mu_{_{T_1}}$ corresponds to the threshold above which 
SEXTRACTOR
detects and measures the object (MU\_THRESHOLD), or to the surface
brightness of the outermost isophote for ELLIPSE. $r_{_{T_1}}$ is the radius
that contains 90 per cent of the light for SEXTRACTOR, the equivalent radius
($r=\sqrt{a\cdot b}=a\cdot \sqrt{1-\epsilon}$) of the most external isophote
for ELLIPSE, or the radius of the aperture considered with
PHOT. $\langle\mu_{\rm eff}\rangle$ is obtained in both cases from $r_{\rm
  eff}$, the radius that contains one-half of the light (see eq. 1 in
Paper\,I). This radius is the output parameter FLUX\_RADIUS for SEXTRACTOR.
Remarks refer to: SE= magnitudes and colours measured with SEXTRACTOR; ELL=
magnitudes and colours obtained with ELLIPSE; PHOT= magnitudes and colours
measured with PHOT; BL= bleeding; NS= nearby a bright star; NG= nearby a
bright galaxy; FG= very faint and diffuse galaxy; PB= possible background.
\end{minipage} 
\end{table*}

\section{ELLIPSE versus SExtractor}
\label{ELL_SEX} 
 
\begin{table*} 
\caption{ELLIPSE photometry of FS90 early-type galaxies placed in the
  central region of Antlia that were presented with SExtractor photometry in
  Paper\,I.  All the photometric relations analyzed in the present paper
  were built with these values. }
\label{tabla_ELL} 
\begin{tabular}{@{}ccccccr@{}c@{}lcr@{}c@{}l} 
\hline 
\multicolumn{1}{c}{FS90} & \multicolumn{1}{c}{FS90} & \multicolumn{1}{c}{FS90} &  \multicolumn{1}{c} {$T_1$} &  \multicolumn{1}{c}{$(C-T_1)$}  &  \multicolumn{1}{c} {$\mu_{_{T_1}}$}&  \multicolumn{3}{c} {$r_{_{T_1}}$}  &  \multicolumn{1}{c} {$\langle\mu_{\rm eff}\rangle$} &  \multicolumn{3}{c} {$r_{\rm eff}$} \\ 
\multicolumn{1}{c}{ID}  &\multicolumn{1}{c}{mor.} &\multicolumn{1}{c}{status}  & \multicolumn{1}{c}{\scriptsize mag}& \multicolumn{1}{c}{\scriptsize mag} & \multicolumn{1}{c}{\scriptsize mag arcsec$^{-2}$} & \multicolumn{3}{c}{\scriptsize arcsec} & \multicolumn{1}{c}{\scriptsize mag arcsec$^{-2}$} & \multicolumn{3}{c}{\scriptsize arcsec}  \\ 
\hline 
 70 &        dE          & 1 &  17.68 (0.02) & 1.73 (0.08) & 26.4 & 12&.&9 & 23.2 &  5&.&1 \\  
 72 &        S0          & 1 &  14.38 (0.01) & 1.93 (0.02) & 26.4 & 29&.&7 & 20.3 &  6&.&1 \\  
 73 &        dE          & 1 &  16.95 (0.01) & 1.72 (0.04) & 27.0 & 15&.&1 & 22.1 &  4&.&3 \\  
 85 &        dE          & 1 &  18.70 (0.01) & 1.63 (0.04) & 25.3 &  6&.&6 & 23.6 &  3&.&8 \\  
 87 &        dE,N        & 1 &  15.85 (0.01) & 1.86 (0.04) & 26.8 & 23&.&4 & 22.3 &  7&.&9 \\  
114 &        dE          & 1 &  20.96 (0.02) & 1.58 (0.05) & 26.4 &  3&.&6 & 24.3 &  1&.&9 \\  
118 &        dE          & 1 &  19.01 (0.03) & 1.77 (0.14) & 26.5 & 11&.&3 & 25.0 &  6&.&4 \\  
123 &        dE,N        & 2 &  16.32 (0.01) & 1.90 (0.03) & 26.4 & 17&.&2 & 21.8 &  5&.&0 \\  
133 &        d:E,N       & 1 &  14.54 (0.01) & 1.86 (0.02) & 26.8 & 29&.&2 & 20.5 &  6&.&3 \\  
136 &        dE,N        & 1 &  16.03 (0.01) & 1.86 (0.04) & 26.6 & 21&.&3 & 21.9 &  5&.&9 \\  
159 &        d:E,N?      & 1 &  16.10 (0.01) & 1.81 (0.03) & 26.7 & 16&.&2 & 21.3 &  4&.&3 \\  
160 &        dE          & 1 &  19.46 (0.02) & 1.82 (0.07) & 27.7 &  9&.&4 & 24.8 &  4&.&7 \\  
162 &        dE,N        & 1 &  17.31 (0.02) & 1.78 (0.04) & 25.8 & 11&.&6 & 22.7 &  4&.&7 \\  
176 &        dE,N        & 1 &  17.33 (0.01) & 1.97 (0.03) & 25.7 & 10&.&1 & 22.2   &  3&.&8 \\   
177 &        d:E,N       & 1 &  15.39 (0.01) & 1.84 (0.03) & 26.0 & 20&.&0 & 20.9   &  5&.&1 \\   
186 &        dE          & 1 &  18.67 (0.02) & 1.68 (0.05) & 26.0 &  7&.&7 & 23.4   &  3&.&5 \\   
188 &        dE          & 1 &  17.96 (0.02) & 1.81 (0.08) & 26.4 & 11&.&1 & 23.4   &  4&.&8 \\   
192 &        E(M32?)     & 3 &  16.66 (0.01) & 2.11 (0.03) & 27.6 & 11&.&4 & 19.9   &  1&.&7 \\ 
201 &        dE          & 1 &  19.43 (0.02) & 1.75 (0.06) & 26.3 &  6&.&5 & 23.9   &  3&.&1 \\   
216 &         E          & 2 &  16.15 (0.01) & 1.77 (0.02) & 27.2 & 13&.&8 & 20.3   &  2&.&7 \\   
228 &        dE,N        & 1 &  17.47 (0.01) & 1.76 (0.06) & 27.0 & 13&.&1 & 22.3   &  3&.&7 \\   
231 &        d:E,N       & 1 &  14.97 (0.01) & 2.19 (0.02) & 26.7 & 21&.&7 & 20.4   &  4&.&9 \\   
241 &        dE,N        & 1 &  16.81 (0.02) & 2.02 (0.08) & 26.7 & 18&.&8 & 23.1   &  7&.&2 \\   
\hline 
\end{tabular} 
\medskip 

Notes .- $\mu_{_{T_1}}$ corresponds to the surface
brightness of the outermost isophote measured with ELLIPSE. $r_{_{T_1}}$ is the 
equivalent radius ($r=\sqrt{a\cdot b}=a\cdot \sqrt{1-\epsilon}$) of the most 
external isophote. $\langle\mu_{\rm eff}\rangle$ is obtained from 
$r_{\rm  eff}$, the radius that contains one-half of the light, following eq. 
1 in Paper\,I. 
\end{table*} 
 
\begin{figure*}
\includegraphics[scale=0.27]{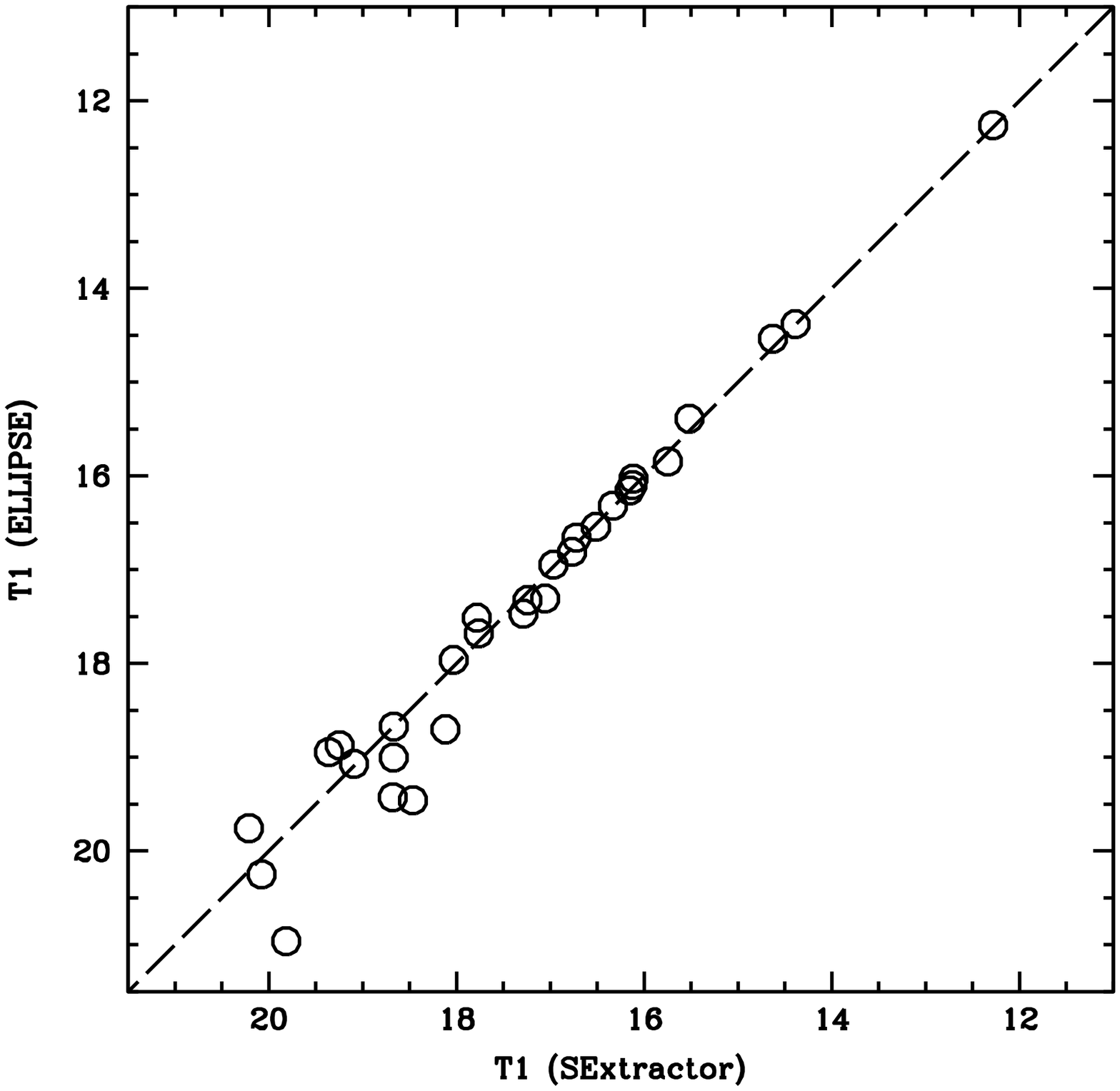}
\includegraphics[scale=0.27]{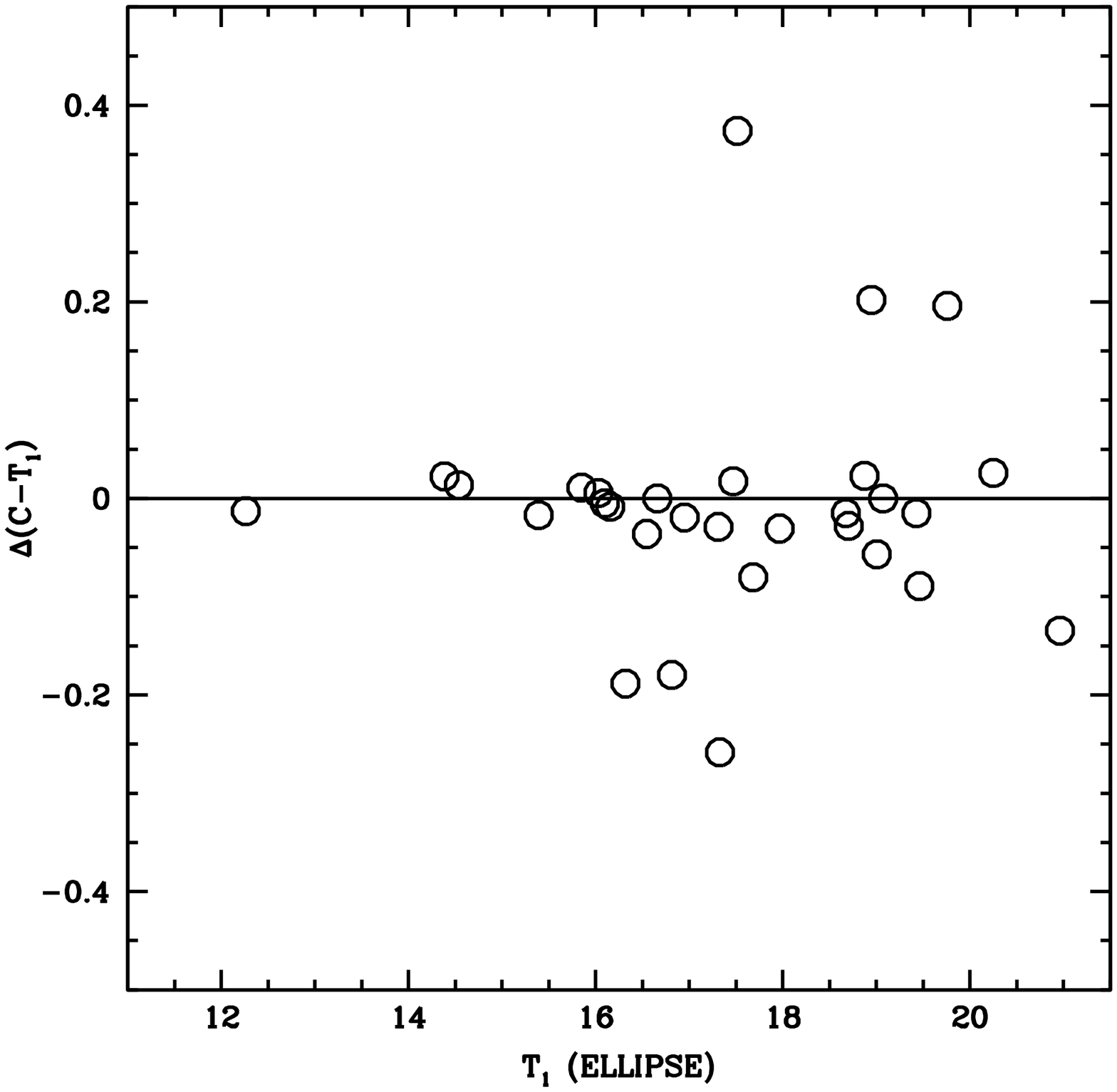}
\includegraphics[scale=0.27]{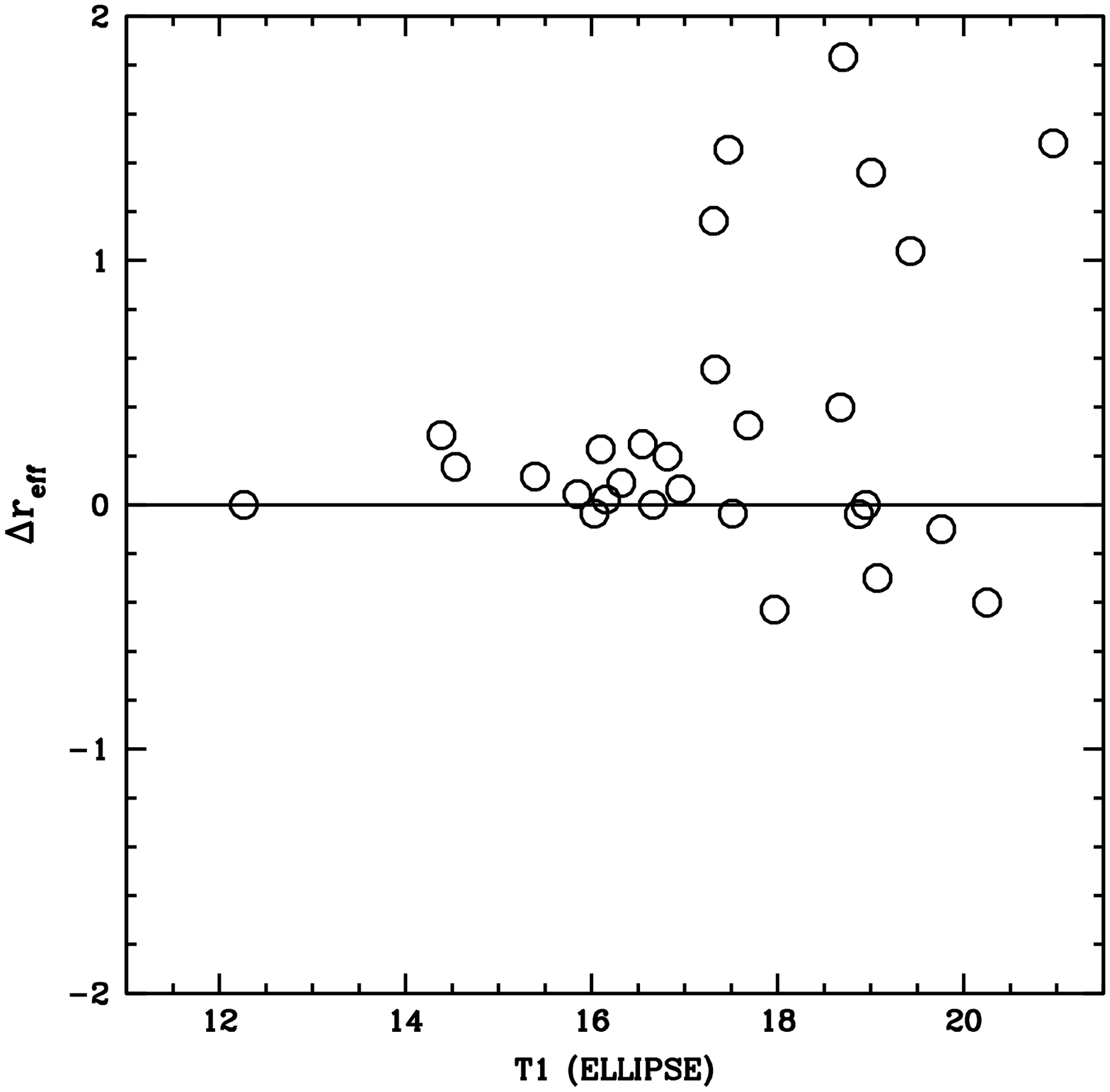}
\caption{Comparison between $T_1$ magnitudes (left), $(C-T_1)$ colours (center) 
and effective radii (right), obtained with ELLIPSE and SExtractor.}
\label{T1_ell_SEx}
\end{figure*}

In Paper\,I we have obtained the CMR from colours and magnitudes measured 
with SExtractor and with ELLIPSE. Therefore, it is interesting to quantify if 
there are significant differences between these two ways of obtaining 
photometry. To this aim, in Figures~\ref{T1_ell_SEx} we show plots comparing 
$T_1$ magnitudes, $(C-T_1)$ colours, and effective radii obtained with 
SExtractor and ELLIPSE.
 
In the left panel of Figure\,\ref{T1_ell_SEx}, we  see that for 
luminosities in the range $12\lesssim T_1 \lesssim 18$, magnitudes obtained 
from SExtractor and ELLIPSE agree well, the mean offset being 
 $|\Delta T_1|=0.08$. However, there are significant differences 
for $T_1 > 18$\,mag, that can be as large as 1.14 mag. The mean offset for 
faint galaxies is $|\Delta T_1|=0.44$ mag. In the case of $C$ magnitudes 
(no plot is shown), considerable offsets arise for $C > 19$\,mag. 
For magnitudes brighter than this limit we obtained $|\Delta C|=0.11$ and for  
fainter ones, $|\Delta C|=0.45$\,mag. 
 
In the central panel of Figure\,\ref{T1_ell_SEx}, we show the differences 
in $(C-T_1)$ colours as 
$\Delta (C-T_1) = (C-T_1)_{\rm SEx} - (C-T_1)_{\rm ELL}$ 
as a function of the ELLIPSE luminosity of the galaxy. For $T_1 > 16$\,mag, 
offsets as large as $|\Delta (C-T_1)|=0.4$ mag might arise for some objects: 
6 galaxies out of 30 (20 per cent) present $|\Delta (C-T_1)|>0.2$\,mag.

When effective radii are compared (righ panel of Figure\,\ref{T1_ell_SEx}), 
those obtained by SExtractor tend to be larger than those found by ELLIPSE. 
In the plot, we considered $\Delta r_{\rm eff}= r_{\rm eff}^{\rm SEx} - 
r_{\rm eff}^{\rm ELL}$. In particular, for $T_1 > 17$\,mag, the mean offset is  
$|\Delta r_{\rm eff}|=0.8$ arcsec and for $T_1 < 17$\,mag,  
$|\Delta r_{\rm eff}|=0.1$ arcsec. 
 
\begin{figure*}
\includegraphics[scale=0.43]{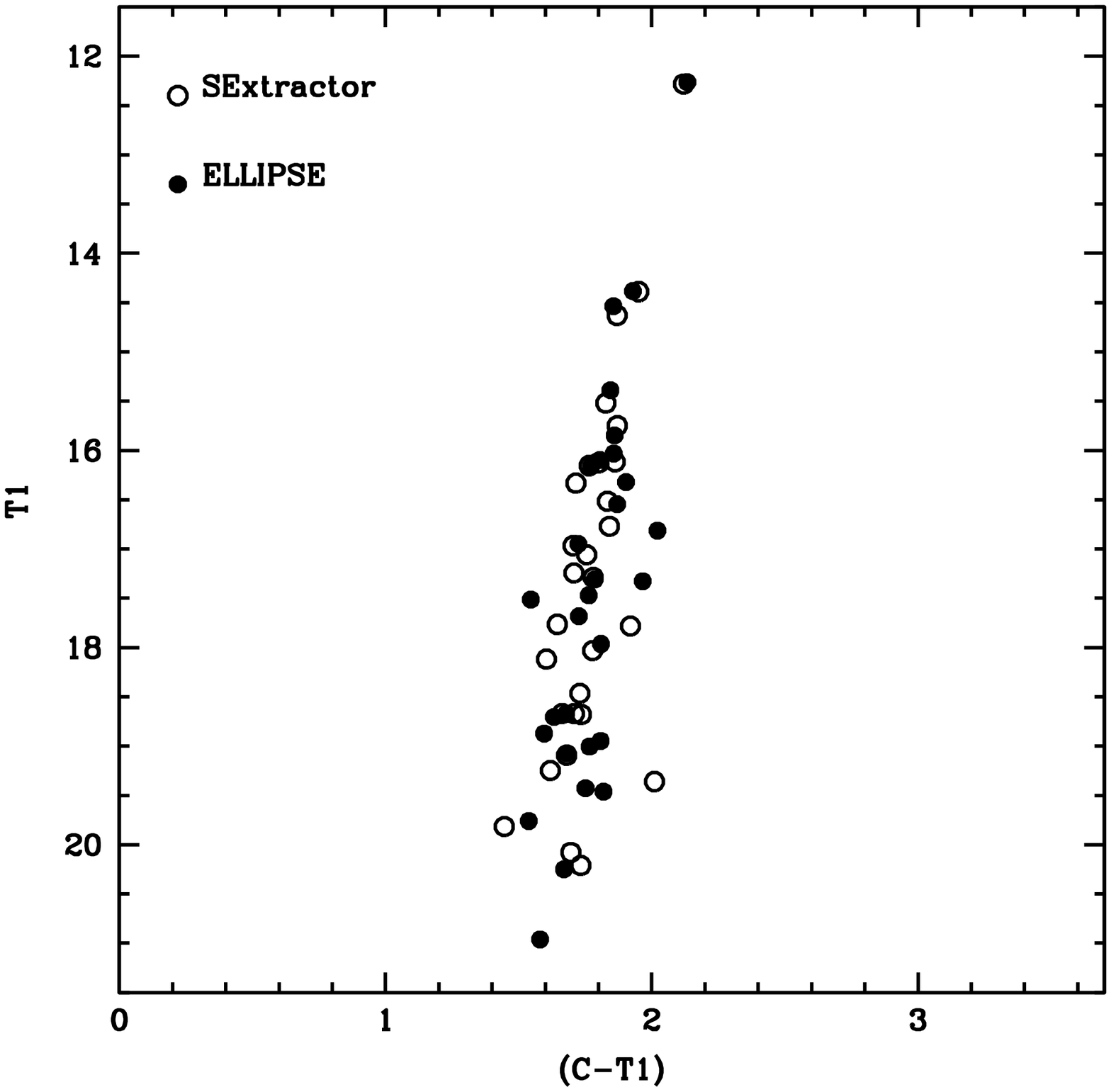}
\includegraphics[scale=0.43]{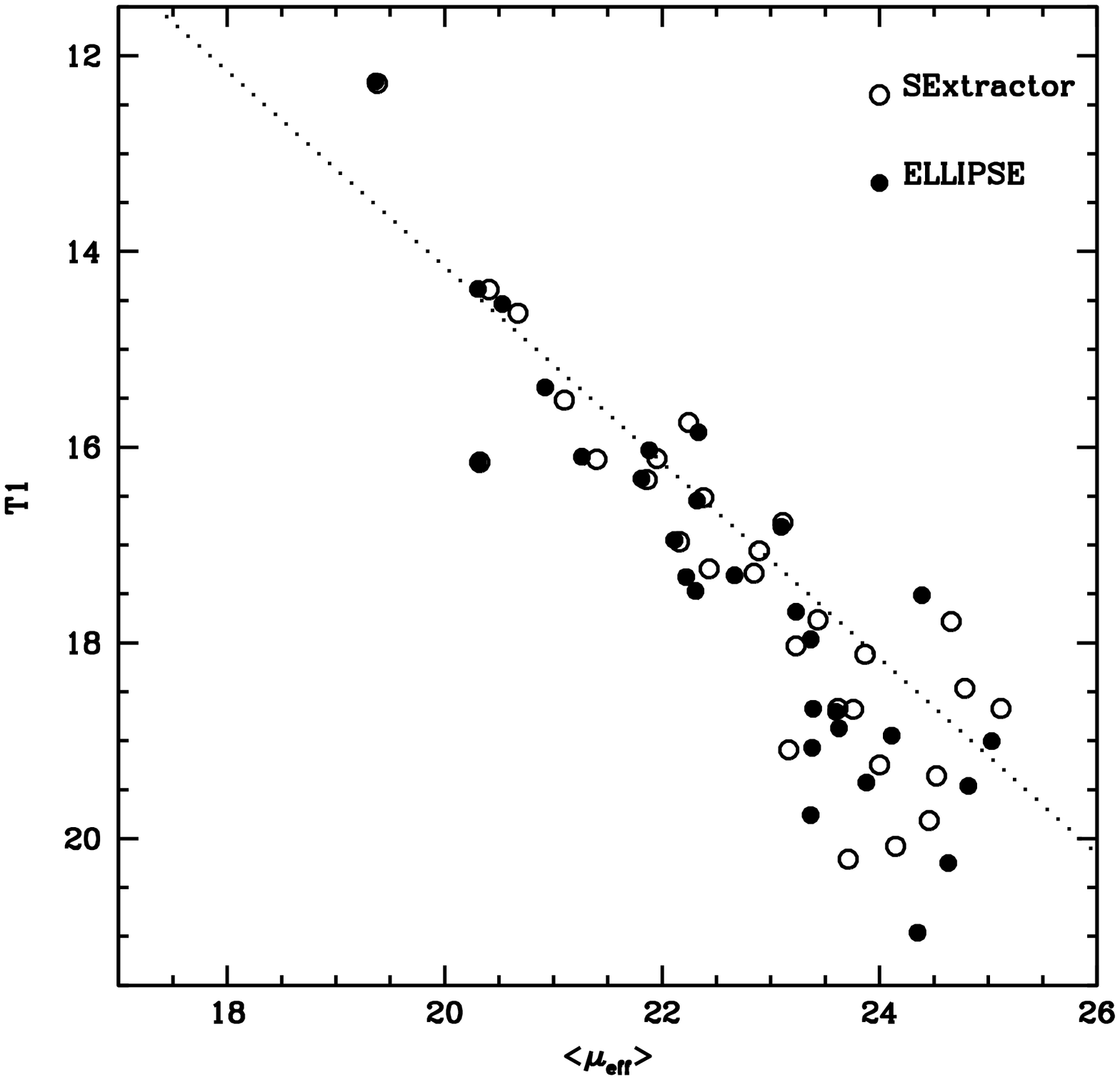}
\caption{Comparison between colour--magnitude and 
$\langle\mu_{\rm eff}\rangle$--luminosity relations obtained with ELLIPSE and 
SExtractor.}
\label{T1_ell_SEx2}
\end{figure*}

Despite these differences, we see from both panels of 
Figure\,\ref{T1_ell_SEx2} that our conclusions about the colour-magnitude 
and luminosity--$\langle\mu_{\rm eff}\rangle$ relations almost do not change by 
considering photometry obtained by one way or another. The slope and 
zero point of both CMRs are similar, and the locus of constant effective 
radius arise naturally in both samples. SExtractor photometry seems to 
introduce some additional scatter at the faint end of the relation, 
while ELLIPSE slightly increases the dispersion at intermediate brightness. 
The luminosity--$\mu_{\rm eff}$ relation presents lower scatter when the 
SExtractor photometry is considered. However, the break at the faint end 
of the relation is found in both samples. 
In the case of FS90\,192, one of the cE galaxies, it can be seen that its 
location away from the locus of constant effective radius towards higher 
values of $\langle\mu_{\rm eff}\rangle$, is similar in both cases. 
 
From our findings we conclude that SExtractor and ELLIPSE are suitable to 
study mean photometric relations with a comparable level of confidence. 
However, for individual studies of low brightness galaxies, some care must be 
taken when considering colours and structural parameters. 

\label{lastpage}

\end{document}